\documentclass{article}

\usepackage{arxiv}
\usepackage[authoryear]{natbib}
\newcommand{\citemaY}[1]{(Y.\ Ma et al.,\ \citeyear{#1})}

\usepackage{multirow}
\newcommand{\CI}[2]{\makecell{(#1,\\#2)}}
\usepackage{tabularx}

\usepackage{rotating}  
\usepackage{float} 
\usepackage{etoolbox}
\usepackage{tabularx}
\usepackage{makecell}
\usepackage{caption}
\usepackage[utf8]{inputenc} 
\usepackage[T1]{fontenc}    
\usepackage{hyperref}       
\usepackage{url}            
\usepackage{booktabs}       
\usepackage{amsfonts}       
\usepackage{nicefrac}       
\usepackage{microtype}      
\usepackage{lipsum}		
\usepackage{graphicx}

\usepackage{doi}
\usepackage{tikz} 
\usetikzlibrary{arrows.meta}
\usetikzlibrary{positioning} 
\usepackage{amsmath}
\usepackage{bm}
\usepackage{tabularx}

\title{A statistical framework for dynamic cognitive diagnosis in digital learning environments}

\date{} 					

\author{
  \href{https://orcid.org/0000-0001-7508-5385}{\includegraphics[scale=0.06]{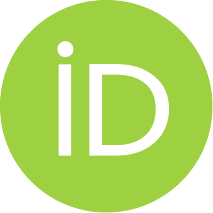}\hspace{1mm}Yawen Ma}\thanks{Corresponding author.} \\
  Centre for Health Informatics, Computing, and Statistics \\
  Lancaster Medical School, Lancaster University \\
  Lancaster, LA1 4YW, United Kingdom \\
  \texttt{y.ma24@lancaster.ac.uk} \\
  \And
  \href{https://orcid.org/0000-0002-0621-5032}{\includegraphics[scale=0.06]{orcid.pdf}\hspace{1mm}Anastasia Ushakova} \\
  Centre for Health Informatics, Computing, and Statistics \\
  Lancaster Medical School, Lancaster University \\
  Lancaster, LA1 4YW, United Kingdom \\
  \texttt{a.ushakova@lancaster.ac.uk} \\
  \And
  \href{https://orcid.org/0000-0003-2780-188X}{\includegraphics[scale=0.06]{orcid.pdf}\hspace{1mm}Kate Cain} \\
  Department of Psychology \\
  Lancaster University \\
  Lancaster, LA1 4YF, United Kingdom \\
  \texttt{k.cain@lancaster.ac.uk} \\
  \And
  \href{https://orcid.org/0000-0002-7930-6701}{\includegraphics[scale=0.06]{orcid.pdf}\hspace{1mm}Gabriel Wallin} \\
  School of Mathematical Sciences \\
  Lancaster University \\
  Lancaster, LA1 4YF, United Kingdom \\
  \texttt{g.wallin@lancaster.ac.uk} \\
}

\renewcommand{\undertitle}{}
\renewcommand{\shorttitle}{Dynamic CDMs in Digital Learning}
\renewcommand{\headeright}{JRSS-A}

\hypersetup{
pdftitle={Dynamic CDMs in Digital Learning},
pdfsubject={q-bio.NC, q-bio.QM},
pdfauthor={Y. Ma et al.},
pdfkeywords={Cognitive Diagnostic Models · Deterministic Input, Noisy “And” Gate (DINA) Models · Educational Game
Application · Log Files · Q-matrix Estimation},
}

\begin{document}
\maketitle

\begin{abstract}
Reading is foundational for educational, employment, and economic outcomes, but a persistent proportion of students globally struggle to develop adequate reading skills. Some countries promote digital tools to support reading development, alongside regular classroom instruction. Such tools generate rich log data capturing students' behaviour and performance. This study proposes a dynamic cognitive diagnostic modeling (CDM) framework based on restricted latent class models to trace students' time-varying skills mastery using log files from digital tools. Unlike traditional CDMs that require expert-defined skill-item mappings ($Q$-matrix), our approach jointly estimates the $Q$-matrix and latent skill profiles, integrates log-derived covariates (e.g., reattempts, response times, counts of mastered items) and individual characteristics, and models transitions in mastery using a Bayesian estimation approach. Applied to real-world data, the model demonstrates practical value in educational settings by effectively uncovering individual skill profiles and the skill-item mappings. Simulation studies confirm robust recovery of $Q$-matrix structures and latent profiles with high accuracy under varied sample sizes, item counts and different sparsity of $Q$-matrices. The framework offers a data-driven, time-dependent restricted latent class modeling approach to understanding early reading development.
\end{abstract}



\maketitle
\section{Introduction}

Early literacy is widely recognized as essential for educational success and lifelong development \citep{cree2023economic}. Yet, despite substantial investments in literacy education, recent data indicate persistent global challenges. For example, according to the Progress in International Reading Literacy Study, 86\% of {9- to 10-year-olds} in England reached the Intermediate International Benchmark, compared to 81\% in the United States, 37\% in Brazil, and only 9\% in South Africa, highlighting substantial global disparities in basic literacy achievement \citep{lindorff2024pirls, mullis2023pirls}. To address gaps in literacy skills, some countries advocate the integration of evidence-based digital reading support in classrooms, to support students with specific education needs and English language learners \citep{usdoe2015}. Building on the demonstrated benefits of individualised and real-time feedback in digital learning environment \citep{maier2022personalized}, this study addresses the critical need to empirically evaluate how digital reading tools can support early literacy development. Although these educational technologies are now widely adopted, relatively few studies have utilised the rich log files they generate to capture fine-grained learning processes. Such log files enable detailed tracking of student interactions, such as response times and learning trajectories, which in turn facilitates statistical modeling of students' learning development.

Many existing approaches to modeling students' learning development employ latent variable frameworks, in which unobservable proficiencies are inferred from observable item responses. A widely used class of such models in educational and psychological assessment is the restricted latent class models, also known as cognitive diagnostic models (CDMs), originally proposed by \cite{haertel1984application} and further {expanded} by \cite{templin2010diagnostic}. \cite{haertel1984application} initially referred to these models as binary skills models, which classify students into one of $2^K$ latent profiles based on their mastery or non-mastery of skills (also called “attributes”), where $K$ denotes the number of skills of interest. These models typically account for guessing (responding correctly by chance despite lacking mastery) and slipping (responding incorrectly despite possessing the skill) behaviours \citep{rupp2008unique, templin2010diagnostic}. CDMs provide diagnostic feedback for teachers to help them make informed decisions regarding targeted instruction or interventions \citep{zhan2018cognitive}. 

While CDMs and their variants have demonstrated effectiveness across diverse applications, traditional implementations face several limitations that researchers have sought to address. First, these models are typically restricted to static, single time point assessments. To overcome this constraint, researchers have developed various extensions to capture learning over time. Different approaches have integrated CDMs with transition models to track temporal skill development. For example, latent transition analysis (LTA; \citet{bray2010modeling}) combined with CDMs can assess shifts in mastery across repeated assessments \citep{kaya2017assessing}. Building on this framework, \citet{liang2023latent} introduced a bias-corrected three-step method for latent transition CDMs to evaluate covariate effects. Several studies have incorporated hidden Markov models with CDMs to track skill acquisition in computer-based spatial learning interventions \citep{wang2018tracking, wang2019joint, chen2018hidden, wang2020development}. More recently, \citet{wayman2025restricted} proposed a Bayesian longitudinal extension of restricted latent class models - structured as a directed graphical model \citep{murphy2012machine} - which accommodates polytomous attributes and allows covariates to influence transitions between latent states.

Second, CDMs incorporate a design matrix, known as the $Q$-matrix, which specifies the mapping between the test items and the attributes of the underlying skills. Each row of the $Q$-matrix corresponds to an item and each column to a skill, with entries of 1 indicating that the item requires the skill and 0 otherwise. Although the $Q$-matrix itself only defines the item-skill relationships, the complete CDM framework uses this structure to generate diagnostic inferences about individual skill mastery. However, a misspecified $Q$-matrix can significantly bias parameter estimates and diagnostic classifications \citep{rupp2008effects}. Many applications of CDMs rely on a predefined $Q$-matrix provided by domain experts \citep{wang2018tracking, zhan2018cognitive}, or focus on validating and modifying pre-specified expert $Q$-matrices (Chiu et al., \citeyear{chiu2009cluster}; de la Torre, \citeyear{delatorre2011}; W.~Ma and de la Torre, \citeyear{ma2020empirical}). To overcome the limitations of such confirmatory approaches, researchers have developed data-driven, exploratory estimation methods. For instance, \cite{chen2015statistical} proposed a regularised estimator for the $Q$-matrix in a cross-sectional setting, whereas \cite{chen2018bayesian} developed a Bayesian approach. \cite{wayman2025restricted} compared their longitudinal model's $Q$-matrix estimates to a prior confirmatory analysis by \cite{tang2021does}. Prior work has furthermore established identifiability conditions for the $Q$-matrix \citep{chen2015statistical, xu2018identifying, chen2018bayesian}.

Third, most CDMs assume that students' attributes are binary, which may be overly simplistic for domains such as reading development, where a broader range of proficiency levels are necessary to sensitively capture ability. Polytomous skill models have been developed to address this limitation, incorporating partial response accuracy \citep{chen2013general, xu2025polytomous, zhan2020partial} or jointly modeling response accuracy and response times to account for accuracy-speed trade-offs \citep{wang2020using, liu2024mixture, liu2025general}. These polytomous approaches have been subsequently adapted to dynamic frameworks to track learning progression over time.

Finally, while there is growing interest in integrating log files into CDMs, most existing applications either target different populations or use alternative modeling approaches. For example, \citet{foldnes2024school} employed machine learning techniques on gameplay data to detect reading difficulties. Other studies have explored process data in university-level reading subskills using CDMs \citep{chen2023investigating}, or adult problem-solving skill within technology rich environments \citep{rajeb2023incorporating}. Although these studies illustrated the value of log files, the application of CDMs to log files in early literacy development remains limited. Yet, log files provide opportunities to track not only response correctness but also dynamic insights into time-related learning processes, such as response times, item durations, and session timestamps. {Thus, the application of CDMs to log file data provides a unique opportunity to understand learning processes at a finer temporal grain size than previously studied.}

Our approach in this study was motivated by {access to} digital learning log files, which included multiple time-points, an unknown $Q$-matrix, log-derived behaviour indicators and individual characteristics for each student. This resulted in a novel framework for time dependent restricted latent class modeling that integrated the various extensions discussed above into a single unified approach. While previous research has addressed individual aspects, such as temporal dynamics, data-driven $Q$-matrix estimation, or integration of log-derived covariates and individual characteristics, our model synthesised these advances within one comprehensive framework. We jointly estimated all components, including the item-skill relationship ($Q$-matrix), time-varying latent skill profiles, and transition parameters, within a single integrated system. The $Q$-matrix was thus inferred directly from the data, based on the dependence structure among item responses rather than assuming that it was known. Simultaneously, the framework modeled the dynamic evolution of skills over time while incorporating rich log-derived behavioural indicators and individual characteristics, including reattempts, response times, counts of mastered items, demographics, and game and learning characteristics, leveraging the full information available in digital learning environments.

The remainder of this paper is structured as follows. Section 2 provides the background and description of the digital learning data used in our analysis. Section 3 details our methodological framework, including the full specification of the time-dependent restricted latent class model {with a model generalisation section that explicitly state the applicability to diverse datasets}. In Section 4, we present an empirical study applying our model to real-world educational data, and Section 5 evaluates the model's performance through simulation studies. Section 6 discusses the implications of our findings, limitations, and directions for future research. Finally, Section 7 provides a summary of our contributions and their significance for both statistical methodology and educational practice.

\section{Data Background}
\begin{figure}
\centering
\caption{The hierarchical structure of the log files. The left column shows the full structure of Boost Reading (skill families, games, levels, and attempts). The right column highlights the subset selected for analysis, including two skill families, one game from each, and relevant levels and attempts.}
\begin{tikzpicture}[
    node distance = 8mm and 30mm,
    cat/.style = {
        rectangle,
        draw,
        rounded corners,
        minimum width = 13em,
        minimum height = 3em,
        align = center,
        font = \small
    },
    var/.style = {
        rectangle,
        draw,
        rounded corners,
        minimum width = 7em,
        minimum height = 2.2em,
        align = center,
        font = \footnotesize
    },
    line/.style = {-{Latex[length=2mm]}}
]
{
\node[font=\bfseries\large] (headerL) {Full structure};
\node[font=\bfseries\large, right=of headerL, xshift=6mm] (headerR) {Subset for analysis};
\draw[line] (headerL.south) -- ++(0,-6mm);
\draw[line] (headerR.south) -- ++(0,-6mm);
}

\node[cat, below=of headerL] (skills)    {11 Skill families\\\scriptsize(Amplify-defined reading-related skills)};
\node[cat, below=of skills]     (games)   {48 Games};
\node[cat, below=of games]      (levels)  {Levels\\\scriptsize(Varies by games)};
\node[cat, below=of levels]     (attempts){Attempts\\\scriptsize(Varied by students)};

\draw[line] (skills.south)   -- (games.north);
\draw[line] (games.south)    -- (levels.north);
\draw[line] (levels.south)   -- (attempts.north);

\node[var, right=of skills]   (varS) {Decoding and vocabulary};
\draw[line] (skills.east) -- (varS.west);

\node[var, right=of games]    (varG) {2 games selected: \\ One decoding and one vocabulary};
\draw[line] (games.east)  -- (varG.west);

\node[var, right=of levels, yshift=7mm]  (varL1) {Decoding game: 69 levels\\
                                                  Vocabulary game: 36 levels};
\node[var, right=of levels, yshift=-5mm] (varL2) {Log-based variables\\\scriptsize(Mastery, attempts, total time across attempts)};
\draw[line] (levels.east) -- (varL1.west);
\draw[line] (levels.east) -- (varL2.west);

\node[var, right=of attempts, yshift=5mm] (varA) {3 consecutive attempts};
\node[var, right=of attempts, yshift=-7mm] (varB) {Log-based variables\\\scriptsize(Scores, elapsed time)};
\draw[line] (attempts.east) -- (varA.west);
\draw[line] (attempts.east) -- (varB.west);
    \label{fig:0}
\end{tikzpicture}
\end{figure}
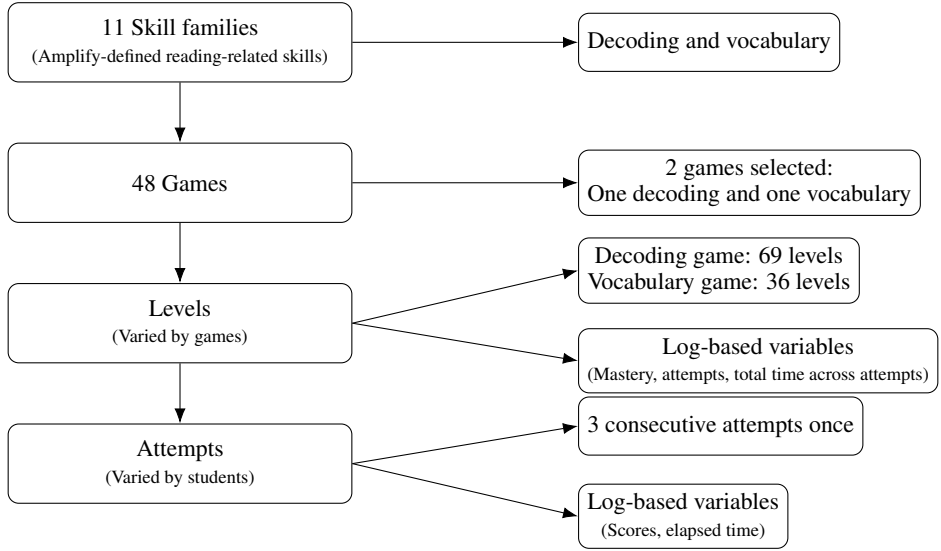

The data for this study originate from the Boost Reading digital program, developed by Amplify, a U.S.-based education technology company (\url{https://amplify.com}). Founded in 2000, Amplify now reaches more than 5,000 school districts, serving over 15 million students in 2024. Amplify offers a wide range of curriculum programmes in literacy (Boost Reading is one of them), science, and mathematics, providing schools and educators with digital tools to support effective teaching and learning. Boost Reading (previously Amplify Reading) is a research-informed classroom-based digital reading supplement consisting of multiple literacy-focused games targeting core reading skills such as phonological awareness, decoding, vocabulary, and sentence comprehension. These games are categorised into distinct research-informed defined skill families, each aligned with a foundational reading skill, specifically designed for students in kindergarten through 5th grade (K-5). The diagram in Figure \ref{fig:0} summarises the hierarchical structure of the log files from Boost. The left column displays the complete structure (11 skill families, 48 games, levels, attempts), while the right column specifies the subset analysed in this paper, focusing on decoding and vocabulary skill families and one game from each. The {positive} impact of Boost games and student interaction in K–2 is reported in \citet{newton2019examining}.

Within each skill family, multiple games include different levels. Students achieve either ``mastery'' or ``no mastery'' at each level, where mastery is defined as achieving approximately 80\% accuracy on the items they engage with. {There are multiple games to practice each skill, each game comprises a different number of levels.} Students may attempt each level an unlimited number of times. However, after three consecutive non-mastery attempts for a level {in any given game} students are directed to other games to support related skills before they reattempt the challenging levels of that game. Students do not choose which game or level to engage with; instead, access to games and levels is determined by their initial ability and on-going in-game performance. Within each game, levels are unlocked sequentially, progressing from foundational to advanced levels. As a result, students follow different learning trajectories and encounter varying sets of levels. In addition to detailed log files for specific games, we also obtained students’ initial reading performance measured by the Dynamic Indicators of Basic Early Literacy Skills (DIBELS; \citet{universityoforegon2018dibels}), administered to children prior to starting the Boost program. {These scores were used for initial placement.}

\section{Methodological Framework}

Using the data structure outlined above, we developed a comprehensive framework for analysing student learning trajectories in digital environments. Our approach addressed the limitations of traditional CDMs by jointly estimating three key components: (1) the underlying item–skill mappings ($Q$-matrix), (2) item-specific parameters and individual skill profiles across time points, and (3) the effects of log-derived covariates and individual characteristics on both initial skills mastery and their transitions.

\subsection{Model Overview}

The proposed model integrated cognitive diagnostic measurement with temporal dynamics in analysing how students' proficiency in specific skills evolved over time. At each time point $t$, student $i$'s observed responses to learning items were denoted by {$\mathbf{Y}_{i,t} = (Y_{i,1,t},\ldots,Y_{i,J,t})^\top$}, where $Y_{i,j,t} \in \{0,1\}$ indicated an incorrect or correct response to item $j$. These observed responses were governed by latent skill mastery patterns and item-specific parameters, as detailed in the following subsections.

\subsection{Cognitive Diagnostic Model}

CDMs are restricted latent class models designed to infer students' mastery of specific skills from their observed responses. Under these models, each student $i$ at time $t$ is characterised by a binary attribute vector $\bm{\alpha}_{i,t} = (\alpha_{i,1,t}, \dots, \alpha_{i,K,t})$, where $\alpha_{i,k,t} \in \{0,1\}$ indicates mastery (1) or non-mastery (0) of attribute $k$.

Among various CDMs, we employed the Deterministic Inputs, Noisy {``AND"} Gate model (DINA; \citet{haertel1984application, junker2001cognitive}). The DINA model assumes a non-compensatory structure, meaning that students must master all required attributes to successfully complete an item. {The binary entry $Q_{jkt}$ of the $Q$-matrix ($\mathbf{Q}$) specifies whether item $j$ requires skill $k$ at time $t$.} Formally, the ideal response indicator $\eta_{i,j,t}$ for student $i$ on item $j$ at time $t$ is given by:
{
\begin{equation}
\label{eq:eta-dina}
\eta_{i,j,t} = \prod_{k=1}^{K} \alpha_{i,k,t}^{Q_{jkt}},
\end{equation}
}
where $\eta_{i,j,t}=1$ indicates full mastery of all required attributes, and $\eta_{i,j,t}=0$ otherwise. 

To account for response uncertainty, the DINA model incorporates two parameters: slipping ($s_{j,t}$) and guessing ($g_{j,t}$). The probability of a correct response is modeled as:
\begin{align}
\label{eq:dina-prob}
& P\bigl(Y_{i,j,t}=1 \mid \bm{\alpha}_{i,t}, \mathbf{Q}\bigr) 
= (1 - s_{j,t})^{\eta_{i,j,t}} g_{j,t}^{1 - \eta_{i,j,t}},  \\ 
& s_{j,t} = P(Y_{i,j,t} = 0 \mid \eta_{i,j,t} = 1), \quad
g_{j,t} = P(Y_{i,j,t} = 1 \mid \eta_{i,j,t} = 0). \label{eq:gs}
\end{align}

Thus, a student who has mastered all required attributes may still respond incorrectly due to slipping ($s_{j,t}$), while a student lacking mastery may respond correctly by guessing ($g_{j,t}$). We selected the DINA model for its straightforward interpretations, smaller sample size requirements for accurate parameter estimation \citep{rojas2012choosing}, and its flexibility for extension to more general CDMs.

\subsection{The $Q$-Matrix}

A critical component of our framework is the $Q$-matrix, a binary matrix that specifies which skills are required for each item:
{
\begin{align}
\label{eq:q_jkt}
Q_{jkt} =
\begin{cases}
1 & \text{if item $j$ requires skill $k$ at time point $t$}, \\
0 & \text{otherwise}.
\end{cases}
\end{align}}

Unlike many applications where this structure is pre-specified by domain experts, in Boost Reading, the mapping between items and skills \footnote{Here, ``skill" is used differently from the definition used in Boost Reading.} is not explicitly defined. We therefore treated each element {$Q_{jkt}$} as a parameter to be estimated from the data. Table \ref{tab:qmatrix_combined_gs} presents an example of an estimated $Q$-matrix from our application. The structure of the $Q$-matrix plays a critical role in model identifiability \citep{fang2019identifiability}. {To ensure identifiability during sampling, we implemented a custom row-wise Gibbs sampler for $Q$-matrix estimation in practice. For example, for $K=3$, each row was proposed from the $2^3 - 1 = 7$ non-zero attribute patterns. At each update, a candidate row was generated and the resulting $Q$-matrix was evaluated against the necessary identifiability conditions under the DINA model. The conditions require each item to measure at least one attribute, and each attribute to be measured by at least three items \citep{chiu2009cluster, gu2021sufficient}. A proposal was accepted only if these conditions were satisfied, otherwise, it was rejected. The MCMC chains were initialized with $Q$-matrices containing two identity submatrices to satisfy the basic sufficiency conditions. After initialization, all entries were freely updated. Consequently, all retained posterior draws correspond to identifiable $Q$-matrices. All $J \times K$ entries of the time-specific $Q$-matrices were treated as free parameters and updated throughout the Markov chain Monte Carlo (MCMC) iterations.}

\subsection{Transition Model with Covariates}

An important aspect of the modeling framework is how students transition between mastery states over time, and how these transitions are influenced by learning behaviours captured in log files and individual characteristics. We employed logistic regression models to link covariates $Z$ to both initial skill mastery and transitions between states.

The initial attribute mastery probabilities at the first time point are modeled as:
\begin{align}
\text{logit}\left(P(\alpha_{i,k,t=1} = 1)\right) &= {\beta_{k,0}} + \sum_{c=1}^C \beta_{k,c} Z_{i,c}, \label{eq:beta}
\end{align}
where {$\beta_{k,0}$} represents the baseline log-odds of mastering attribute $k$ initially, $\beta_{k,c}$ quantifies the effect of covariate $c$ on this probability, and $Z_{i,c}$ denotes the value of covariate $c$ for student $i$. For subsequent time points, we modeled transitions between latent attribute states from time $t$ to $t+1$ as:
\begin{align}
\text{logit}\left(P(\alpha_{i,k,t+1} = 1 \mid \alpha_{i,k,t} = 0)\right) &= \gamma_{01,k,0} + \sum_{c=1}^C \gamma_{01,k,c} Z_{i,c}, \label{eq:gamma01} \\
\text{logit}\left(P(\alpha_{i,k,t+1} = 0 \mid \alpha_{i,k,t} = 1)\right) &= \gamma_{10,k,0} + \sum_{c=1}^C \gamma_{10,k,c} Z_{i,c}, \label{eq:gamma10}
\end{align}

Here, $\gamma_{01,k,0}$ and $\gamma_{01,k,c}$ quantify the baseline and covariate effects on transitioning from non-mastery to mastery of attribute $k$, while $\gamma_{10,k,0}$ and $\gamma_{10,k,c}$ quantify the effects on transitions from mastery back to non-mastery. By incorporating log-derived covariates and individual characteristics, our model was able to identify which specific covariates were most predictive of skill acquisition and retention. {For the $\gamma_{10,k,0}$ and $\gamma_{10,k,c}$ parameters, which correspond to transitions from mastery to non-mastery, we imposed a soft monotonicity constraint to reflect the substantive assumption that learning is largely cumulative and that mastered attributes are unlikely to be unlearned within a short period. This assumption is substantively plausible in many educational domains, such as reading development, where foundational skills support later comprehension-related skills \citep{hoover1990simple}. It also provides a practically useful structural restriction by reducing the number of possible learning trajectories and simplifying the transition states \citep{chen2020multivariate}, and is commonly adopted in educational applications of longitudinal cognitive diagnosis \citep{liang2023, Yigit2021FirstOrder}. In practice, we specified more informative priors for the {intercept parameters $\gamma_{10,k,0}$, centred} at $-2$ on the logit scale, corresponding to a low baseline probability of skill loss, while retaining weakly informative priors centred at zero for covariate effects. We adopt the monotonicity assumption as a theoretically motivated modelling choice to improve interpretability. Importantly, the proposed framework can be estimated without imposing monotonicity.}

\begin{figure}[http]
    \centering
    \caption{{The relationships between covariates $Z$, latent variables $\alpha$, and responses $Y$ across three time points.}}
\label{fig:model_diagram}
\begin{tikzpicture}[>=stealth, node distance=2cm, thick]
{
    \node[draw, rectangle] (Z) at (0,2)   {$Z$};
    \node[draw, circle]    (Lt-1) at (-4,0)   {$\alpha(t-1)$};
    \node[draw, circle]    (Lt) at (0,0)   {$\alpha(t)$};
    \node[draw, circle]    (Lt+1) at (4,0)   {$\alpha(t+1)$};
    \node[draw, rectangle] (Yt) at (0,-2) {$Y(t)$};
    \node[draw, rectangle] (Yt+1) at (4,-2) {$Y(t+1)$};
    \node[draw, rectangle] (Yt-1) at (-4, -2) {$Y(t-1)$};
    \node[draw, circle] (Qt-1) at (-4,-4) {$Q(t-1)$ };
    \node[draw, circle] (Qt) at (0,-4) {$Q(t)$};
    \node[draw, circle] (Qt+1) at (4,-4) {$Q(t+1)$};
    \draw[->] (Lt-1) -- node[midway, above] (Pt) {$P(t{-}1 \to t)$} (Lt);
    \draw[->] (Z) -- (Pt);
    \draw[->] (Lt) -- node[midway, above] (Pt+1) {$P(t \to t{+}1)$} (Lt+1);
    \draw[->] (Z) -- (Pt+1);
    \draw[->] (Lt-1)  -- (Lt);
    \draw[->] (Lt)  -- (Lt+1);
    \draw[->] (Lt-1) -- (Yt-1);
    \draw[->] (Lt) -- (Yt);
    \draw[->] (Lt+1) -- (Yt+1);
    \draw[->] (Yt) -- (Yt+1);
    \draw[->] (Yt-1) -- (Yt);
    \draw[dashed, ->] (Qt-1) -- (Yt-1);
    \draw[dashed, ->] (Qt) -- (Yt);
    \draw[dashed, ->] (Qt+1) -- (Yt+1);
   }
\end{tikzpicture}
\end{figure}
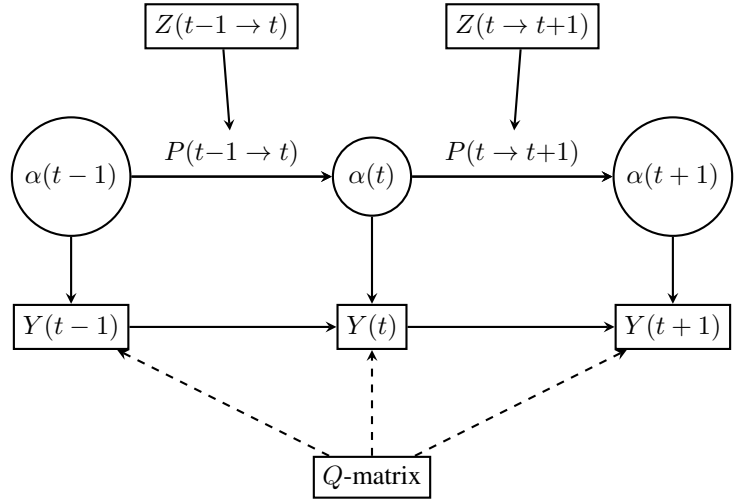

The relationships between covariates $Z$, latent attributes $\alpha$, and observed responses $Y$ across time points are summarised in Figure \ref{fig:model_diagram}.

\subsection{Inference Procedure}

To estimate all model parameters simultaneously, we employed a Bayesian approach using MCMC. This enabled us to quantify uncertainty in all parameters while incorporating prior knowledge where available.

\subsubsection{Prior Specifications}

To encourage sparsity in the $Q$-matrix, we adopted a hierarchical Bernoulli–Beta prior:
{
\begin{align}
Q_{jkt} &\sim \mathrm{Bernoulli}(\theta),\label{eq:$Q$-matrix} \\ 
\theta &\sim \mathrm{Beta}(\alpha,\beta), \nonumber
\end{align}
}
where the prior mean $\frac{\alpha}{\alpha+\beta}$ equals the non–zero proportion of the true $Q$-matrix and the concentration defined as $\alpha+\beta$. This structure permits the data to inform the overall sparsity level. For the empirical analysis, the $\mathrm{Beta}(24,6)$ prior had a mean of $0.8$ and a variance of approximately $0.0052$, placing roughly two-thirds of its mass between 0.65 and 0.92. Additionally, we conducted a sensitivity analysis with alternative priors reported in supplementary material A (posterior means of $g$ and $s$ differed by less than 0.04), finding that $\mathrm{Beta}(24,6)$ provided the best balance between model fit and interpretability of the resulting $Q$-matrix structure. This prior is informative enough to guide estimation toward meaningful structures while remaining flexible enough to adapt to the data. 

For the guessing and slipping parameters, we used {non-informative priors (flat priors)} following \citet{wang2018ho_hmm}:
\begin{align}
g_{j,t} &\sim \text{Beta}(1, 1) \label{eq:g}\\
s_{j,t} &\sim \text{Beta}(1, 1) \label{eq:s}
\end{align}

We initialised these parameters with draws from $\mathrm{Uniform}(0, 0.3)$ to reflect the empirical observation that item-level guessing and slipping parameters rarely exceed 0.30 in applied settings \citep{zhang2018modeling,culpepper2016revisiting}. {For the regression coefficients in the attribute and transition models in Equations \eqref{eq:beta}, \eqref{eq:gamma01} and \eqref{eq:gamma10}, we assigned normal priors $N(0,1)$ to {$\beta_{k,0}$}, $\beta_{k,c}$, $\gamma_{01,k,0}$, $\gamma_{01,k,c}$ and $\gamma_{10,k,c}$. For the transition intercept parameter $\gamma_{10,k,0}$ in Equation \eqref{eq:gamma10}, we specified a more informative prior centred at $-2$ on the logit scale, reflecting the absorbing learning assumption}. Prior to analysis, all continuous covariates were standardized to have a mean of zero and a standard deviation of one. In order to test the sensitivity of the choice of priors, we varied them and re-ran the analysis. In doing so, we found little difference in both classification performance and posterior densities of hyperparameters.

\subsubsection{Joint Posterior Distribution}

The joint posterior distribution of all parameters given the observed data is:
{
\begin{align}
& P(\mathbf{Q}, \mathbf{g}, \mathbf{s}, \boldsymbol{\beta}, \boldsymbol{\gamma_{01}}, \boldsymbol{\gamma_{10}}, \boldsymbol{\alpha}_{1:T} \mid \mathbf{Y}_{1:T}, \mathbf{Z}) \nonumber\\   \propto 
& \Big(\prod_{t=1}^{T}\prod_{j=1}^{J} \prod_{k=1}^{K} P(\mathbf{Y}_t \mid \mathbf{Q}_{jkt}, \mathbf{g}_{j,t}, \mathbf{s}_{j,t}, \mathbf{\alpha}_t)\Big) \cdot \notag
 \Big(\prod_{t=2}^T 
       P(\mathbf{\alpha}_t \mid \mathbf{\alpha}_{t-1}, \gamma_{01,k,c}, \gamma_{10,k,c}, \mathbf{Z})\Big) \cdot \notag \nonumber\\ 
&  P(\mathbf{\alpha}_1 \mid {\boldsymbol{\beta}}, \mathbf{Z}) \cdot P(\mathbf{Q}_{jkt}) \cdot 
  P(\mathbf{g}_{j,t}, \mathbf{s}_{j,t}) \cdot P({\boldsymbol{\beta}}, \gamma_{01,k,c}, \gamma_{10,k,c}) \label{eq:inference}
\end{align}
}

{The vectors $\mathbf{g}$ and $\mathbf{s}$ denote the item-level guessing and slipping parameters, respectively, with elements of the vectors defined in Equations \eqref{eq:g} and \eqref{eq:s}. {The coefficient vector ($\boldsymbol{\beta}$ = ($\beta_{k,0}$, $\beta_{k,c}$)) governs the initial mastery probability for attribute (k) at (t=1) through the logistic regression in Equation \eqref{eq:beta}, where ($\beta_{k,0}$) denotes the intercept and ($\beta_{k,c}$) denotes the vector of covariate effects.} The coefficient vector $\boldsymbol{\beta}$ governs the initial mastery probabilities at $t=1$ through the logistic regression in Equation \eqref{eq:beta}. The vectors $\boldsymbol{\gamma_{01}}$ and $\boldsymbol{\gamma_{10}}$ represent the regression coefficients for transitions from non-mastery to mastery and from mastery to non-mastery, respectively, with the elements defined in Equations \eqref{eq:gamma01} and \eqref{eq:gamma10}. The latent attribute vectors $\boldsymbol{\alpha}_{1:T} = \{\boldsymbol{\alpha}_1, \dots, \boldsymbol{\alpha}_T\}$ denote the time-varying attribute profiles. The response vectors $\boldsymbol{Y}_{1:T} = \{\boldsymbol{Y}_1, \dots, \boldsymbol{Y}_T\}$ denote the item response data for all individuals across time points. Finally, time-invariant $\mathbf{Z}$ represents the covariates included in both the initial mastery and transition models.} The component, {$P(\mathbf{Y}_t \mid \mathbf{Q}_{jkt}, \mathbf{g}_{jt}, \mathbf{s}_{jt}, \mathbf{\alpha}_t)$}, represents the likelihood of observed responses at time $t$ based on the DINA model defined in Equation \ref{eq:dina-prob}. The second component, $P(\mathbf{\alpha}_t \mid \mathbf{\alpha}_{t-1}, {\boldsymbol{\gamma_{01}}}, {\boldsymbol{\gamma_{10}}}, \mathbf{Z})$, captures the transition probabilities between attribute states from time $t-1$ to $t$ as defined in Equations \ref{eq:gamma01} and \ref{eq:gamma10}. The third component, $P(\mathbf{\alpha}_1 \mid {\boldsymbol{\beta}}, \mathbf{Z})$, models the initial attribute mastery probabilities at time $t=1$ using the logistic regression in Equation \ref{eq:beta}. The remaining terms, {$P(\mathbf{Q}_{jkt})$, $P(\mathbf{g}_{j,t}, \mathbf{s}_{j,t})$, and $P({\boldsymbol{\beta}}, \gamma_{01,k,c}, \gamma_{10,k,c})$}, represent the prior distributions for the $Q$-matrix, the guessing and slipping parameters, and the regression coefficients, respectively, as specified in the previous section.

\subsubsection{Computation}

{To maintain identifiability during sampling, we implemented a custom row-wise Gibbs sampler in the \texttt{nimble} R package \citep{deValpine2017NIMBLE, deValpine2026nimble}.} For each model, we ran three parallel chains with different starting values. The complete code for reproducing all results in this study is available on GitHub\footnote{\url{https://github.com/Yawen-Ma/$Q$-matrix.git}}. Additionally, the study was preregistered via the Open Science Framework at \url{https://osf.io/5mv3r}.

{
\subsection{Model Generalisation}
The proposed dynamic CDM framework can be adapted to large-scale educational assessment datasets. We outline several extensions that enhance the model's flexibility and applicability, including relaxing the constraints on the $Q$-matrix, extending the measurement model from DINA to more general CDMS (e.g., generalized DINA), enriching the transition model to incorporate action sequences, and examining measurement invariance through differential item functioning.
\subsubsection{Known and time-invariant $Q$-matrix}
In many large-scale assessments, such as the Trends in International Mathematics and Science Study (TIMSS; \citealp{mullis2023}), the Programme for International Student Assessment (PISA; \citealp{oecd2023}), and the Programme for the International Assessment of Adult Competencies (PIAAC; \citealp{oecd2013}), the item-skill mapping is pre-specified and refined through pilot studies \citep{andersson2022, vondavier2019}. In such settings, the $Q$-matrix can be treated as known and time-invariant. Formally, the model replaces the time-varying structure $\mathbf{Q}_{jkt}$ with a fixed matrix $\mathbf{Q}_{jk}$ in Equation \eqref{eq:q_jkt}, where
\begin{align*}
\mathbf{Q}_{jk} =
\begin{cases}
1 & \text{if item $j$ requires skill $k$}, \\
0 & \text{otherwise}.
\end{cases}
\end{align*}
Equation \eqref{eq:$Q$-matrix} is no longer needed and the likelihood in \eqref{eq:inference} remains unchanged, but the joint posterior simplifies because $\mathbf{Q}_{jkt}$ is no longer sampled. When the $Q$-matrix is fixed, the joint posterior excludes  $\mathbf{Q}_{jkt}$, reducing the dimension of the parameter space and improving estimation {stability}. Under this setting, the covariate effect $\boldsymbol{\beta}$ and the transition parameter $\boldsymbol{\gamma_{01}}$ indicates high recovery across sample sizes in Supplementary Material E. This suggests that the part of the regression bias observed in the unknown $Q$-matrix setting originates from the uncertainty in $Q$-matrix estimation.}

{
\subsubsection{Partially fixed $Q$-matrix}
In some settings, a subset of rows of the $Q$-matrix is pre-specified based on expert knowledge. These rows can be treated as fixed and denoted by $Q^{\mathrm{fix}}$. Formally, for item $j \in \mathcal{S}_{\mathrm{fix}}$ and skill $k$,
\begin{align*}
Q^{\mathrm{fix}}_{jk} =
\begin{cases}
1 & \text{if item $j$ requires skill $k$},\\[3pt]
0 & \text{otherwise}.
\end{cases}
\end{align*}
The remaining rows, denoted by $Q^{\mathrm{\star}}_{jk}$ for $j \in \mathcal{S}_{\mathrm{free}}$, retain the Bernoulli--Beta prior,
\begin{align*}
Q^{\mathrm{\star}}_{jk} &\sim \mathrm{Bernoulli}(\theta),\\
\theta &\sim \mathrm{Beta}(\alpha,\beta),
\end{align*}
with
$$
\mathcal{S}_{\mathrm{fix}} \cup \mathcal{S}_{\mathrm{free}} = \{1,\dots,J\},\qquad
\mathcal{S}_{\mathrm{fix}} \cap \mathcal{S}_{\mathrm{free}} = \emptyset.
$$
This specification incorporates pre-defined constraints through $Q^{\mathrm{fix}}$ while allowing data-driven refinement of the unspecified $Q^{\mathrm{\star}}$ through posterior sampling.}

{
\subsubsection{Extension to generalized DINA models}
The proposed dynamic CDM framework can be extended to the generalized DINA (GDINA) model \citep{delatorre2011}. Unlike the DINA model, which assumes a non-compensatory conjunctive structure, the GDINA model specifies a saturated linear predictor for the ideal response probability. For item $j$ requiring a subset of attributes, the conditional success probability is
\begin{align*}
P(Y_{ij}=1 \mid \boldsymbol{\alpha}_i) 
&= \lambda_{j,0} 
+ \sum_{k \in \mathcal{K}_j} \lambda_{j,k} \alpha_{i,k}
+ \sum_{k<k'} \lambda_{j,kk'} \alpha_{i,k} \alpha_{i,k'}
+ \cdots
+ \lambda_{j,1\cdots K_j} \prod_{k \in \mathcal{K}_j} \alpha_{i,k},
\end{align*} where $\lambda_{j,0}$ represents the intercept (the probability of success when none of the required attributes are mastered); $\lambda_{j,k}$ is the main effect due to the mastery of attribute $k$; $\lambda_{j,kk'}$ is the interaction effect representing the change in success probability due to the simultaneous mastery of attributes $k$ and $k'$; $\lambda_{j,1 \dots K_j}$ is the highest-order interaction effect of all $K_j$ attributes and $K_j = |\mathcal{K}_j|$. Under the GDINA framework, the DINA and Deterministic Input, Noisy ``OR" Gate (DINO) models arise as special cases under specific constraints on $\lambda_{j,\cdot}$. The DINA model imposes a non-compensatory conjunctive structure such that only the highest-order interaction term is retained:
\begin{align*}
\lambda_{j,k}=0,\quad
\lambda_{j,kk'}=0,\quad
\dots,\quad
\lambda_{j,1\dots K_j}=\delta_j,
\end{align*}
thus
\begin{align*}
P(Y_{ij}=1\mid \boldsymbol{\alpha}_i)=
\begin{cases}
\lambda_{j,0}, & \prod_{k\in\mathcal{K}_j} \alpha_{i,k}=0,\\[6pt]
\lambda_{j,0}+\delta_j, & \prod_{k\in\mathcal{K}_j} \alpha_{i,k}=1,
\end{cases}
\end{align*}
where $\lambda_{j,0}=g_j$ and $\lambda_{j,0}+\delta_j=1-s_j$ correspond to the guessing and slipping parameters, respectively. In contrast, the DINO model adopts a compensatory disjunctive structure by retaining only main effects:
\begin{align}
\label{eq:lamda_dino}
\lambda_{j,kk'}=0,\quad
\dots,\quad
\lambda_{j,1\dots K_j}=0,
\end{align}
thus
\begin{align}
\label{eq:lamda_dino1}
P(Y_{ij}=1\mid \boldsymbol{\alpha}_i)=\lambda_{j,0}
+ \sum_{k\in\mathcal{K}_j}\lambda_{j,k}\alpha_{i,k},
\end{align}
where $\lambda_{j,0}=g_j$ and $\lambda_{j,k}=(1-s_j)-g_j$. Under this specification, the GDINA model replaces the DINA measurement component in Equations \eqref{eq:eta-dina}-\eqref{eq:gs}. Similarly, the DINO model can be obtained following Equations \eqref{eq:lamda_dino} and \eqref{eq:lamda_dino1}, allowing the proposed dynamic framework to accommodate alternative measurement structures without altering the transition or $Q$-matrix components.}

{\subsubsection{Integration of unstructured action sequences}
The framework may also be extended to incorporate unstructured process data, such as action sequences in assessments (e.g., PIAAC). These sequences (e.g., including mouse clicks, navigation paths, and time-stamped events) can first be transformed into lower-dimensional representations of strategies or action patterns through feature extraction \citep{He2021ProcessData}. The resulting extracted features then serve as time-varying covariates within our dynamic CDM framework, allowing the framework to capture both skill mastery and problem-solving strategies over time without modifying the model's transition or $Q$-matrix components.}

{
\subsubsection{Differential item functioning}
The proposed model can be extended to incorporate person-level covariates to ensure measurement invariance. Specifically, the model can be extended to detect differential item functioning (DIF) across demographic groups, such as gender, socio-economic status, English language status, and present of any special education needs. For item $j$ and student $i$ in group $Z_i$, where $Z_i \in \{1,\dots,G\}$ denotes the group membership of student $i$, DIF can be examined by modifying the item response function as:
\begin{align*}
\text{logit}\big(P(Y_{ij}=1 \mid \boldsymbol{\alpha}_i, Z_i)\big)
= \delta_{j,0} + \delta_{j,1}\eta_{ij} + \delta_{j,g} Z_i,
\end{align*}
where $g \in \{1,\dots,G\}$, $\eta_{ij}$ is the ideal response indicator defined in Equation \eqref{eq:eta-dina}, $\delta_{j,0}$ is the baseline log odds of a correct response for the reference group ($Z_i=0$) when $\eta_{ij}=0$, $\delta_{j,1}$ represents the effect of mastering all required attributes in the reference group, and $\delta_{j,g}$ captures the group specific shift in log odds. Measurement invariance for item $j$ corresponds to the null hypothesis
\begin{align*}
H_0:\delta_{j,g}=0.
\end{align*}
The transition component can be modified as
\begin{align*}
\text{logit}\big(P(\alpha_{i,k,t+1}=1 \mid \alpha_{i,k,t}=0, Z_i)\big)
= \gamma_{01,k,0} + \gamma_{01,k,g} Z_i,
\end{align*}
where $\gamma_{01,k,0}$ denotes the baseline log odds of transitioning from non-mastery to mastery of attribute $k$ for the reference group, and $\gamma_{01,k,g}$ represents the group specific effect on the learning rate. Group invariance in the transitions is then assessed by testing
\begin{align*}
H_0:\gamma_{01,k,g}=0.
\end{align*}
This formulation enables subgroup comparisons while leaving the $Q$-matrix and transition structure unchanged.}

\section{Empirical Study}

\subsection{Data}

We selected two games that support the development of two core components of reading comprehension: decoding skills, which support word recognition, and vocabulary, which supports language comprehension \citep{gough1986}. Theoretically, word recognition and language comprehension are considered as separate constructs, and earlier analysis of a related dataset confirms that student engagement and performance on each are separable \citemaY{ma2024}. In our empirical study, we treated decoding and vocabulary as two target attributes. To ensure sufficient engagement and a balanced sample, we included 263 students who had interacted with both games corresponding to these skills. In our analysis, we examined the beginning of Year 1 and Year 2 of the study dataset ($T = 2$ in Equation \ref{eq:inference}). We extracted six responses per time point from the student log files from the first level that they interacted with for the decoding game (items 1-3) and for the vocabulary game (items 4-6). Detailed descriptions of the game-level content are provided in the Appendix. 

{In the empirical application, students did not always encounter the same set of items either across individuals or across the two time points. This variation arises naturally from the adaptive design of the learning environment, where item exposure depends on each learner’s progression. Although the specific level numbers differed, exploratory analyses of the level content (Supplementary Material B) showed that many levels within each game assess highly similar or identical instructional content. For example, in the decoding game (Game 1), level 7 and level 8 both assess the same set of individual letter-sound correspondences, and several other commonly encountered levels share comparable content sets. {We assume that items aligned to the same attribute were designed to assess highly similar instructional content and were therefore treated as exchangeable.} The observed similarity in level content across the most frequently encountered level combinations provide empirical support for this item exchangeability assumption in the present dataset. {Nevertheless, different game levels assessing highly similar instructional content may still not share identical guessing and slipping parameters. Therefore, the guessing and slipping parameters estimated under this framework should be interpreted as averages over the exchangeable item set, rather than as properties of a specific game level. Such} differences in item difficulty or discrimination across the selected items may also introduce additional variability in the estimation of attribute profiles over time.}

To control for individual differences, we considered 12 covariates informed by prior research \citemaY{ma2024}. These comprised six log-based variable—average number of attempts, number of levels mastered, and average response times—calculated separately for each of the two games, as presented in Table \ref{tab:summary_continuous}. Additionally, we included six individual characteristics—four demographic variables (gender, race, English language status, and special educational needs) and two variables related to game and learning characteristics (initial literacy ability and engagement group)—as summarised in Table \ref{tab:summary_categorical}. The engagement group variable reflected students’ overall participation across multiple literacy games in Boost, based on a set of theory- and data-driven engagement indicators \citemaY{ma2024}. That study identified nine distinct engagement profiles, and students in the current study sample were assigned to three of these profiles. 
Specifically, students in the high vocabulary group engaged more with games targeting vocabulary skills, those in the high decoding group engaged more with decoding-focused games, and the balanced group demonstrated comparable engagement across both skill families.

Table \ref{tab:summary_continuous} presents the summary statistics for the log-based variables. In each academic year, students completed six items in our sample, comprising three levels per game, with a maximum of three levels mastered in each game. Compared to the vocabulary game, students in the decoding game exhibited slightly higher average levels mastered, longer average response times, and lower variability across measures. In contrast, students showed higher average reattempts in the vocabulary game than in the decoding game, suggesting greater effort was needed in the vocabulary game.

\begin{table}[ht]
\caption{Summary statistics for log-based continuous variables.\\
\textit{Note}: NLM = average number of levels mastered; NRA = average number of reattempts; RT = average response time. For NRA and RT, individual means were first computed across game levels for each student and then averaged across the entire sample. For NLM, the number of levels mastered was first counted at the individual level before calculating the group average.}
\label{tab:summary_continuous}
\centering
\begin{tabular}{lll}
\hline
\textbf{Variable} & \textbf{Mean (SD)} & \textbf{Quantiles: (1st, 3rd)} \\
\hline
\multicolumn{3}{l}{\textit{Log-based Variables}} \\
\hline
NLM decoding game & 1.87 (0.90) & (1, 3) \\
NLM vocabulary game & 1.76 (0.93) & (1, 2) \\
NRA decoding game & 2.14 (1.41) & (1, 3) \\
NRA vocabulary game & 2.41 (1.55) & (1, 3) \\
RT decoding game & 2.14 (0.58) & (1.78, 2.47)\\
RT vocabulary game & 1.80 (1.25) & (1.13, 2.06) \\
\hline
\end{tabular}
\end{table}

As shown in Table \ref{tab:summary_categorical}, approximately half of the students belonged to the high vocabulary group. Most students in this group were non-English language learners (non-ELL) and did not have special education needs (non-SEN). The gender distribution was relatively balanced, with a slightly higher proportion of male students. Regarding initial literacy ability, students were distributed across different benchmark levels on DIBELS, with a notable proportion classified as Well Below Benchmark or At Benchmark defined by Amplify criteria. The sample included students from diverse racial and ethnic backgrounds, with Hispanic or Latino, White, and Black or African American students comprising the largest groups. 

\begin{table}[ht]
\centering
\caption{Summary statistics for categorical demographic and game and learning characteristics variables. Values represent the number of students in each category, with the corresponding proportion.\\
\textit{Note}: ELL = English language learner; SEN = special educational needs; B = Black or African American; H = Hispanic or Latino; M = Multiracial/Other; W = White; NS = Not Specified; AI = American Indian; AN = Alaskan Native; AS = Asian.}
\label{tab:summary_categorical}
\begin{tabular}{ll}
\hline
\textbf{Variable} & \textbf{Summary} \\
\hline
\multicolumn{2}{l}{\textit{Demographic Variables}} \\
\hline
ELL & non-ELL: 196 (74.5\%); ELL: 67 (25.5\%) \\
SEN & non-SEN: 233 (88.6\%); SEN: 30 (11.4\%) \\
Gender & Female: 117 (44.5\%); Male: 146 (55.5\%) \\
Race   & AS: 15 (5.7\%); B: 34 (12.9\%); H: 63 (24\%); M: 27 (10.3\%); \\
       & W: 58 (22.1\%); NS: 47 (17.9\%); Other (AI \& AN): 3 (1.2\%); \\
       & Missing: 16 (6.1\%) \\ 
\hline
\multicolumn{2}{l}{\textit{Game and Learning Characteristics}} \\
\hline
Engagement Group & High Vocabulary Group: 142 (54\%);\\ 
                 & High Decoding Group: 26 (9.9\%); \\
                 & Balanced Group: 95 (36.1\%) \\
Initial Literacy Ability & Above Benchmark: 53 (20.2\%); At Benchmark: 57 (21.7\%);\\
    & Below Benchmark: 38 (14.4\%); \\
    & Well Below Benchmark: 75 (28.5\%); NA: 40 (15.2\%) \\
\hline
\end{tabular}
\end{table}

\subsection{Model Diagnostics}

We applied the proposed model to the dataset and assessed convergence following \citet{vehtari2021rank}. From 60,000 total iterations (3 chains with 20,000 for each), the first half were discarded as warm-up. Diagnostics indicated the maximum potential scale reduction factor ($\hat{R}$) with {1.01}. The minimum effective sample size (ESS) were {2690}, indicating efficient sampling. Running time for the empirical analysis (MCMC chain length = 30,000) was approximately {6} minutes {using one core}, conducted on a MacBook Pro (13-inch, M1, 2020) equipped with an Apple M1 chip (8-core: 4 performance and 4 efficiency cores) and 16 GB of unified memory.

\subsection{Estimation and Evaluation of the $Q$-matrix and Item Parameters}
The estimated $Q$-matrices at Time 1 and Time 2 are shown in Table \ref{tab:qmatrix_combined_gs}. At Time 1, the $Q$-matrix aligned with the test design: Items 1–3 targeted decoding, and Items 4–6 targeted vocabulary, reflecting the game structure. Note that the game structure was treated as unknown when we estimated the model, adding evidence that the model was able to correctly identify the item-skill relationship. At Time 2, Items {1,} 2, 3, and 6 loaded on both decoding and vocabulary, indicating increased skill integration. 

To evaluate the estimated $Q$-matrices, we adopted the proportion of variance accounted for (PVAF), a criterion based on the G-DINA discrimination index (GDI; \citealp[pp. 257–261]{de2016general}). For item $j$, we let $K_j^* = \sum_{k=1}^K q_{jk}$ denote the number of required attributes, and let $\boldsymbol{\alpha}_{lj}^*$ denote the $l$th reduced attribute pattern among the $L = 2^{K_j^*}$ possible combinations. The GDI is then defined as the variance of success probabilities across these patterns given a possible $q$-vector $\mathbf{q}$:

\begin{equation*}
\zeta_j^2(\mathbf{q}) = \sum_{l=1}^{2^{K_j^*}} p(\boldsymbol{\alpha}_{lj}^* \mid \mathbf{q}) \left[ P(Y = 1 \mid \boldsymbol{\alpha}_{lj}^*, \mathbf{q}) - \bar{P}(Y = 1 \mid \mathbf{q}) \right]^2,
\end{equation*}

where the average success probability is:

\begin{equation*}
\bar{P}(Y = 1 \mid \mathbf{q}) = \sum_{l=1}^{2^{K_j^*}} p(\boldsymbol{\alpha}_{lj}^*) P(Y = 1 \mid \boldsymbol{\alpha}_{lj}^*, \mathbf{q}).
\end{equation*}

The PVAF was computed as the ratio of the GDI for a particular $q$-vector to the maximum GDI. As noted by \citet[p. 261]{de2016general}, the maximum GDI is achieved when all the attributes are specified. Thus, the PVAF expresses how well a candidate $q$-vector accounts for the variability in success probabilities relative to this maximum. In our analysis, two Time 1 items met revision criteria (PVAF $<$ 0.8). After modification, {only one} exceeded the threshold: {Item 2} was updated to include vocabulary. No changes were indicated at Time 2. As validation methods for dynamic models with covariates are limited, additional simulation studies were conducted to support these findings.

Table \ref{tab:qmatrix_combined_gs} summarises the guessing ($g$), and slipping ($s$) parameters for each item across Time 1 and Time 2. Compared to real-data analyses reported by \cite{liu2025general}, where guessing rates often exceeded 0.5 while slipping rates were often below 0.4, the {average} guessing and slipping parameters estimated in the current study were moderate. {Notably, item 3 exhibited slightly higher slipping rates at both time points (see Table \ref{tab:qmatrix_combined_gs}). Observed changes in the estimated $Q$-matrix may reflect time-varying item characteristics across years, in addition to potential changes in students' underlying skill mastery. Accordingly, interpretation of changes in $Q$-matrix should take into account contextual variation across years.}

\begin{table}[ht]
\centering
\caption{Estimated $Q$-matrix, guessing ($g$), and slipping ($s$) parameters across Time 1 and Time 2. $Q$-matrix values indicate whether each item requires decoding ($A_1$) and/or ($A_2$) vocabulary skills. Bolded values in the table highlight the maximum estimates of $g$ and $s$ across items.\\
{\textit{Note}: Because students did not necessarily encounter identical game levels, the estimated $g$ and $s$ parameters should be interpreted as average guessing and slipping parameters over the exchangeable item set, rather than as properties of specific game levels.}}
\label{tab:qmatrix_combined_gs}
{
\begin{tabular}{c|ccrr|ccrr}
\toprule
\multirow{2}{*}{Item} 
& \multicolumn{4}{c|}{\textbf{Time 1}} 
& \multicolumn{4}{c}{\textbf{Time 2}} \\
\cmidrule(lr){2-5} \cmidrule(lr){6-9}
& $A_1$ & $A_2$ & $g$ & $s$ 
& $A_1$ & $A_2$ & $g$ & $s$ \\
\midrule
1 & 1 & 0 & 0.344 & 0.228 & 1 & 1 & \textbf{0.416} & 0.272 \\
2 & 1 & 0 & 0.259 & 0.144 & 1 & 1 & 0.338 & 0.225 \\
3 & 1 & 0 & 0.274 & \textbf{0.448} & 1 & 1 & 0.302 & \textbf{0.396} \\
4 & 0 & 1 & 0.158 & 0.376 & 1 & 0 & 0.327 & 0.275 \\
5 & 0 & 1 & \textbf{0.41}0 & 0.368 & 0 & 1 & 0.399 & 0.367 \\
6 & 0 & 1 & 0.140 & 0.316 & 1 & 1 & 0.396 & 0.343 \\
\bottomrule
\end{tabular}}
\end{table}

\subsection{Identified Attribute Profiles, Initial Attribute Mastery and Attribute Transition with Covariates} 
Table \ref{tab:profile_distribution_attribute_transition_matrix} presents the distribution and transitions of attribute profiles across two time points. {More than one third of students who had mastered neither skill at Time 1 mastered at least one skill at Time 2. Under the monotonicity assumption, mastery was treated as an absorbing state, thus, no transitions from mastery to non-mastery were observed. Overall, vocabulary showed a higher mastery rate than decoding, as more students transitioned to profiles involving vocabulary mastery (01 and 11).}

\begin{table}[ht]
\centering
\caption{{Transitions of attribute profiles from Time 1 to Time 2 (N = 263). Each row represents a specific profile transition and reports the number (percentage of students). Profile labels: 00 = no mastery; 10 = decoding only; 01 = vocabulary only; 11 = mastery of both skills.}}
\label{tab:profile_distribution_attribute_transition_matrix}
{
\begin{tabular}{lcc}
\toprule
\makecell[l]{\textbf{Transition} \\ \footnotesize (Time 1 $\rightarrow$ Time 2)}
& \textbf{N(\%)}
& \textbf{Sum N(\%)} \\
\midrule
00 $\rightarrow$ 00 & 26(9.89)  & \multirow{4}{*}{93(35.37)} \\
00 $\rightarrow$ 10 & 2(0.76)   &  \\
00 $\rightarrow$ 01 & 33(12.55) &  \\
00 $\rightarrow$ 11 & 32(12.17) &  \\
\midrule
10 $\rightarrow$ 10 & 10(3.80)  & \multirow{2}{*}{40(15.21)} \\
10 $\rightarrow$ 11 & 30(11.41) &  \\
\midrule
01 $\rightarrow$ 01 & 46(17.49) & \multirow{2}{*}{77(29.28)} \\
01 $\rightarrow$ 11 & 31(11.79) &  \\
\midrule
11 $\rightarrow$ 11 & 53(20.15) & 53(20.15) \\
\midrule
\textbf{Total} & 263 & 100.00 \\
\bottomrule
\end{tabular}}
\end{table}

Table \ref{tab:betaz_gamma01_gamma10_significant} presents the posterior means of odds ratios for initial mastery ({$\boldsymbol{\beta}$}) and transition probabilities ({\boldsymbol{$\gamma_{01}$}}), estimated separately by attribute $K$. Covariates included six log-based variables, four demographics, and two game and learning characteristics listed in Tables \ref{tab:summary_continuous} and \ref{tab:summary_categorical}. Covariates without statistically significant effects are not shown the {tables}. All odds ratios and confidence intervals for all covariates are provided in supplementary material C. A sparsity criterion was applied following \citet{wayman2025restricted}, whereby odds ratios with 95\% credible intervals including 1 were considered inactive.

{
For initial mastery ({$\boldsymbol{\beta}$}), a greater number of reattempts in the decoding game was negatively associated with mastery of decoding ($k = 1$), whereas higher initial literacy ability (above benchmark) and longer time spent on decoding were positively associated with decoding mastery. In addition, mastering fewer vocabulary levels was associated with a higher likelihood of decoding mastery. For transitions from non-mastery to mastery ({\boldsymbol{$\gamma_{01}$}}), more reattempts in vocabulary were negatively associated with mastering vocabulary, while longer time spent on vocabulary increased the likelihood of transition. Boys were less likely than girls to transition to decoding mastery, and mastering more decoding levels was positively associated with transition.
}

\begin{table}[http]
\centering
\caption{Significant covariates for {$\boldsymbol{\beta}$} (initial mastery), {\boldsymbol{$\gamma_{01}$}} by attribute ($K$). Only covariates with statistically significant odds ratios (OR) are shown.
\textit{Note}: $K$ = attribute; (Intercept) = model intercept; NRA = number of reattempts; NLM = number of levels mastered; RT = response time; Gender = female (0), male (1); ILA = initial literacy ability (at benchmark: reference level); WB = well below benchmark; Decoding = decoding game; Vocabulary = vocabulary game.
}
\label{tab:betaz_gamma01_gamma10_significant}
{
\begin{tabular}{ccc|ccc}
\toprule
\multicolumn{3}{c|}{Initial mastery} & \multicolumn{3}{c}{Transition probabilities} \\
\multicolumn{3}{c|}{{$\boldsymbol{\beta}$}} & \multicolumn{3}{c}{{\boldsymbol{$\gamma_{01}$}}}  \\
$K$ & Covariates & OR & $K$ & Covariates & OR \\
\midrule
1 & NRA Decoding & 0.067 & 1 & Gender & 0.894    \\
1 & NLM Vocabulary & 0.450 & 1 & NLM Decoding & 4.147   \\
1 & ILA-AB  & 3.816 & 2 & NRA Vocabulary & 0.371 \\
1 & RT Decoding & 12.974 & 2 & RT Vocabulary & 3.116  
\\
\bottomrule
\end{tabular}}
\end{table}

\section{Simulation Study}
The main objectives of this simulation study were to assess the model performance under various conditions, and to evaluate the model's robustness under settings that can be present within the area of application. We designed 18 simulation settings across three experimental factors: sample size ($N$), the number of items ($J$), and $Q$-matrix sparsity level ($\theta$). 

\subsection{Simulation Design and Validations}
{The simulation study was designed not only to reflect aspects of the empirical setting but also to evaluate the model under a more general and structurally complex scenario. We considered a dynamic CDM with three attributes measured across three time points. Sample size varied across conditions $N \in \{200, 400, 600\}$, and the number of items administered at each time point varied as $J \in \{6, 18, 30\}$.} The true $Q$-matrices used under each simulation condition are provided in supplementary material D. We considered both a sparse scenario and a dense scenario. The density level adopted in the dense condition was close to the upper limit reported in prior studies \citep{rupp2008effects, chen2015statistical, xu2018identifying}. {The prior distribution for the sparsity of the $Q$-matrix under each simulation condition is described in supplementary material F, together with sensitivity analyses under alternative prior specifications.} Overall recovery accuracy was defined as the proportion of correctly classified $Q$-matrix entries. {Within each replication, posterior probabilities of candidate $Q$-matrix row patterns were estimated from the retained MCMC draws.}

{The final $Q$-matrix estimate was obtained using a row-wise maximum a posteriori (MAP) rule by selecting the pattern with the highest posterior probability. The proportion of correctly recovered entries was computed within each replication and averaged across replications. Posterior uncertainty for the $Q$-matrix was summarized by the posterior frequencies of ones and zeros for each entry of $Q$ across posterior samples after discarding the burn-in samples. We presented the posterior mean (PM) for ones and zeros in Table \ref{tab:recovery_Q_full}}. For comparison, simulation results assuming a known $Q$-matrix for a small sample size ($N$=200) are provided in supplementary material E. 

{We assessed model performance using both parameter estimation accuracy and classification accuracy. 
For continuous parameters, 95\% credible intervals were obtained from posterior samples. Selected interval results under the smallest and largest sample conditions are reported in Supplementary Material I.} For item parameters and regression coefficients in the attribute and transition models, we computed the mean {absolute error (MAE)} and root mean square error (RMSE) across replications, given by
$$
{\text{MAE} = \frac{1}{R} \sum_{r=1}^{R} |\hat{p}^{(r)} - p|, \quad }
\text{RMSE} = \sqrt{ \frac{1}{R} \sum_{r=1}^{R} (\hat{p}^{(r)} - p)^2 },
$$
where $p$ is the true parameter and $R$ is the number of replications.

{
The classification accuracy of students’ attribute profiles, in terms of each individual attribute ($k$) at each time point ($t$), was calculated as follows:
$$
\text{AAR}_{k,t} = \frac{1}{NR} \sum_{r=1}^{R} \sum_{i=1}^{N} \mathbb{I} (\hat{\boldsymbol{\alpha}}^{(r)}_{ikt} = \boldsymbol{\alpha}_{ikt}),
$$
} {
where $\alpha_{ikt}\in\{0,1\}$ denote the true mastery state of student $i = 1, \dots, N$ on attribute $k = 1, \dots, K$ at time point $t$, and $\hat{\alpha}^{(r)}_{ikt}$ denoted its estimate obtained from the $r$th-replication ($r = 1, \dots, R$). The indicator function takes the value 1 if the condition holds and 0 otherwise. The estimated attribute profiles ($\hat{\alpha}$) were obtained by applying a 0.5 threshold to their respective posterior means. This approach ensures that each student {was} assigned to the mastery state that {had} the highest posterior probability.} 





\subsection{Results}
{
For each simulation condition, we conducted 100 independent replications. The model estimation was implemented using four independent Markov chains per replication, each initialized with different starting values to ensure broad coverage of the parameter space. Each chain consisted of 5,000 burn-in iterations, followed by 5,000 monitored iterations for posterior inference. Convergence was assessed using the potential scale reduction factor ($\hat{R}$), as proposed by \citet{gelman1992inference}. All parameters yielded $\hat{R}$ values below 1.1, indicating acceptable convergence. To quantify simulation uncertainty, we applied a nonparametric bootstrap procedure to the 100 replications. Specifically, 1,000 bootstrap samples were drawn with replacement, and standard errors for all performance metrics were derived from these bootstrap distributions. Computational time increased with both sample size and test length, with average runtime per replication ranging from 7 to 90 minutes across conditions.
}

For the sparsity prior on the $Q$-matrix, the prior mean $\frac{\alpha}{\alpha+\beta}$ matched the true sparsity level ({0.3 for {sparse $Q$-matrices}, 0.5 for dense {$Q$-matrices}}), and the concentration $\alpha+\beta$ was scaled with the number of items to stabilise estimation ({10, 15, and 20} for $J=6$, $18$, and $30$, respectively). Sensitivity analyses further varied the prior specifications, and recovery of the $Q$-matrix and attribute classifications showed little difference (see supplementary material F), confirming the robustness of the model to reasonable prior misspecification. 

{Table \ref{tab:recovery_alpha_updated} presents attribute agreement rates (AARs) with bootstrap standard errors across conditions. Across all combinations of $N$ and $J_t$, AARs generally exceeded 0.87 even in out most challenging small sample conditions ($N =200$, $J_t = 6$). For a fixed $N$, AARs tended to increase slightly as $J_t$  increased from 6 to 18, though gains beyond $J_t = 18$ were modest. As both sample size and test length increased, recovery improved {consistently}, with AARs frequently exceeding 0.99 when $N \ge 400$ and $J_t \ge 18$.}

\begin{table}[h!]
\centering
{
\caption{{Recovery of attribute profiles measured by attribute agreement rates ($\text{AAR}_1, \text{AAR}_2, \text{AAR}_3$) across time points ($T$), under varying sparsity levels, sample sizes ($N$), and number of items ($J_t$). Standard errors (SE) are bootstrap-based (1,000 resamples). Bolded values in the table highlight the highest values.}}
\label{tab:recovery_alpha_updated}
\footnotesize
\setlength{\tabcolsep}{2.5pt} 
\begin{tabular}{cccc | ccc | ccc}
\toprule
\multicolumn{4}{c}{} & \multicolumn{3}{c}{Sparse $Q$} & \multicolumn{3}{c}{Dense $Q$} \\
\cmidrule(lr){5-7} \cmidrule(lr){8-10}
$N$ & $J_t$ & $T$ & & AAR$_1$ (SE) & AAR$_2$ (SE) & AAR$_3$ (SE) & AAR$_1$ (SE) & AAR$_2$ (SE) & AAR$_3$ (SE) \\
\midrule
200 & 6  & 1 & & .913 (.010) & .885 (.010) & .865 (.016) & .917 (.005) & .902 (.007) & .884 (.009) \\
    &    & 2 & & .911 (.011) & .924 (.011) & .893 (.018) & .929 (.004) & .926 (.005) & .915 (.004) \\
    &    & 3 & & .942 (.007) & .908 (.010) & .953 (.006) & .943 (.004) & .939 (.004) & .947 (.004) \\
\addlinespace[0.3em]
    & 18 & 1 & & .996 (.001) & .992 (.004) & .979 (.009) & .981 (.003) & .953 (.006) & .972 (.005) \\
    &    & 2 & & .998 (.001) & .988 (.005) & .991 (.004) & .985 (.002) & .978 (.007) & .949 (.009) \\
    &    & 3 & & \textbf{1.000 (.000)} & .995 (.003) & .992 (.005) & .988 (.004) & .996 (.002) & .982 (.004) \\
\addlinespace[0.3em]
    & 30 & 1 & & .995 (.002) & .985 (.002) & .976 (.004) & .990 (.004) & .977 (.002) & .900 (.019) \\
    &    & 2 & & .998 (.001) & .980 (.004) & .964 (.012) & .997 (.002) & .973 (.011) & .911 (.016) \\
    &    & 3 & & .985 (.014) & .995 (.001) & .970 (.015) & .993 (.002) & \textbf{.999 (.001)} & \textbf{.990 (.006)} \\
\midrule
400 & 6  & 1 & & .910 (.027) & .897 (.021) & .907 (.004) & .923 (.003) & .916 (.004) & .908 (.004) \\
    &    & 2 & & .946 (.010) & .936 (.007) & .884 (.017) & .932 (.003) & .930 (.002) & .920 (.003) \\
    &    & 3 & & .941 (.005) & .903 (.016) & .932 (.005) & .944 (.003) & .941 (.002) & .950 (.002) \\
\addlinespace[0.3em]
    & 18 & 1 & & .993 (.001) & .990 (.001) & .982 (.002) & .972 (.003) & .958 (.004) & .949 (.004) \\
    &    & 2 & & .997 (.001) & .994 (.001) & .989 (.001) & .993 (.001) & .981 (.003) & .962 (.004) \\
    &    & 3 & & .998 (.001) & .995 (.001) & \textbf{.994 (.001)} & .991 (.001) & .996 (.001) & .985 (.002) \\
\addlinespace[0.3em]
    & 30 & 1 & & .997 (.001) & .986 (.003) & .976 (.003) & .993 (.002) & .974 (.006) & .865 (.013) \\
    &    & 2 & & .998 (.001) & .983 (.005) & .971 (.005) & \textbf{.999 (.000)} & .970 (.008) & .896 (.015) \\
    &    & 3 & & \textbf{1.000 (.000)} & \textbf{.996 (.001)} & .985 (.004) & .998 (.001) & .998 (.001) & .985 (.008) \\
\midrule
600 & 6  & 1 & & .940 (.013) & .935 (.008) & .882 (.012) & .927 (.002) & .915 (.003) & .905 (.003) \\
    &    & 2 & & .924 (.003) & .934 (.004) & .896 (.015) & .932 (.002) & .929 (.002) & .921 (.002) \\
    &    & 3 & & .935 (.006) & .926 (.003) & .950 (.006) & .945 (.002) & .942 (.002) & .952 (.002) \\
\addlinespace[0.3em]
    & 18 & 1 & & .994 (.001) & .991 (.001) & .981 (.001) & .970 (.002) & .962 (.003) & .949 (.003) \\
    &    & 2 & & .997 (.000) & .994 (.001) & .991 (.001) & .991 (.001) & .980 (.002) & .956 (.003) \\
    &    & 3 & & .998 (.000) & .994 (.001) & \textbf{.994 (.001)} & .992 (.001) & .995 (.001) & .986 (.001) \\
\addlinespace[0.3em]
    & 30 & 1 & & .979 (.017) & .940 (.042) & .968 (.004) & .988 (.006) & .963 (.005) & .880 (.006) \\
    &    & 2 & & .977 (.019) & .950 (.033) & .967 (.008) & .994 (.003) & .959 (.010) & .882 (.011) \\
    &    & 3 & & .990 (.009) & .990 (.005) & .973 (.006) & .991 (.004) & .988 (.008) & .961 (.010) \\
\bottomrule
\end{tabular}
}
\end{table}

{As shown in Table \ref{tab:recovery_Q_full}, $Q$-matrix recovery improved with both sample size ($N$) and test length ($J_t$) under sparse and dense structures. Accuracy was generally higher at Time 2 and Time 3 than at Time 1, particularly with smaller samples. Posterior mean (PM) estimates of the binary $Q$ entries increased consistently with $N$ and $J_t$. For larger samples, PM$_1$ values were typically above 0.98. Bootstrap standard errors (SEs) were uniformly small across conditions, and decreased as $N$ and $J_t$ increased.}


\begin{table}[h!]
\centering
\caption{{Recovery accuracy of $Q$-matrix across three time points ($T=1, 2, 3$) under varying sample sizes ($N$) and item lengths ($J_t$). Results are grouped by Sparse and Dense $Q$ matrices. PM$_1$ and PM$_0$ denote the average posterior means of $Q_{jkt}$ over entries where the true value equals 1 and 0, respectively. Standard errors (SE) are bootstrap-based (1,000 resamples). Bolded values in the table highlight the highest values.}}
\label{tab:recovery_Q_full}
\footnotesize
\setlength{\tabcolsep}{1pt}
{
\begin{tabular}{cccc | ccccc |ccccc}
\toprule
\multicolumn{4}{c}{} & \multicolumn{5}{c|}{Sparse $Q$} & \multicolumn{5}{c}{Dense $Q$} \\
\cmidrule(lr){5-9} \cmidrule(lr){10-14}
$N$ & $J_t$ & $T$ & & Acc.(SE) & FNR(SE) & FPR(SE) & $\text{PM}_1$ & $\text{PM}_0$  & Acc.(SE) & FNR(SE) & FPR(SE) & $\text{PM}_1$ & $\text{PM}_0$ \\
\midrule
200 & 6 & 1 & & .976(.017) & .032(.021) & .016(.016) & .861  & .060
& .930(.017) & .082(.026) & .056(.014) & .857 & .060\\
 &  & 2 & & .897(.019) & .032(.021) & .175(.022) & .900 & .144
 & .971(.012) & .026(.014) & .032(.016) & .893 & .055\\
 &  & 3 & & .937(.008) & .000(.000) & .127(.016) & .919 & .132 
 & .977(.007) & .019(.008) & .028(.012) & .869 & .053 \\
 & 18 & 1 & & .977(.010) & .011(.002) & .021(.011) & .986 & .060
 & .972(.012) & .031(.018) & .023(.013) & .967 & .024  \\
 &  & 2 & & .981(.001) & .013(.003) & .011(.000) & .997 & .021
 & .977(.020) & .014(.004) & .011(.002) & .961 & .022 \\
 &  & 3 & & .998(.002) & .009(.002) & .081(.003) & .999  & .019
 & 1.000(.000) & .000(.000) & .000(.000) & 1.000 & .000\\
 & 30 & 1 & & .997(.001) & .001(.001) & .001(.001) & .976 & .009
 & .998(.002) & .004(.004) & .000(.000) & .971 & .005 \\
 &  & 2 & & .998(.001) & .001(.001) & .001(.001) & .989 & .012
 & .999(.001) & .002(.001) & .000(.000) & .996 & .002\\
 &  & 3 & & .984(.016) & .016(.016) & .016(.016) & .988 & .003
 & \textbf{1.000(.000)} & .000(.000) & .000(.000) & .998 & .001 \\
\midrule
400 & 6 & 1 & & .870(.067) & .037(.037) & .222(.111) & .805 & .195
& .972(.006) & .026(.007) & .030(.007) & .879  & .057 \\
 &  & 2 & & .944(.000) & .000(.000) & .111(.000) & .907 & .139
 & .972(.007) & .015(.006) & .044(.011) & .891 & .066\\
 &  & 3 & & .907(.037) & .000(.000) & .185(.074) & .833 & .148
 & .977(.007) & .024(.009) & .021(.007) & .836 & .069 \\
 & 18 & 1 & & .995(.002) & .003(.001) & .001(.001) & .990 & .014
 & .999(.001) & .002(.002) & .000(.000) & .995 & .003\\
 &  & 2 & & .992(.001) & .017(.002) & .011(.003) & .971 & .057
 & .997(.001) & .012(.008) & .013(.003) & .987 & .066 \\
 &  & 3 & & .986(.016) & .016(.006) & .016(.003) & .998 & .003 
 & .997(.008) & .009(.001) & .006(.002) & .990 & .005 \\
 & 30 & 1 & & .998(.001) & .001(.001) & .001(.001) & .983 &  .014
 & \textbf{1.000(.000)} & .000(.000) & .000(.000) & .976 & .003 \\
 &  & 2 & & .980(.004) & .002(.001) & .000(.000) & .990 & .001
 & \textbf{1.000(.000)} & .000(.000) & .000(.000) & .996 & .033 \\
 &  & 3 & & 1.000(.000) & .000(.000) & .000(.000) & 1.000 & .000
 & 1.000(.000) & .000(.000) & .000(.000) & 1.000 & .000 \\
\midrule
600 & 6 & 1 & & \textbf{1.000(.000)} & .000(.000) & .000(.000) & .903 & .084
& .975(.005) & .023(.006) & .027(.007) & .887 & .084\\
 &  & 2 & & .861(.028) & .056(.056) & .222(.000) & .778 & .084
 & .974(.005) & .009(.005) & .048(.008) & .858 & .084\\
 &  & 3 & & .944(.056) & .000(.000) & .111(.011) & .875 & .125
 & .977(.006) & .019(.007) & .027(.006) & .868 & .064\\
 & 18 & 1 & & .999(.001) & .000(.000) & .000(.000) & .988 & .008
 & .999(.001) & .001(.001) & .000(.000) & .996 & .005\\
 &  & 2 & & \textbf{1.000(.000)} & .000(.000) & .000(.000) & .994 & .004
 & .999(.000) & .001(.001) & .000(.000) & .998 & .002\\
 &  & 3 & & .999(.000) & .002(.001) & .000(.000) & .998  & .002 
 & \textbf{1.000(.000)} & .000(.000) & .000(.000) & 1.000  & .000\\
 & 30 & 1 & & .951(.049) & .000(.000) & .098(.098) & .949 & .049
 & .989(.008) & .009(.006) & .015(.011) & .968 & .022 \\
 &  & 2 & & .967(.033) & .000(.000) & .067(.067) & .967 & .033
 & .980(.011) & .008(.008) & .039(.020) & .976 & .030 \\
 &  & 3 & & \textbf{1.000(.000)} & .000(.000) & .000(.000) & .980 & .023
 & .988(.009) & .007(.007) & .019(.012) & .988 & .019\\
\bottomrule
\end{tabular}
}
\end{table}

{Table \ref{tab:gs_bias} summarises the estimation accuracy of guessing and slipping parameters ($g_{j,t}$ and $s_{j,t}$) under the sparse $Q$. For $g_{j,t}$, MAE was lower than 0.075 and RMSE lower than 0.098. For $s_{j,t}$, MAE slightly higher than guessing parameter, with the highest values under the smallest sample and shortest test condition. Accuracy improved with larger sample sizes, and estimates remained stable across time points. Results for the dense $Q$-matrix are presented in supplementary material G, where MAE remained minimal across all conditions ($g_{j,t} \leq 0.065$, $s_{j,t} \leq 0.208$). Item-level RMSE comparisons for $g_{j,t}$ and $s_{j,t}$ under different simulation conditions are reported in supplementary material H.}

\begin{table}[h!]
\centering
{
\caption{{Estimation accuracy of item parameters ($g_{j,t}$ and $s_{j,t}$), evaluated by MAE and RMSE, under sparse $Q$-matrix, sample sizes ($N$), and number of items ($J_t$). Standard errors (SE) are bootstrap-based (1,000 resamples). Bolded values in the table highlight the highest values.}}
\label{tab:gs_bias}
\centering
\begin{tabular}[t]{cccc|cc|cc}
\toprule
$N$ & $J_t$ & $T$ & Sparsity & $g_{\text{MAE}}$ (SE) & $g_{\text{RMSE}}$ (SE) & $s_{\text{MAE}}$ (SE) & $s_{\text{RMSE}}$ (SE) \\
\midrule
200 & 6  & 1 & Sparse & .027 (.004) & .035 (.003) & \textbf{.209 (.018)} & .241 (.018) \\
    &    & 2 &        & .050 (.006) & .063 (.006) & .073 (.009) & .094 (.009) \\
    &    & 3 &        & \textbf{.075 (.010)} & \textbf{.098 (.010)} & .047 (.006) & .060 (.006) \\
\addlinespace[0.3em]
    & 18 & 1 &        & .021 (.002) & .027 (.002) & .046 (.005) & .062 (.005) \\
    &    & 2 &        & .026 (.003) & .035 (.003) & .028 (.003) & .035 (.002) \\
    &    & 3 &        & .036 (.003) & .047 (.004) & .021 (.002) & .027 (.002) \\
\addlinespace[0.3em]
    & 30 & 1 &        & .026 (.001) & .033 (.001) & .050 (.002) & .072 (.002) \\
    &    & 2 &        & .042 (.002) & .059 (.002) & .026 (.001) & .035 (.001) \\
    &    & 3 &        & .042 (.002) & .056 (.002) & .024 (.001) & .030 (.001) \\
\midrule
400 & 6  & 1 & Sparse & .031 (.007) & .042 (.007) & .202 (.033) & \textbf{.246 (.032) }\\
    &    & 2 &        & .049 (.007) & .057 (.007) & .080 (.015) & .104 (.016) \\
    &    & 3 &        & .067 (.012) & .084 (.012) & .065 (.012) & .082 (.012) \\
\addlinespace[0.3em]
    & 18 & 1 &        & .016 (.001) & .020 (.001) & .032 (.001) & .042 (.001) \\
    &    & 2 &        & .020 (.001) & .026 (.001) & .018 (.001) & .023 (.001) \\
    &    & 3 &        & .024 (.001) & .030 (.001) & .015 (.001) & .019 (.001) \\
\addlinespace[0.3em]
    & 30 & 1 &        & .020 (.001) & .027 (.001) & .036 (.002) & .051 (.002) \\
    &    & 2 &        & .033 (.002) & .046 (.002) & .021 (.001) & .028 (.001) \\
    &    & 3 &        & .032 (.002) & .043 (.002) & .014 (.001) & .018 (.001) \\
\midrule
600 & 6  & 1 & Sparse & .029 (.007) & .038 (.007) & .093 (.016) & .110 (.017) \\
    &    & 2 &        & .058 (.009) & .065 (.009) & .092 (.022) & .121 (.024) \\
    &    & 3 &        & .064 (.012) & .076 (.012) & .046 (.013) & .066 (.013) \\
\addlinespace[0.3em]
    & 18 & 1 &        & .013 (.001) & .017 (.001) & .026 (.001) & .036 (.001) \\
    &    & 2 &        & .016 (.001) & .021 (.001) & .015 (.001) & .019 (.001) \\
    &    & 3 &        & .020 (.001) & .026 (.001) & .013 (.001) & .017 (.001) \\
\addlinespace[0.3em]
    & 30 & 1 &        & .019 (.001) & .024 (.001) & .031 (.002) & .041 (.002) \\
    &    & 2 &        & .033 (.003) & .044 (.002) & .019 (.001) & .024 (.001) \\
    &    & 3 &        & .033 (.003) & .048 (.003) & .013 (.001) & .017 (.001) \\
\bottomrule
\end{tabular}
}
\end{table}

{
Table \ref{tab:beta_metrics} presents the estimation accuracy for initial mastery probabilities ({$\beta_{k,0}$}) and covariate effects ({$\beta_{k,c}$}). 
Across conditions, RMSEs for {$\beta_{k,0}$} and {$\beta_{k,c}$} were below 0.444. Estimation accuracy of both parameters improved as both $N$ and $J_t$ increased, with notably smaller errors when item length increased. Bootstrap standard errors were consistently small, indicating stable estimation across replications. Table \ref{tab:gamma01_12_23_metrics} summarises the estimation accuracy for covariate effects on mastery acquisition transitions. We report results separately for transitions from time 1 to time 2 ($\gamma_{01,k}^{12}$) and from time 2 to time 3 ($\gamma_{01,k}^{23}$). In both sparse and dense $Q$-matrix conditions, estimation accuracy improved with larger sample sizes and greater numbers of items. Bootstrap standard errors remained small across all conditions.
As expected, RMSEs for transition parameters were generally larger than those for initial mastery parameters for small sample size, reflecting the additional uncertainty introduced by modelling longitudinal change. This pattern is consistent with previous findings showing larger estimation errors for transition parameters relative to initial mastery parameters \citep{liang2023latent}.
}

\begin{table}[h!]
\centering
{
\caption{{Estimation accuracy of initial covariate effects on attribute ($k$) mastery ($\boldsymbol{\beta}$), evaluated by mean bias (MBias) and root mean square error (RMSE), under varying sparsity levels ($\theta$), sample sizes ($N$), and number of items ($J_t$). Each $\beta_{k,z}$ quantifies the impact of covariate $z$ on the initial mastery probability of attribute $k$ at time $T = 1$. Standard errors (SE) are bootstrap-based (1,000 resamples). Bolded values in the table highlight the highest values.}}
\label{tab:beta_metrics}
\footnotesize
\setlength{\tabcolsep}{2pt}
\begin{tabular}{ccc| ccc ccc |ccc ccc}
\toprule
\multicolumn{3}{c}{} & \multicolumn{6}{c|}{Sparse $Q$} & \multicolumn{6}{c}{Dense $Q$} \\
\cmidrule(lr){4-9} \cmidrule(lr){10-15}
$N$ & $J_t$ & Metric & \makecell{{$\beta_{1,0}$}\\(SE)} & \makecell{{$\beta_{2,0}$}\\(SE)} & \makecell{{$\beta_{3,0}$}\\(SE)} & \makecell{{$\beta_{1,c}$}\\(SE)} & \makecell{{$\beta_{2,c}$}\\(SE)} & \makecell{{$\beta_{3,c}$}\\(SE)} & \makecell{{$\beta_{1,0}$}\\(SE)} & \makecell{{$\beta_{2,0}$}\\(SE)} & \makecell{{$\beta_{3,0}$}\\(SE)} & \makecell{{$\beta_{1,c}$}\\(SE)} & \makecell{{$\beta_{2,c}$}\\(SE)} & \makecell{{$\beta_{3,c}$}\\(SE)} \\
\midrule
200 & 6 & MAE 
& \makecell{.255 \\ (.055)} & \makecell{.223 \\ (.055)} & \makecell{\textbf{.444} \\ (.056)} 
& \makecell{.222 \\ (.035)} & \makecell{.226 \\ (.015)} & \makecell{.186 \\ (.026)}
& \makecell{.248 \\ (.039)} & \makecell{.227 \\ (.040)} & \makecell{.243 \\ (.033)} 
& \makecell{.195 \\ (.010)} & \makecell{.233 \\ (.015)} & \makecell{.237 \\ (.015)} \\
    &   & RMSE 
& \makecell{.255 \\ (.057)} & \makecell{.223 \\ (.056)} & \makecell{\textbf{.444} \\ (.056)} 
& \makecell{.287 \\ (.042)} & \makecell{.267 \\ (.013)} & \makecell{.214 \\ (.026)}
& \makecell{.248 \\ (.039)} & \makecell{.227 \\ (.040)} & \makecell{.243 \\ (.033)} 
& \makecell{.237 \\ (.013)} & \makecell{.281 \\ (.020)} & \makecell{.285 \\ (.019)} \\
200 & 18 & MAE 
& \makecell{.167 \\ (.035)} & \makecell{.191 \\ (.088)} & \makecell{.129 \\ (.038)} 
& \makecell{.181 \\ (.011)} & \makecell{.140 \\ (.018)} & \makecell{.195 \\ (.042)}
& \makecell{.138 \\ (.083)} & \makecell{.230 \\ (.094)} & \makecell{.215 \\ (.058)} 
& \makecell{.178 \\ (.040)} & \makecell{.160 \\ (.007)} & \makecell{.145 \\ (.027)} \\
    &    & RMSE 
& \makecell{.167 \\ (.035)} & \makecell{.191 \\ (.091)} & \makecell{.129 \\ (.038)} 
& \makecell{.214 \\ (.012)} & \makecell{.169 \\ (.018)} & \makecell{.222 \\ (.047)}
& \makecell{.138 \\ (.086)} & \makecell{.230 \\ (.090)} & \makecell{.215 \\ (.057)} 
& \makecell{.222 \\ (.049)} & \makecell{.197 \\ (.013)} & \makecell{.159 \\ (.030)} \\
\addlinespace[0.3em]
200 & 30 & MAE 
& \makecell{.144 \\ (.021)} & \makecell{.150 \\ (.019)} & \makecell{.297 \\ (.047)} 
& \makecell{.158 \\ (.012)} & \makecell{.141 \\ (.008)} & \makecell{.141 \\ (.011)}
& \makecell{.107 \\ (.025)} & \makecell{\textbf{.296} \\ (.075)} & \makecell{.238 \\ (.097)} 
& \makecell{.185 \\ (.022)} & \makecell{.116 \\ (.019)} & \makecell{.287 \\ (.052)} \\
    &    & RMSE 
& \makecell{.144 \\ (.020)} & \makecell{.150 \\ (.019)} & \makecell{.297 \\ (.046)} 
& \makecell{.184 \\ (.012)} & \makecell{.164 \\ (.009)} & \makecell{.172 \\ (.013)}
& \makecell{.107 \\ (.025)} & \makecell{\textbf{.296} \\ (.077)} & \makecell{.238 \\ (.105)} 
& \makecell{.211 \\ (.022)} & \makecell{.144 \\ (.024)} & \makecell{.340 \\ (.057)} \\
\midrule
400 & 6 & MAE 
& \makecell{.376 \\ (.096)} & \makecell{.237 \\ (.085)} & \makecell{.314 \\ (.183)} 
& \makecell{.163 \\ (.013)} & \makecell{.114 \\ (.014)} & \makecell{.137 \\ (.015)}
& \makecell{.214 \\ (.018)} & \makecell{.212 \\ (.020)} & \makecell{.232 \\ (.022)} 
& \makecell{.155 \\ (.006)} & \makecell{.142 \\ (.006)} & \makecell{.130 \\ (.006)} \\
    &   & RMSE 
& \makecell{.376 \\ (.095)} & \makecell{.237 \\ (.087)} & \makecell{.314 \\ (.187)} 
& \makecell{.195 \\ (.016)} & \makecell{.148 \\ (.019)} & \makecell{.177 \\ (.027)}
& \makecell{.214 \\ (.019)} & \makecell{.212 \\ (.020)} & \makecell{.232 \\ (.023)} 
& \makecell{.186 \\ (.008)} & \makecell{.171 \\ (.008)} & \makecell{.162 \\ (.008)} \\
400 & 18 & MAE 
& \makecell{.104 \\ (.019)} & \makecell{.114 \\ (.017)} & \makecell{.110 \\ (.012)} 
& \makecell{.105 \\ (.008)} & \makecell{.107 \\ (.009)} & \makecell{.094 \\ (.008)}
& \makecell{.110 \\ (.020)} & \makecell{.257 \\ (.039)} & \makecell{.301 \\ (.044)} 
& \makecell{.106 \\ (.008)} & \makecell{.133 \\ (.010)} & \makecell{.106 \\ (.008)} \\
    &    & RMSE 
& \makecell{.104 \\ (.019)} & \makecell{.114 \\ (.017)} & \makecell{.110 \\ (.012)} 
& \makecell{.130 \\ (.009)} & \makecell{.123 \\ (.010)} & \makecell{.113 \\ (.010)}
& \makecell{.110 \\ (.019)} & \makecell{.257 \\ (.038)} & \makecell{.301 \\ (.042)} 
& \makecell{.137 \\ (.011)} & \makecell{.157 \\ (.011)} & \makecell{.135 \\ (.010)} \\
400 & 30 & MAE 
& \makecell{.110 \\ (.028)} & \makecell{.127 \\ (.035)} & \makecell{.217 \\ (.036)} 
& \makecell{.099 \\ (.010)} & \makecell{.105 \\ (.014)} & \makecell{.106 \\ (.010)}
& \makecell{.179 \\ (.041)} & \makecell{.282 \\ (.077)} & \makecell{.294 \\ (.083)} 
& \makecell{.120 \\ (.010)} & \makecell{.110 \\ (.011)} & \makecell{.162 \\ (.018)} \\
    &    & RMSE 
& \makecell{.110 \\ (.030)} & \makecell{.127 \\ (.035)} & \makecell{.217 \\ (.034)} 
& \makecell{.123 \\ (.009)} & \makecell{.126 \\ (.016)} & \makecell{.127 \\ (.011)}
& \makecell{.179 \\ (.040)} & \makecell{.282 \\ (.075)} & \makecell{.294 \\ (.088)} 
& \makecell{.147 \\ (.009)} & \makecell{.127 \\ (.011)} & \makecell{.200 \\ (.022)} \\
\midrule
600 & 6 & MAE 
& \makecell{.099 \\ (.012)} & \makecell{.120 \\ (.063)} & \makecell{.326 \\ (.058)} 
& \makecell{.102 \\ (.011)} & \makecell{.080 \\ (.018)} & \makecell{.131 \\ (.028)}
& \makecell{.162 \\ (.012)} & \makecell{.188 \\ (.017)} & \makecell{.211 \\ (.018)} 
& \makecell{.126 \\ (.005)} & \makecell{.113 \\ (.004)} & \makecell{.107 \\ (.004)} \\
    &   & RMSE 
& \makecell{.099 \\ (.012)} & \makecell{.120 \\ (.062)} & \makecell{.326 \\ (.057)} 
& \makecell{.108 \\ (.012)} & \makecell{.101 \\ (.024)} & \makecell{.153 \\ (.029)}
& \makecell{.162 \\ (.013)} & \makecell{.188 \\ (.017)} & \makecell{.211 \\ (.020)} 
& \makecell{.149 \\ (.006)} & \makecell{.133 \\ (.005)} & \makecell{.127 \\ (.005)} \\
600 & 18 & MAE 
& \makecell{.104 \\ (.019)} & \makecell{.111 \\ (.017)} & \makecell{.067 \\ (.014)} 
& \makecell{.090 \\ (.007)} & \makecell{.097 \\ (.008)} & \makecell{.083 \\ (.005)}
& \makecell{.131 \\ (.022)} & \makecell{.207 \\ (.026)} & \makecell{.264 \\ (.031)} 
& \makecell{.101 \\ (.008)} & \makecell{.091 \\ (.007)} & \makecell{.084 \\ (.006)} \\
    &    & RMSE 
& \makecell{.104 \\ (.019)} & \makecell{.111 \\ (.018)} & \makecell{.067 \\ (.014)} 
& \makecell{.111 \\ (.011)} & \makecell{.114 \\ (.008)} & \makecell{.099 \\ (.005)}
& \makecell{.131 \\ (.023)} & \makecell{.207 \\ (.023)} & \makecell{.264 \\ (.032)} 
& \makecell{.127 \\ (.011)} & \makecell{.112 \\ (.008)} & \makecell{.102 \\ (.007)} \\
600 & 30 & MAE 
& \makecell{.128 \\ (.058)} & \makecell{.130 \\ (.031)} & \makecell{.262 \\ (.061)} 
& \makecell{.144 \\ (.025)} & \makecell{.120 \\ (.042)} & \makecell{.105 \\ (.009)}
& \makecell{.103 \\ (.013)} & \makecell{.270 \\ (.032)} & \makecell{.245 \\ (.037)} 
& \makecell{.094 \\ (.005)} & \makecell{.103 \\ (.005)} & \makecell{.143 \\ (.007)} \\
    &    & RMSE 
& \makecell{.128 \\ (.054)} & \makecell{.130 \\ (.031)} & \makecell{.262 \\ (.063)} 
& \makecell{.174 \\ (.025)} & \makecell{.157 \\ (.050)} & \makecell{.132 \\ (.014)}
& \makecell{.103 \\ (.012)} & \makecell{.270 \\ (.031)} & \makecell{.245 \\ (.038)} 
& \makecell{.113 \\ (.006)} & \makecell{.125 \\ (.006)} & \makecell{.177 \\ (.010)} \\
\bottomrule
\end{tabular}}
\end{table}

\begin{table}[h!]
\centering
\caption{{Estimation accuracy of covariate effects on attribute ($k$) mastery acquisition transitions ($\boldsymbol{\gamma}_{01}$), evaluated by mean absolute error (MAE) and root mean square error (RMSE). Results are reported for two transitions, from time 1 to time 2 $\gamma_{01,k}^{12}$ for attribute $k$, from time 2 to time 3 $\gamma_{01,k}^{23}$ for attribute $k$, under varying sample sizes ($N$) and numbers of items ($J_t$), for both Sparse and Dense $Q$-matrices. Standard errors (SE) are bootstrap-based (1,000 resamples). Bolded values in the table highlight the highest values.}}
\label{tab:gamma01_12_23_metrics}
\footnotesize
\setlength{\tabcolsep}{2pt}
{
\begin{tabular}[t]{ccc|cccccc|cccccc}
\toprule
\multicolumn{3}{c}{} &
\multicolumn{6}{c|}{Sparse $Q$} &
\multicolumn{6}{c}{Dense $Q$} \\
\cmidrule(lr){4-9}\cmidrule(lr){10-15}
$N$ & $J_t$ & Metric &
\makecell{$\gamma_{01,1}^{12}$\\(SE)} &
\makecell{$\gamma_{01,2}^{12}$\\(SE)} &
\makecell{$\gamma_{01,3}^{12}$\\(SE)} &
\makecell{$\gamma_{01,1}^{23}$\\(SE)} &
\makecell{$\gamma_{01,2}^{23}$\\(SE)} &
\makecell{$\gamma_{01,3}^{23}$\\(SE)} &
\makecell{$\gamma_{01,1}^{12}$\\(SE)} &
\makecell{$\gamma_{01,2}^{12}$\\(SE)} &
\makecell{$\gamma_{01,3}^{12}$\\(SE)} &
\makecell{$\gamma_{01,1}^{23}$\\(SE)} &
\makecell{$\gamma_{01,2}^{23}$\\(SE)} &
\makecell{$\gamma_{01,3}^{23}$\\(SE)} \\
\midrule
200 & 6 & MAE
& \makecell{.209 \\ (.018)} & \makecell{.272 \\ (.040)} & \makecell{.281 \\ (.036)}
& \makecell{.318 \\ (.046)} & \makecell{.326 \\ (.028)} & \makecell{.382 \\ (.032)}
& \makecell{.269 \\ (.016)} & \makecell{.293 \\ (.019)} & \makecell{.296 \\ (.017)}
& \makecell{.316 \\ (.017)} & \makecell{.366 \\ (.012)} & \makecell{.351 \\ (.019)} \\
& & RMSE
& \makecell{.289 \\ (.031)} & \makecell{.324 \\ (.038)} & \makecell{.351 \\ (.041)}
& \makecell{.400 \\ (.065)} & \makecell{.409 \\ (.032)} & \makecell{\textbf{.442} \\ (.028)}
& \makecell{.331 \\ (.018)} & \makecell{.350 \\ (.023)} & \makecell{.367 \\ (.020)}
& \makecell{.387 \\ (.020)} & \makecell{\textbf{.440} \\ (.016)} & \makecell{.412 \\ (.020)} \\
200 & 18 & MAE
& \makecell{.157 \\ (.017)} & \makecell{.132 \\ (.015)} & \makecell{.196 \\ (.014)}
& \makecell{.228 \\ (.026)} & \makecell{.260 \\ (.050)} & \makecell{.219 \\ (.045)}
& \makecell{.130 \\ (.021)} & \makecell{.120 \\ (.018)} & \makecell{.185 \\ (.024)}
& \makecell{.259 \\ (.080)} & \makecell{.167 \\ (.028)} & \makecell{.279 \\ (.025)} \\
& & RMSE
& \makecell{.190 \\ (.019)} & \makecell{.161 \\ (.019)} & \makecell{.241 \\ (.020)}
& \makecell{.282 \\ (.016)} & \makecell{.320 \\ (.062)} & \makecell{.259 \\ (.049)}
& \makecell{.157 \\ (.028)} & \makecell{.154 \\ (.022)} & \makecell{.218 \\ (.029)}
& \makecell{.309 \\ (.100)} & \makecell{.209 \\ (.034)} & \makecell{.336 \\ (.034)} \\
200 & 30 & MAE
& \makecell{.178 \\ (.013)} & \makecell{.168 \\ (.008)} & \makecell{.175 \\ (.019)}
& \makecell{.205 \\ (.014)} & \makecell{.223 \\ (.016)} & \makecell{.190 \\ (.017)}
& \makecell{.206 \\ (.016)} & \makecell{.173 \\ (.027)} & \makecell{.174 \\ (.028)}
& \makecell{.205 \\ (.019)} & \makecell{.176 \\ (.025)} & \makecell{.261 \\ (.030)} \\
& & RMSE
& \makecell{.215 \\ (.015)} & \makecell{.206 \\ (.010)} & \makecell{.213 \\ (.023)}
& \makecell{.251 \\ (.017)} & \makecell{.280 \\ (.019)} & \makecell{.231 \\ (.021)}
& \makecell{.250 \\ (.018)} & \makecell{.217 \\ (.028)} & \makecell{.226 \\ (.042)}
& \makecell{.251 \\ (.026)} & \makecell{.213 \\ (.026)} & \makecell{.307 \\ (.032)} \\
\midrule
400 & 6 & MAE
& \makecell{.255 \\ (.022)} & \makecell{.161 \\ (.033)} & \makecell{.218 \\ (.035)}
& \makecell{.230 \\ (.033)} & \makecell{.258 \\ (.025)} & \makecell{.312 \\ (.022)}
& \makecell{.186 \\ (.009)} & \makecell{.202 \\ (.011)} & \makecell{.183 \\ (.008)}
& \makecell{.249 \\ (.012)} & \makecell{.249 \\ (.010)} & \makecell{.255 \\ (.011)} \\
& & RMSE
& \makecell{.298 \\ (.018)} & \makecell{.188 \\ (.034)} & \makecell{.268 \\ (.034)}
& \makecell{.301 \\ (.046)} & \makecell{.311 \\ (.014)} & \makecell{.412 \\ (.034)}
& \makecell{.232 \\ (.011)} & \makecell{.249 \\ (.013)} & \makecell{.230 \\ (.010)}
& \makecell{.295 \\ (.015)} & \makecell{.303 \\ (.011)} & \makecell{.313 \\ (.014)} \\
400 & 18 & MAE
& \makecell{.108 \\ (.006)} & \makecell{.104 \\ (.007)} & \makecell{.105 \\ (.010)}
& \makecell{.148 \\ (.008)} & \makecell{.140 \\ (.012)} & \makecell{.140 \\ (.009)}
& \makecell{.127 \\ (.009)} & \makecell{.106 \\ (.008)} & \makecell{.129 \\ (.013)}
& \makecell{.141 \\ (.009)} & \makecell{.133 \\ (.014)} & \makecell{.143 \\ (.009)} \\
& & RMSE
& \makecell{.127 \\ (.006)} & \makecell{.124 \\ (.008)} & \makecell{.129 \\ (.011)}
& \makecell{.181 \\ (.010)} & \makecell{.169 \\ (.013)} & \makecell{.171 \\ (.011)}
& \makecell{.157 \\ (.012)} & \makecell{.129 \\ (.009)} & \makecell{.154 \\ (.015)}
& \makecell{.172 \\ (.009)} & \makecell{.168 \\ (.017)} & \makecell{.178 \\ (.010)} \\
400 & 30 & MAE
& \makecell{.135 \\ (.017)} & \makecell{.095 \\ (.009)} & \makecell{.133 \\ (.011)}
& \makecell{.152 \\ (.015)} & \makecell{.149 \\ (.013)} & \makecell{.187 \\ (.009)}
& \makecell{.110 \\ (.006)} & \makecell{.119 \\ (.022)} & \makecell{.137 \\ (.013)}
& \makecell{.126 \\ (.015)} & \makecell{.137 \\ (.011)} & \makecell{.157 \\ (.014)} \\
& & RMSE
& \makecell{.163 \\ (.020)} & \makecell{.113 \\ (.010)} & \makecell{.164 \\ (.011)}
& \makecell{.179 \\ (.017)} & \makecell{.171 \\ (.015)} & \makecell{.225 \\ (.011)}
& \makecell{.133 \\ (.009)} & \makecell{.150 \\ (.026)} & \makecell{.183 \\ (.021)}
& \makecell{.154 \\ (.017)} & \makecell{.171 \\ (.013)} & \makecell{.211 \\ (.025)} \\
\midrule
600 & 6 & MAE
& \makecell{.158 \\ (.010)} & \makecell{.189 \\ (.034)} & \makecell{.134 \\ (.023)}
& \makecell{.258 \\ (.052)} & \makecell{.219 \\ (.019)} & \makecell{.168 \\ (.015)}
& \makecell{.145 \\ (.005)} & \makecell{.147 \\ (.007)} & \makecell{.174 \\ (.008)}
& \makecell{.216 \\ (.010)} & \makecell{.203 \\ (.008)} & \makecell{.191 \\ (.008)} \\
& & RMSE
& \makecell{.200 \\ (.018)} & \makecell{.249 \\ (.050)} & \makecell{.163 \\ (.027)}
& \makecell{.308 \\ (.065)} & \makecell{.266 \\ (.021)} & \makecell{.223 \\ (.039)}
& \makecell{.177 \\ (.006)} & \makecell{.181 \\ (.009)} & \makecell{.228 \\ (.011)}
& \makecell{.268 \\ (.012)} & \makecell{.249 \\ (.010)} & \makecell{.233 \\ (.009)} \\
600 & 18 & MAE
& \makecell{.085 \\ (.005)} & \makecell{.090 \\ (.007)} & \makecell{.087 \\ (.005)}
& \makecell{.124 \\ (.007)} & \makecell{.120 \\ (.007)} & \makecell{.116 \\ (.007)}
& \makecell{.098 \\ (.006)} & \makecell{.094 \\ (.006)} & \makecell{.089 \\ (.007)}
& \makecell{.120 \\ (.008)} & \makecell{.113 \\ (.007)} & \makecell{.131 \\ (.009)} \\
& & RMSE
& \makecell{.104 \\ (.007)} & \makecell{.108 \\ (.008)} & \makecell{.104 \\ (.006)}
& \makecell{.153 \\ (.008)} & \makecell{.143 \\ (.008)} & \makecell{.141 \\ (.008)}
& \makecell{.121 \\ (.008)} & \makecell{.111 \\ (.006)} & \makecell{.110 \\ (.008)}
& \makecell{.145 \\ (.008)} & \makecell{.136 \\ (.008)} & \makecell{.159 \\ (.010)} \\
600 & 30 & MAE
& \makecell{.090 \\ (.021)} & \makecell{.107 \\ (.024)} & \makecell{.106 \\ (.009)}
& \makecell{.111 \\ (.016)} & \makecell{.091 \\ (.013)} & \makecell{.129 \\ (.008)}
& \makecell{.097 \\ (.005)} & \makecell{.122 \\ (.007)} & \makecell{.131 \\ (.006)}
& \makecell{.110 \\ (.005)} & \makecell{.112 \\ (.005)} & \makecell{.136 \\ (.005)} \\
& & RMSE
& \makecell{.108 \\ (.022)} & \makecell{.139 \\ (.034)} & \makecell{.123 \\ (.009)}
& \makecell{.142 \\ (.019)} & \makecell{.109 \\ (.015)} & \makecell{.158 \\ (.012)}
& \makecell{.114 \\ (.005)} & \makecell{.151 \\ (.009)} & \makecell{.169 \\ (.009)}
& \makecell{.135 \\ (.006)} & \makecell{.141 \\ (.005)} & \makecell{.176 \\ (.007)} \\
\bottomrule
\end{tabular}}
\end{table}

\section{Discussions and Future Directions}
This study introduces a temporal cognitive diagnostic modeling framework for analysing student learning from digital educational tools. Application of the framework to log data from a research-informed digital reading program demonstrates its flexibility, interpretability, and alignment with theoretical models of early reading development. Simulation studies further confirm the accuracy and robustness of the proposed model across various conditions.

In this study, we examined decoding and vocabulary as two subcomponents of the critical constructs proposed in the Simple View of Reading \citep{gough1986}: word recognition and language comprehension, respectively. Decoding supports the development of word recognition, while vocabulary supports language comprehension. These are theoretically considered distinct constructs. Our findings reinforce this distinction: transitions from non-mastery to mastery within each skill (decoding or vocabulary) followed expected developmental trajectories, and some students mastered one skill without the other. The theoretical independence, together with our empirical observation that a substantial proportion of students mastered only one of the two skills, justified the use of a non-compensatory modeling framework. 

However, the estimation of the $Q$-matrix for this specific dataset also revealed evidence of interrelations between word recognition and vocabulary skills. Specifically, students appeared to require vocabulary knowledge to support their performance in word recognition skill games, consistent with findings reported by \citet{LARRC2015}. Students who demonstrated mastery in both skills may have acquired them in parallel or may have benefited from prior familiarity with the digital learning environment. Log-based covariates and demographics contributed meaningfully to profile classification. Higher achievement (i.e., more levels mastered), fewer reattempts, and faster response times were positively associated with mastery, in line with prior evidence on the roles of practice and fluency in early literacy development \citep{laberge1974toward, nichols2008fluency}. Students with weaker initial literacy skills were less likely to master decoding, possibly due to persistent challenges in acquiring letter-sound correspondence {knowledge}. Furthermore, boys exhibited a lower likelihood of mastering vocabulary compared to girls, consistent with previous findings on gender differences in vocabulary development \citep{huttenlocher1991early}. By integrating multiple log-derived behavioural indicators and individual characteristics, rather than relying solely on a single measure, our modeling approach more effectively captures the dynamic, multidimensional nature of early reading development, thus better reflecting the complexity of real-world cognitive processes. Additionally, an important contribution of our study {comes from} in its ability to empirically validate theoretically specified item-skill mappings ($Q$-matrix). Thus, our approach aligns with expert design while improving the robustness and validity of digital reading tools in practice.

In simulations, the proposed MCMC algorithm reliably recovered latent profiles, item-skill mapping ($Q$-matrix), and parameters across a range of conditions, supporting the model’s practical utility in real-world applications where the $Q$-matrix structure is unknown. Notably, $Q$-matrix recovery remained consistently high even under relatively small designs (e.g., 200 students and 6 items), diverging from earlier recommendations (e.g., \citealt{delatorre2011}; \citealt{liu2025general}) and highlighting the model’s robustness in recovering item–skill mappings from limited data. However, static validation methods such as PVAF \citep{de2016general} appear less suited for dynamic models that incorporate covariates and attribute transitions. In contrast, increases in sample size or item numbers—regardless of the sparsity of the $Q$ matrix—consistently improved parameter recovery, in line with prior findings by \citet{liu2025general} and \citet{yu2023dont}. Furthermore, our classification accuracy for attribute profiles closely matched the accuracy of recovery of the unknown parameters reported in \citet{liu2025general}, despite not assuming the $Q$-matrix to be known. These results indicate that high accuracy can still be achieved even when the $Q$-matrix is estimated rather than fixed.

{Several directions may extend the current modeling framework. 
One avenue concerns the incorporation of richer data structures commonly available in digital learning environments. 
While the present application focused on time-invariant covariates, future work could incorporate time-varying predictors to capture dynamic influences on learning trajectories and behavioral indicators derived from process data (e.g., number of attempts, response time, and level mastery). 
Because such indicators may contain measurement error, future research could treat these predictors as latent variables to partition measurement error from true score variance \citep{cole2014manifest} or incorporate hierarchical Bayesian approaches to explicitly account for uncertainty in behavioral metrics \citep{gelman2013bayesian}. 
The diversity of covariates available in digital environments also suggests opportunities to incorporate covariate selection methods.}

A second direction involves further methodological developments in the measurement model. 
Future work may explore approaches for estimating the dimensionality of skills, such as parallel analysis. Related extensions may also consider more flexible skill representations, such as polytomous mastery states reflecting partial proficiency (e.g., fluency), and modeling strategies that distinguish item-level contextual variation from latent skill development when estimating time-varying $Q$-matrices. 
Variations in estimated $Q$-matrices across time points should therefore not necessarily be interpreted as shifts in students' cognitive integration, as such changes may reflect differences in instructional content or game levels encountered over time, related work examining time-invariant $Q$-matrices provides a complementary perspective \citep{Ma2026JointStepwise}. 
Extending the framework to longer-term or more frequent assessments may also reveal richer developmental patterns and potential change points in learning. In the present study, the model assumed item exchangeability across game levels. Introducing greater item variation and modeling item-level heterogeneity could further improve generalisability. {An additional extension related to the dynamic CDMs is the temporal dependence between time-varying $Q$-matrices. In the current framework, $Q$-matrices at different time points are treated as independent. This represents a limitation, as adjacent $Q$-matrices are plausibly temporally correlated in longitudinal learning settings. Future work could consider incorporating temporal dependence structures across $Q$-matrices, which may improve the stability and interpretability of estimated skill requirements over time.} 

{Finally, as demographic information such as gender and race is incorporated into the model, an important next step is to ensure measurement invariance across subgroups. 
Potential differential item functioning may arise if certain items or game levels favor specific demographic groups, which could lead to biased estimates of attribute profiles. 
The current Bayesian framework could be extended to detect such biases through multi-group dynamic CDM formulations \citep{delatorre2011}, supporting the fairness and validity of instructional recommendations.}

Lastly, although decoding and vocabulary were modeled using a non-compensatory framework, many other skills learning contexts may involve compensatory or partially compensatory relationships. Extensions of this framework to models such as {Deterministic Input, Noisy ``OR" Gate} (DINO) model \citep{templin2006measurement}, or the generalized DINA model \citep{de2016general} could reveal diverse patterns of skill integration and individual differences.

\section{Conclusion}

This study presents a temporal cognitive diagnostic modeling framework designed for analysing student learning through digital educational tools. By applying this framework to log files from a research-informed digital reading program, we demonstrated its flexibility, interpretability, and strong alignment with established theoretical models of early reading development. Simulation studies further confirmed the model's accuracy and robustness in jointly estimating the unknown item-skill mappings, latent attribute profiles, item parameters, and covariate effects across diverse conditions. From a policy perspective, our approach supports educational initiatives such as {Every student succeeds act (ESSA) \citep{usdoe2015}}, emphasising personalized instruction and evidence-based digital tools. Practically, the ability to dynamically track student transitions between non-mastery and mastery using response accuracy, log-derived behavioural indicators, and individual characteristics reflects the multidimensional complexity of reading development in real-world settings. The capacity to identify item-skill relationships solely from response data is an important contribution that can help educators to validate targeted educational tools, monitor their effectiveness longitudinally, and support students’ reading development. 

\section{Acknowledgments}

The authors acknowledge the support of Boost Reading at Amplify for providing the dataset used in this analysis.

\section{Funding}

This research was supported by the Engineering and Physical Sciences Research Council, which funded the first author through a PhD studentship.

\section{Conflict of Interest}

The authors declare no competing interests.

\section{Data Availability}

Due to the commercial sensitivity of the data, the data-sharing agreement with the company that provided the dataset requires the raw data to remain confidential and therefore they cannot be shared.

\bibliographystyle{chicago-doi}
\bibliography{biblio-3}

@article{He2021ProcessData,
  author  = {He, Qiwei and Borgonovi, Francesca and Paccagnella, Marco},
  title   = {Leveraging Process Data to Assess Adults’ Problem-Solving Skills: Using Sequence Mining to Identify Behavioral Patterns across Digital Tasks},
  journal = {Computers \& Education},
  year    = {2021},
  volume  = {166},
  pages   = {104170},
  doi     = {10.1016/j.compedu.2021.104170}
}

@article{deValpine2017NIMBLE,
  author  = {de Valpine, Perry and Turek, Daniel and Paciorek, Christopher J. and Anderson-Bergman, Cliff and Lang, Duncan Temple and Bodik, Rastislav},
  title   = {Programming with Models: Writing Statistical Algorithms for General Model Structures with NIMBLE},
  journal = {Journal of Computational and Graphical Statistics},
  year    = {2017},
  volume  = {26},
  number  = {2},
  pages   = {403--413},
  doi     = {10.1080/10618600.2016.1172487}
}

@misc{Ma2026JointStepwise,
  author       = {Ma, Yawen and Wallin, Gabriel and Ushakova, Anastasia and Cain, Kate},
  title        = {A Comparison of Joint and Stepwise Dynamic Cognitive Diagnostic Models},
  howpublished = {Proceedings of the International Meeting of the Psychometric Society (IMPS2025)},
  year         = {2026},
  note         = {Manuscript under review}
}

@article{Yigit2021FirstOrder,
  author  = {Yigit, Hilal D. and Douglas, Jeffrey A.},
  title   = {First-Order Learning Models with the GDINA: Estimation with the EM Algorithm and Applications},
  journal = {Applied Psychological Measurement},
  year    = {2021},
  volume  = {45},
  number  = {3},
  pages   = {143--158},
  doi     = {10.1177/0146621621990746}
}

@Manual{deValpine2026nimble,
  title        = {{NIMBLE}: MCMC, Particle Filtering, and Programmable Hierarchical Modeling},
  author       = {Perry de Valpine and Christopher Paciorek and Daniel Turek and Nick Michaud and Cliff Anderson-Bergman and Fritz Obermeyer and Claudia Wehrhahn Cortes and Abel Rodrìguez and Duncan Temple Lang and Sally Paganin},
  year         = {2026},
  note         = {R package version 1.4.1},
  url          = {https://CRAN.R-project.org/package=nimble},
  doi          = {10.5281/zenodo.1211190}
}

@article{andersson2022,
  title={A review of the use of cognitive diagnostic models in international large-scale assessments},
  author={Andersson, Bj{\"o}rn and Shin, Hyo Jeong and Chiu, Chia-Yi},
  journal={Journal of Educational and Behavioral Statistics},
  year={2022},
  note={Focuses on the integration of CDMs in assessments like PISA and TIMSS},
  doi={10.3102/10769986221123456}
}

@book{mullis2023,
  title={TIMSS 2023 Assessment Frameworks},
  author={Mullis, Ina V.S. and Martin, Michael O. and von Davier, Matthias},
  year={2023},
  publisher={Boston College, TIMSS \& PIRLS International Study Center},
  url={https://timssandpirls.bc.edu/timss2023/frameworks/}
}

@book{oecd2023,
  title={PISA 2022 Results (Volume I): The State of Learning and Equity in Education},
  author={OECD},
  year={2023},
  publisher={OECD Publishing},
  address={Paris},
  doi={10.1787/53f2383c-en}
}

@book{oecd2013,
  title={Technical Report of the Survey of Adult Skills (PIAAC)},
  author={OECD},
  year={2013},
  publisher={OECD Publishing},
  address={Paris},
  url={https://www.oecd.org/skills/piaac/PIAAC_Technical_Report_2nd_Edition.pdf}
}

@article{vondavier2019,
  title={Item response theory and cognitive diagnostic models: A perspective on recent developments and future directions},
  author={von Davier, Matthias and Lee, Young-Sun},
  journal={Journal of Educational and Behavioral Statistics},
  volume={44},
  number={6},
  pages={645--653},
  year={2019},
  publisher={SAGE Publications},
  doi={10.3102/1076998619881825}
}

@article{delatorre2011,
  title={The generalized DINA model framework},
  author={de la Torre, Jimmy},
  journal={Psychometrika},
  volume={76},
  number={2},
  pages={179--199},
  year={2011},
  publisher={Springer},
  doi={10.1007/s11336-011-9207-7}
}

@article{wang2018ho_hmm,
  title={Using response times to assess learning progress: A joint model for responses and response times},
  author={Wang, Shiyu and Zhang, Susu and Douglas, Jeff and Culpepper, Steven},
  journal={Measurement: Interdisciplinary Research and Perspectives},
  volume={16},
  number={1},
  pages={45--58},
  year={2018},
  publisher={Taylor \& Francis},
  doi = {10.1080/15366367.2018.1435105}
}

@article{chen2015statistical,
  title={Statistical analysis of Q-matrix based diagnostic classification models},
  author={Chen, Yunxiao and Liu, Jingchen and Xu, Gongjun and Ying, Zhiliang},
  journal={Journal of the American Statistical Association},
  volume={110},
  number={510},
  pages={850--866},
  year={2015},
  publisher={Taylor \& Francis},
  doi = {10.1080/01621459.2014.934827}
}

@article{wang2018tracking,
  title={Tracking skill acquisition with cognitive diagnosis models: A higher-order, hidden Markov model with covariates},
  author={Wang, Shiyu and Yang, Yan and Culpepper, Steven Andrew and Douglas, Jeffrey A},
  journal={Journal of Educational and Behavioral Statistics},
  volume={43},
  number={1},
  pages={57--87},
  year={2018},
  publisher={SAGE Publications Sage CA: Los Angeles, CA},
  doi = {10.3102/1076998617719727}
}

@article{liang2023latent,
  title={Latent transition cognitive diagnosis model with covariates: A three-step approach},
  author={Liang, Qianru and la Torre, Jimmy de and Law, Nancy},
  journal={Journal of Educational and Behavioral Statistics},
  volume={48},
  number={6},
  pages={690--718},
  year={2023},
  publisher={SAGE Publications Sage CA: Los Angeles, CA},
  doi = {10.3102/10769986231163320}
}

@misc{usdoe2015,
  author       = {{U.S. Department of Education}},
  title        = {Every Student Succeeds Act (ESSA)},
  year         = {2015},
  howpublished = {Public Law No. 114-95},
  url          = {https://www.ed.gov/essa}
}

@article{gough1986,
  title     = {Decoding, reading, and reading disability},
  author    = {Gough, Philip B. and Tunmer, William E.},
  journal   = {Remedial and Special Education},
  volume    = {7},
  number    = {1},
  pages     = {6--10},
  year      = {1986},
  doi       = {10.1177/074193258600700104}
}

@article{liang2023,
  title     = {Latent transition cognitive diagnosis model with covariates: A three-step approach},
  author    = {Liang, Lili and Others},
  journal   = {Journal of Educational and Behavioral Statistics},
  volume    = {48},
  number    = {3},
  pages     = {324--345},
  year      = {2023},
  publisher = {SAGE Publications},
  doi       = {10.3102/10769986231163320}
}

@article{maier2022personalized,
  title={Personalized feedback in digital learning environments: Classification framework and literature review},
  author={Maier, Uwe and Klotz, Christian},
  journal={Computers and Education: Artificial Intelligence},
  volume={3},
  pages={100080},
  year={2022},
  publisher={Elsevier},
doi = {10.1016/j.caeai.2022.100080}
}

@article{ma2024,
  title={Application of cluster analysis to identify different reader groups through their engagement with a digital reading supplement},
  author={Ma, Yawen and Cain, Kate and Ushakova, Anastasia},
  journal={Computers \& Education},
  volume={214},
  pages={105025},
  year={2024},
  publisher={Elsevier},
  doi = {10.1016/j.compedu.2024.105025}
}

@article{chen2018bayesian,
  title={Bayesian estimation of the DINA Q matrix},
  author={Chen, Yinghan and Culpepper, Steven Andrew and Chen, Yuguo and Douglas, Jeffrey},
  journal={Psychometrika},
  volume={83},
  pages={89--108},
  year={2018},
  publisher={Springer},
  doi = {10.1007/s11336-017-9579-4}
}

@article{fang2019identifiability,
  title={On the identifiability of diagnostic classification models},
  author={Fang, Guanhua and Liu, Jingchen and Ying, Zhiliang},
  journal={Psychometrika},
  volume={84},
  number={1},
  pages={19--40},
  year={2019},
  publisher={Cambridge University Press \& Assessment},
  doi = {10.1007/s11336-018-09658-x}
}

@article{gelman1992inference,
  title={Inference from iterative simulation using multiple sequences},
  author={Gelman, Andrew and Rubin, Donald B},
  journal={Statistical Science},
  volume={7},
  number={4},
  pages={457--472},
  year={1992},
  publisher={Institute of Mathematical Statistics},
  doi = {10.1214/ss/1177011136}
}

@article{de2016general,
  title={A general method of empirical Q-matrix validation},
  author={de la Torre, Jimmy and Chiu, Chia-Yi},
  journal={Psychometrika},
  volume={81},
  number={2},
  pages={253--273},
  year={2016},
  publisher={Cambridge University Press \& Assessment},
  doi = {10.1007/s11336-015-9467-8}
}

@inproceedings{rojas2012choosing,
  title={Choosing between general and specific cognitive diagnosis models when the sample size is small},
  author={Rojas, G and de la Torre, J and Olea, J},
  booktitle={annual meeting of the National Council of Measurement in Education, Vancouver, British Columbia, Canada},
  year={2012}
}

@book{gelman2013bayesian,
  title={Bayesian Data Analysis},
  author={Gelman, Andrew and Carlin, John B. and Stern, Hal S. and Dunson, David B. and Vehtari, Aki and Rubin, Donald B.},
  edition={3rd},
  year={2013},
  publisher={Chapman and Hall/CRC},
  doi={10.1201/b16018}
}

@article{cole2014manifest,
  title={Manifest variable path analysis: Potentially serious and misleading consequences due to uncorrected measurement error},
  author={Cole, David A. and Preacher, Kristopher J.},
  journal={Psychological Methods},
  volume={19},
  number={2},
  pages={300--315},
  year={2014},
  publisher={American Psychological Association},
  doi={10.1037/a0033805}
}

@article{rupp2008effects,
  title={The effects of Q-matrix misspecification on parameter estimates and classification accuracy in the DINA model},
  author={Rupp, Andre A and Templin, Jonathan},
  journal={Educational and Psychological Measurement},
  volume={68},
  number={1},
  pages={78--96},
  year={2008},
  publisher={Sage Publications Sage CA: Los Angeles, CA},
  doi = {10.1177/0013164407301545}
}

@article{rupp2008unique,
  title={Unique characteristics of diagnostic classification models: A comprehensive review of the current state-of-the-art},
  author={Rupp, Andr{\'e} A and Templin, Jonathan L},
  journal={Measurement},
  volume={6},
  number={4},
  pages={219--262},
  year={2008},
  publisher={Taylor \& Francis},
  doi = {10.1080/15366360802490866}
}

@book{templin2010diagnostic,
  title={Diagnostic measurement: Theory, methods, and applications},
  author={Templin, Jonathan and Henson, Robert A and others},
  year={2010},
  publisher={Guilford press}
}

@article{liu2025general,
  title={A general dynamic learning model framework for cognitive diagnosis},
  author={Liu, Zichu and Wang, Shiyu and Xiao, Houping and Zhang, Shumei and Qiu, Tao},
  journal={British Journal of Mathematical and Statistical Psychology},
  year={2025},
  publisher={Wiley Online Library},
  doi = {10.1111/bmsp.12384}
}

@article{vehtari2021rank,
  title={Rank-normalization, folding, and localization: An improved $\hat{R}$ for assessing convergence of MCMC (with discussion)},
  author={Vehtari, Aki and Gelman, Andrew and Simpson, Daniel and Carpenter, Bob and B{\"u}rkner, Paul-Christian},
  journal={Bayesian Analysis},
  volume={16},
  number={2},
  pages={667--718},
  year={2021},
  publisher={International Society for Bayesian Analysis},
  doi = {10.1214/20-BA1221}
}

@misc{wayman2025restricted,
  author       = {Wayman, Eric Alan and Culpepper, Steven Andrew and Douglas, Jeff and Bowers, Jesse},
  title        = {A restricted latent class hidden Markov model for polytomous responses, polytomous attributes, and covariates: Identifiability and application},
  year         = {2025},
  eprint       = {2503.20940},
  archivePrefix = {arXiv},
  primaryClass = {stat.ME},
  note         = {arXiv preprint},
  url          = {https://arxiv.org/abs/2503.20940}
}

@article{hoover1990simple,
  title={The simple view of reading},
  author={Hoover, Wesley A. and Gough, Philip B.},
  journal={Reading and Writing: An Interdisciplinary Journal},
  volume={2},
  number={2},
  pages={127--160},
  year={1990},
  publisher={Springer},
  doi={10.1007/BF00401799}
}

@article{bray2010modeling,
  title={Modeling relations among discrete developmental processes: A general approach to associative latent transition analysis},
  author={Bray, Bethany C. and Lanza, Stephanie T. and Collins, Linda M.},
  journal={Structural Equation Modeling},
  volume={17},
  number={4},
  pages={541--569},
  year={2010},
  publisher={Taylor \& Francis},
  doi={10.1080/10705511.2010.510043}
}

@article{kaya2017assessing,
  title={Assessing change in latent skills across time with longitudinal cognitive diagnosis modeling: An evaluation of model performance},
  author={Kaya, Yasemin and Leite, Walter L.},
  journal={Educational and Psychological Measurement},
  volume={77},
  number={3},
  pages={369--388},
  year={2017},
  publisher={SAGE Publications},
  doi={10.1177/0013164416659314}
}

@article{wang2019joint,
  title={A joint modeling framework of responses and response times to assess learning outcomes},
  author={Wang, Shu and Zhang, Si and Shen, Yanjun},
  journal={Multivariate Behavioral Research},
  volume={55},
  pages={49--68},
  year={2019},
  publisher={Taylor \& Francis},
  doi={10.1080/00273171.2019.1607238}
}

@article{chen2013general,
  title={A general cognitive diagnosis model for expert-defined polytomous attributes},
  author={Chen, Jingchen and de la Torre, Jimmy},
  journal={Applied Psychological Measurement},
  volume={37},
  number={6},
  pages={419--437},
  year={2013},
  publisher={SAGE Publications},
  doi={10.1177/0146621613479818}
}

@article{xu2025polytomous,
  title={A polytomous extension of the higher-order, hidden Markov model with covariates and hierarchical learning trajectories},
  author={Xu, Xinxin and Ren, Siyu and Shan, Xiaoyu and Zhang, Dan},
  journal={Journal of Educational and Behavioral Statistics},
  volume={0},
  number={0},
  year={2025},
  publisher={SAGE Publications},
  doi={10.3102/10769986251313769}
}

@article{zhan2020partial,
  title={A partial mastery, higher-order latent structural model for polytomous attributes in cognitive diagnostic assessments},
  author={Zhan, Peng and Wang, Wei-Chung and Li, Xuelong},
  journal={Journal of Classification},
  volume={37},
  pages={328--351},
  year={2020},
  publisher={Springer},
  doi={10.1007/s00357-019-09323-7}
}

@article{wang2020using,
  title={Using response times and response accuracy to measure fluency within cognitive diagnosis models},
  author={Wang, Shu and Chen, Yue},
  journal={Psychometrika},
  volume={85},
  number={3},
  pages={600--629},
  year={2020},
  publisher={Springer},
  doi={10.1007/s11336-020-09717-2}
}

@techreport{newton2019examining,
  title        = {{\em Examining the impact of Amplify Reading on student literacy in Grades K--2: 2019 report}},
  author       = {Newton, Stephen and Gamble, Harrison and Su, Yu and Zoski, Jennifer and Damico, Danielle},
  year         = {2019},
  institution  = {ERIC},
  number       = {ED604917},
  note         = {Available from ERIC (Education Resources Information Center)},
}

@article{haertel1984application,
  title={An application of latent class models to assessment data},
  author={Haertel, Edward},
  journal={Applied Psychological Measurement},
  volume={8},
  number={3},
  pages={333--346},
  year={1984},
  publisher={Sage Publications Sage CA: Thousand Oaks, CA},
  doi = {10.1177/014662168400800311}
}

@article{junker2001cognitive,
  title={Cognitive assessment models with few assumptions, and connections with nonparametric item response theory},
  author={Junker, Brian W and Sijtsma, Klaas},
  journal={Applied Psychological Measurement},
  volume={25},
  number={3},
  pages={258--272},
  year={2001},
  publisher={Sage Publications Sage CA: Thousand Oaks, CA},
  doi = {10.1177/01466210122032064}
}

@article{templin2006measurement,
  title={Measurement of psychological disorders using cognitive diagnosis models.},
  author={Templin, Jonathan L and Henson, Robert A},
  journal={Psychological Methods},
  volume={11},
  number={3},
  pages={287},
  year={2006},
  publisher={American Psychological Association},
  doi = {10.1037/1082-989X.11.3.287}
}

@article{zhang2018modeling,
  title={Modeling learner heterogeneity: A mixture learning model with responses and response times},
  author={Zhang, Susu and Wang, Shiyu},
  journal={Frontiers in psychology},
  volume={9},
  pages={2339},
  year={2018},
  publisher={Frontiers Media SA},
  doi = {10.3389/fpsyg.2018.02339}
}

@article{culpepper2016revisiting,
  title={Revisiting the 4-parameter item response model: Bayesian estimation and application},
  author={Culpepper, Steven Andrew},
  journal={Psychometrika},
  volume={81},
  number={4},
  pages={1142--1163},
  year={2016},
  publisher={Cambridge University Press \& Assessment},
  doi = {10.1007/s11336-015-9477-6}
}

@article{gu2021sufficient,
  title={Sufficient and necessary conditions for the identifiability of the Q-matrix},
  author={Gu, Yuqi and Xu, Gongjun},
  journal={Statistica Sinica},
  volume={31},
  number={1},
  pages={449--472},
  year={2021},
  publisher={JSTOR},
  doi = {10.5705/ss.202018.0410}
}

@article{xu2018identifying,
  title={Identifying latent structures in restricted latent class models},
  author={Xu, Gongjun and Shang, Zhuoran},
  journal={Journal of the American Statistical Association},
  volume={113},
  number={523},
  pages={1284--1295},
  year={2018},
  publisher={Taylor \& Francis},
  doi = {10.1080/01621459.2017.1340889}
}

@article{liu2024mixture,
  title={A Mixture Fluency model using responses and response times with cognitive diagnosis model framework},
  author={Liu, Zichu and Wang, Shiyu and Zhang, Shumei and Qiu, Tao},
  journal={Behavior Research Methods},
  volume={56},
  number={4},
  pages={3396--3451},
  year={2024},
  publisher={Springer},
  doi = {10.3758/s13428-023-02113-5}
}

@article{chen2018hidden,
  title={A hidden Markov model for learning trajectories in cognitive diagnosis with application to spatial rotation skills},
  author={Chen, Yinghan and Culpepper, Steven Andrew and Wang, Shiyu and Douglas, Jeffrey},
  journal={Applied Psychological Measurement},
  volume={42},
  number={1},
  pages={5--23},
  year={2018},
  publisher={SAGE Publications Sage CA: Los Angeles, CA},
  doi = {10.1177/0146621617721250}
}

@article{wang2020development,
  title={The development of a multidimensional diagnostic assessment with learning tools to improve 3-D mental rotation skills},
  author={Wang, Shiyu and Hu, Yiling and Wang, Qi and Wu, Bian and Shen, Yawei and Carr, Martha},
  journal={Frontiers in Psychology},
  volume={11},
  pages={305},
  year={2020},
  publisher={Frontiers Media SA},
  doi = {10.3389/fpsyg.2020.00305}
}

@book{murphy2012machine,
  title={Machine learning: a probabilistic perspective},
  author={Murphy, Kevin P},
  year={2012},
  publisher={MIT press}
}

@article{zhan2018cognitive,
  title={Cognitive diagnosis modelling incorporating item response times},
  author={Zhan, Peida and Jiao, Hong and Liao, Dandan},
  journal={British Journal of Mathematical and Statistical Psychology},
  volume={71},
  number={2},
  pages={262--286},
  year={2018},
  publisher={Wiley Online Library},
  doi = {10.1111/bmsp.12114}
}

@article{ma2020empirical,
  title={An empirical Q-matrix validation method for the sequential generalized DINA model},
  author={Ma, Wenchao and de la Torre, Jimmy},
  journal={British Journal of Mathematical and Statistical Psychology},
  volume={73},
  number={1},
  pages={142--163},
  year={2020},
  publisher={Wiley Online Library},
  doi = {10.1111/bmsp.12156}
}

@article{tang2021does,
  title={Does diagnostic feedback promote learning? Evidence from a longitudinal cognitive diagnostic assessment},
  author={Tang, Fang and Zhan, Peida},
  journal={AERA Open},
  volume={7},
  pages={23328584211060804},
  year={2021},
  publisher={SAGE Publications Sage CA: Los Angeles, CA},
  doi = {10.1177/23328584211060804}
}

@article{chiu2009cluster,
  title={Cluster analysis for cognitive diagnosis: Theory and applications},
  author={Chiu, Chia-Yi and Douglas, Jeffrey A and Li, Xiaodong},
  journal={Psychometrika},
  volume={74},
  pages={633--665},
  year={2009},
  publisher={Springer},
  doi = {10.1007/s11336-009-9125-0}
}

@article{laberge1974toward,
  title={Toward a theory of automatic information processing in reading},
  author={LaBerge, David and Samuels, S Jay},
  journal={Cognitive psychology},
  volume={6},
  number={2},
  pages={293--323},
  year={1974},
  publisher={Elsevier},
  doi = {10.1016/0010-0285(74)90015-2}
}

@article{nichols2008fluency,
  title={Fluency in learning to read for meaning: Going beyond repeated readings},
  author={Nichols, William Dee and Rupley, William H and Rasinski, Timothy},
  journal={Literacy Research and Instruction},
  volume={48},
  number={1},
  pages={1--13},
  year={2008},
  publisher={Taylor \& Francis},
  doi = {10.1080/19388070802161906}
}

@article{yu2023dont,
  title={Don't worry about the anchor-item setting in longitudinal learning diagnostic assessments},
  author={Yu, Xiaolin and Zhan, Peng and Chen, Qian},
  journal={Frontiers in Psychology},
  volume={14},
  pages={1112463},
  year={2023},
  publisher={Frontiers},
  doi={10.3389/fpsyg.2023.1112463}
}

@article{huttenlocher1991early,
  title={Early vocabulary growth: relation to language input and gender.},
  author={Huttenlocher, Janellen and Haight, Wendy and Bryk, Anthony and Seltzer, Michael and Lyons, Thomas},
  journal={Developmental psychology},
  volume={27},
  number={2},
  pages={236},
  year={1991},
  publisher={American Psychological Association},
  doi = {10.1037/0012-1649.27.2.236}
}

@misc{cree2023economic,
  title        = {The economic \& social cost of illiteracy: A snapshot of illiteracy in a global context},
  author       = {Cree, Anthony and Kay, Andrew and Steward, June},
  year         = {2023},
  month        = {September},
  howpublished = {World Literacy Foundation},
  note         = {Final report},
  url          = {https://worldliteracyfoundation.org/wp-content/uploads/2023/09/The-Economic-Social-Cost-of-Illiteracy-2023.pdf}
}

@techreport{lindorff2024pirls,
  title        = {PIRLS 2021: National report for England. research report},
  author       = {Lindorff, Ariel and Stiff, Jamie and Kayton, Heather},
  year         = {2024},
  institution  = {UK Department for Education},
  type         = {Research Report},
  month        = {April},
  url          = {https://assets.publishing.service.gov.uk/government/uploads/system/uploads/attachment_data/file/1234567/PIRLS_2021_National_Report_England.pdf},
  note         = {Published by the University of Oxford for the UK Department for Education},
}

@book{mullis2023pirls,
  author       = {Mullis, Ina V. S. and von Davier, Matthias and Foy, Pierre and Fishbein, Bethany and Reynolds, Katherine A. and Wry, Erin},
  title        = {PIRLS 2021 international results in reading},
  year         = {2023},
  publisher    = {TIMSS \& PIRLS International Study Center, Lynch School of Education and Human Development, Boston College, and IEA},
  address      = {Chestnut Hill, MA},
  doi          = {10.6017/lse.tpisc.tr2103.kb5342},
  isbn         = {978-1-889938-67-7},
  url          = {https://pirls2021.org/results}
}

@article{LARRC2015,
  title={Learning to read: Should we keep things simple?},
   author    = {{Language and Reading Research Consortium}},
  journal={Reading Research Quarterly},
  volume={50},
  number={2},
  pages={151--169},
  year={2015},
  publisher={Wiley Online Library},
  doi       = {10.1002/rrq.99},
  url       = {https://doi.org/10.1002/rrq.99}
}

@inproceedings{foldnes2024school,
  title={School entry detection of struggling readers using gameplay data and machine learning},
  author={Foldnes, Nj{\aa}l and Uppstad, Per Henning and Gr{\o}nneberg, Steffen and Thomson, Jenny M},
  booktitle={Frontiers in Education},
  volume={9},
  pages={1487694},
  year={2024},
  organization={Frontiers Media SA},
  doi = {10.3389/feduc.2024.1487694}
}

@article{rajeb2023incorporating,
  title={Incorporating process information into cognitive diagnostic models: A four-component joint modeling approach},
  author={Rajeb, Mehdi and Ma, Wenchao and He, Qiwei and Shi, Qingzhou},
  journal={Journal of Educational and Behavioral Statistics},
  pages={10769986251334788},
  year={2023},
  publisher={SAGE Publications Sage CA: Los Angeles, CA},
  doi = {10.3102/10769986251334788}
}

@article{chen2023investigating,
  title={Investigating second language (L2) reading subskill associations: A cognitive diagnosis approach},
  author={Chen, Huilin and Cai, Yuyang and de la Torre, Jimmy},
  journal={Language Assessment Quarterly},
  volume={20},
  number={2},
  pages={166--189},
  year={2023},
  publisher={Taylor \& Francis},
  doi = {10.1080/15434303.2022.2140050}
}

@misc{universityoforegon2018dibels,
  author       = {{University of Oregon}},
  title        = {8th edition of dynamic indicators of basic early literacy skills (DIBELS)},
  year         = {2018},
  publisher    = {Center on Teaching and Learning},
  address      = {Eugene, Oregon},
  howpublished = {\url{https://dibels.uoregon.edu}},
  note         = {Accessed: 2025-05-29}
}

@article{chen2020multivariate,
  title = {A multivariate probit model for learning trajectories: A fine-grained evaluation of an educational intervention},
  author = {Chen, Yinyin and Culpepper, Steven A.},
  journal = {Applied Psychological Measurement},
  volume = {44},
  number = {7--8},
  pages = {515--530},
  year = {2020},
  doi = {10.1177/0146621620920928}
}

\setcounter{table}{0}
\setcounter{figure}{0}

\renewcommand{\thetable}{S\arabic{table}}
\renewcommand{\thefigure}{S\arabic{figure}}

\section{Supplementary Material}
Supplementary material is available online at \textit{Journal of the Royal Statistical Society: Series A}.

\section{Supplementary Material A}

For sparsity in the $Q$-matrix, 
\begin{align*}
Q_{jk} &\sim \mathrm{Bernoulli}(\theta), \\
\theta &\sim \mathrm{Beta}(\alpha,\beta),
\end{align*}
where the prior mean $\frac{\alpha}{\alpha+\beta}$ equals the non–zero proportion of the true $Q$-matrix and the concentration defined as $\alpha+\beta$.

To evaluate the sensitivity of model estimates to the prior specification of $\theta$, we compared item parameters results from the empirical analysis under varied concentration levels ($\alpha+\beta$) in Table \ref{Stab:prior_concentration_sensitivity} and varied priors mean ($\frac{\alpha}{\alpha+\beta}$) in Table \ref{Stab:prior_mean_sensitivity}, relative to the reference prior $\mathrm{Beta}(24,6)$ used in the main analysis. Tables \ref{Stab:item_gs_sensitivity_consentration} and \ref{Stab:item_gs_sensitivity_mean} summarises the changes in the posterior means of the guessing ($g$) and slipping ($s$) parameters under alternative priors.

Results showed that item parameters remained unchanged across different prior means, supporting the robustness of the proposed approach. Across all alternative settings, changes in the posterior means of $g$ and $s$ parameters were generally less than {0.03}, with most differences below 0.02. The sensitivity analysis indicates that the choice of prior concentration and mean for $\theta$ had minimal impact on the estimation of item parameters. The $\text{Beta}(24,6)$ prior thus provides a reasonable balance between interpretability and flexibility for empirical applications.

\begin{table}[htbp]
\centering
\caption{Alternative prior specifications for $\theta$ with fixed means $\frac{\alpha}{\alpha+\beta}=0.8$ and varying concentration ($\alpha+\beta$).}
\label{Stab:prior_concentration_sensitivity}
\begin{tabular}{lcccc}
\toprule
Prior & Prior Mean & Concentration & Variance & 95\% Interval \\
\midrule
$\textrm{Beta}(24,6)$ & 0.8 & 30 & 0.0052 & (0.65, 0.92) \\
$\textrm{Beta}(8,2)$ & 0.8 & 10 & 0.0145 & (0.55, 0.93) \\
$\textrm{Beta}(40,10)$ & 0.8 & 50 & 0.0031 & (0.70, 0.90) \\
\bottomrule
\end{tabular}
\end{table}

\begin{table}[ht]
\centering
\caption{Alternative prior specifications for $\theta$ with fixed concentration ($\alpha+\beta=30$) and varying means $\frac{\alpha}{\alpha+\beta}$.}
\label{Stab:prior_mean_sensitivity}
\begin{tabular}{lcccc}
\toprule
Prior & Prior Mean & Concentration & Variance & 95\% Interval \\
\midrule
$\textrm{Beta}(24,6)$  & 0.8 & 30 & 0.0052 & (0.65, 0.92) \\
$\textrm{Beta}(21,9)$  & 0.7 & 30 & 0.0067 & (0.55, 0.85) \\
$\textrm{Beta}(27,3)$  & 0.9 & 30 & 0.0031 & (0.78, 0.96) \\
\bottomrule
\end{tabular}
\end{table}

\begin{table}[ht]
\centering
\caption{Posterior means of guessing ($g$) and slipping ($s$) parameters for each item at Time 1 and Time 2 under three alternative Beta priors (concentration $10$, $30$, and $50$) whereas fixed mean 0.8. Bold values indicate the highest estimates across conditions.}
\label{Stab:item_gs_sensitivity_consentration}
{
\begin{tabular}{c|l|cc|cc}
\toprule
\multirow{2}{*}{Item} & \multirow{2}{*}{Prior} &
\multicolumn{2}{c|}{Time 1} & \multicolumn{2}{c}{Time 2} \\
\cmidrule(lr){3-4}\cmidrule(lr){5-6}
& & $g$ & $s$ & $g$ & $s$ \\
\midrule
\multirow{3}{*}{1}
 & $\textrm{Beta}(24,6)$  & 0.344 & 0.228 & 0.416 & 0.272 \\ 
& $\textrm{Beta}(8,2)$   & 0.345 & 0.234 & 0.418 & 0.275 \\ 
& $\textrm{Beta}(40,10)$ & 0.342 & 0.238 & 0.416 & 0.273 \\ \addlinespace
\multirow{3}{*}{2}
& $\textrm{Beta}(24,6)$  & 0.259 & 0.144 & 0.338 & 0.225 \\ 
& $\textrm{Beta}(8,2)$   & 0.214 & 0.176 & 0.327 & 0.225 \\ 
& $\textrm{Beta}(40,10)$ & 0.211 & 0.179 & 0.327 & 0.223 \\ \addlinespace
\multirow{3}{*}{3}
& $\textrm{Beta}(24,6)$  & 0.274 & \textbf{0.448} & 0.302 & \textbf{0.396} \\ 
& $\textrm{Beta}(8,2)$   & 0.263 & 0.475 & 0.284 & 0.384 \\ 
& $\textrm{Beta}(40,10)$ & 0.263 & 0.478 & 0.285 & 0.383 \\ \addlinespace
\multirow{3}{*}{4}
& $\textrm{Beta}(24,6)$  & 0.158 & 0.376 & 0.327 & 0.275 \\ 
& $\textrm{Beta}(8,2)$   & 0.130 & 0.392 & 0.320 & 0.265 \\ 
& $\textrm{Beta}(40,10)$ & 0.130 & 0.394 & 0.317 & 0.274 \\ \addlinespace
\multirow{3}{*}{5}
& $\textrm{Beta}(24,6)$  & 0.410 & 0.368 & \textbf{0.399} & 0.367 \\ 
& $\textrm{Beta}(8,2)$   & \textbf{0.427} & 0.369 & 0.371 & 0.355 \\ 
& $\textrm{Beta}(40,10)$ & \textbf{0.427} & 0.368 & 0.373 & 0.356 \\ \addlinespace
\multirow{3}{*}{6}
& $\textrm{Beta}(24,6)$  & 0.140 & 0.316 & 0.396 & 0.343 \\ 
& $\textrm{Beta}(8,2)$   & 0.114 & 0.341 & 0.452 & 0.352 \\ 
& $\textrm{Beta}(40,10)$ & 0.113 & 0.341 & 0.450 & 0.357 \\ 
\bottomrule
\end{tabular}}
\end{table}

\begin{table}[ht]
\centering
\caption{Posterior means of guessing ($g$) and slipping ($s$) parameters for each item at Time~1 and Time~2 under three alternative Beta priors (mean $0.7$, $0.8$, and $0.9$; concentration fixed at $30$). Bold values indicate the highest estimates across conditions.}
\label{Stab:item_gs_sensitivity_mean}
{
\begin{tabular}{c|l|cc|cc}
\toprule
\multirow{2}{*}{Item} & \multirow{2}{*}{Prior} &
\multicolumn{2}{c|}{Time 1} & \multicolumn{2}{c}{Time 2} \\
\cmidrule(lr){3-4}\cmidrule(lr){5-6}
& & $g$ & $s$ & $g$ & $s$ \\
\midrule
\multirow{3}{*}{1}
& $\textrm{Beta}(24,6)$  & 0.344 & 0.228 & 0.416 & 0.272 \\ 
& $\textrm{Beta}(21,9)$  & 0.346 & 0.234 & 0.417 & 0.275 \\ 
& $\textrm{Beta}(27,3)$  & 0.343 & 0.232 & \textbf{0.419} & 0.274 \\ \addlinespace
\multirow{3}{*}{2}
& $\textrm{Beta}(24,6)$  & 0.259 & 0.144 & 0.338 & 0.225 \\ 
& $\textrm{Beta}(21,9)$  & 0.254 & 0.152 & 0.319 & 0.227 \\ 
& $\textrm{Beta}(27,3)$  & 0.253 & 0.155 & 0.333 & 0.223 \\ \addlinespace
\multirow{3}{*}{3}
& $\textrm{Beta}(24,6)$  & 0.274 & 0.448 & 0.302 & \textbf{0.396} \\ 
& $\textrm{Beta}(21,9)$  & 0.266 & \textbf{0.475} & 0.290 & 0.388 \\ 
& $\textrm{Beta}(27,3)$  & 0.258 & \textbf{0.475} & 0.286 & 0.381 \\ \addlinespace
\multirow{3}{*}{4}
& $\textrm{Beta}(24,6)$  & 0.158 & 0.376 & 0.327 & 0.275 \\ 
& $\textrm{Beta}(21,9)$  & 0.131 & 0.392 & 0.330 & 0.252 \\ 
& $\textrm{Beta}(27,3)$  & 0.131 & 0.396 & 0.336 & 0.242 \\ \addlinespace
\multirow{3}{*}{5}
& $\textrm{Beta}(24,6)$  & 0.410 & 0.368 & 0.399 & 0.367 \\ 
& $\textrm{Beta}(21,9)$  & 0.446 & 0.369 & 0.387 & 0.363 \\ 
& $\textrm{Beta}(27,3)$  & \textbf{0.448} & 0.367 & 0.393 & 0.349 \\ \addlinespace
\multirow{3}{*}{6}
& $\textrm{Beta}(24,6)$  & 0.140 & 0.316 & 0.396 & 0.343 \\ 
& $\textrm{Beta}(21,9)$  & 0.114 & 0.340 & 0.369 & 0.363 \\ 
& $\textrm{Beta}(27,3)$  & 0.111 & 0.342 & 0.389 & 0.347 \\ 
\bottomrule
\end{tabular}}
\end{table}

\clearpage
\section{Supplementary Material B}

The exploratory data analysis in this section identifies and visualizes the ten most frequent three levels that students completed in each game and year. These panels highlight the variety of levels engaged, driven by differences in students’ initial starting points and in-game progression. Panels A and B display the top ten combinations from decoding game (Years 1 and 2), while Panels C and D show the corresponding results for vocabulary game (Years 1 and 2). 

\begin{figure}[ht]
  \centering
   \caption{Top ten three-level combinations engaged by students in each game and year. For example, panel A shows the most frequent level sequences for decoding game in Year 1. Each bar represents the number of students who completed the corresponding three-level combination.}
  \label{fig:eda_combined_panel}
  \includegraphics[width=\textwidth,
                   height=0.8\textheight,
                   keepaspectratio]{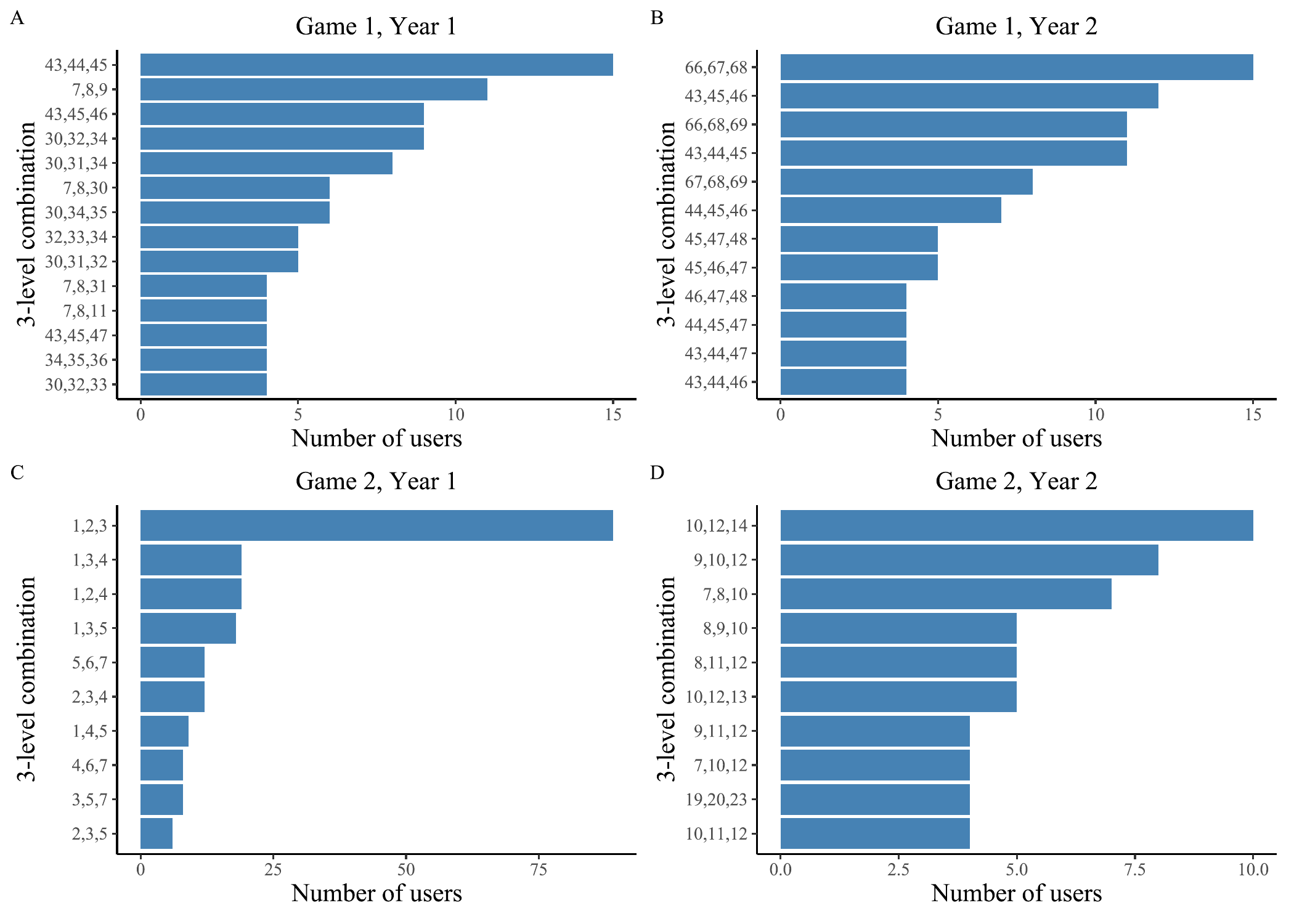}
 
\end{figure}

\begin{table}[!h]
\centering
\caption{Level content of the most frequently encountered level in Figure \ref{fig:eda_combined_panel}.}
\label{tab:level_content}
\begin{tabular}{llp{10cm}}
\toprule
\textbf{Game} & \textbf{Levels} & \textbf{Level content} \\
\midrule
Game 1 & 7--8 & Letters: m, a, t, o, d, c, g, i, n, s \\
Game 1 & 30--33 & Letters: m, a, t, o, d, c, g, i, n, s, h, f, v, z, p, e, b, l, r, u, w, k, j, y, x \\
Game 1 & 34--35 & Review of all individual letter sounds \\
Game 1 & 43--44 & Letter combinations: ch, qu, th (voiced), ng, ee, oo, sh, ou, th (unvoiced), aw \\
Game 1 & 45--47 & Letter combinations: ch, qu, th (voiced), ng, ee, oo, sh, ou, th (unvoiced), aw, er, ck; review of all combinations learned \\
Game 1 & 48 & Letter combinations: ch, qu, th (voiced), ng, ee, oo, sh, ou, th (unvoiced), aw, er, ck, oi, ar \\
Game 1 & 49 & Letter combinations: ch, qu, th (voiced), ng, ee, oo, sh, ou, th (unvoiced), aw, er, ck, oi, ar, tch, wr \\
Game 1 & 52 & Letter combinations: ch, qu, th (voiced), ng, ee, oo, sh, ou, th (unvoiced), aw, er, ck, oi, ar, tch, wr, ai, kn; review of all combinations learned \\
Game 1 & 66--67 & Letter combinations: ch, qu, th (voiced), ng, ee, oo, sh, ou, th (unvoiced), aw, er, ck, oi, ar, tch, wr, ai, kn, or, wh, ow, igh, ay, oy, al, ph, ow, ea, au, oa, ir, ol \\
Game 1 & 68 & Review of all letter combinations: ch, qu, th (voiced), ng, ee, oo, sh, ou, th (unvoiced), aw, er, ck, oi, ar, tch, wr, ai, kn, or, wh, ow, igh, ay, oy, al, ph, ow, ea, au, oa, ir, ol \\
\midrule
Game 2 & 1--18 & Extended vocabulary instruction across multiple contexts (grade K/1 words) \\
Game 2 & 19--36 & Extended vocabulary instruction across multiple contexts (grade 1/2 words) \\
\bottomrule
\end{tabular}
\end{table}

\clearpage
\section{Supplementary Material C}

The posterior means and 95\% confidence intervals (CIs) of the odds ratios (ORs) for $\beta_z$ (initial mastery) by attribute ($K$), as well as the transition probabilities $\gamma_{01}$, are reported in Tables \ref{Stab:betaZ_OR_part1}-\ref{Stab:gamma01_OR_part1}.

\begin{sidewaystable}[htbp]
\centering
\caption{{
Posterior means of odds ratios (OR) for $\beta_0$ and $\beta_z$ (initial mastery) by attribute ($K$), with 95\% confidence intervals (CI). Statistically significant results (CI excluding 1) are shown in \textbf{bold}. Part 1 of 2.\\
\textit{Note}: $K$ = attribute; (Intercept) = model intercept; nra = number of reattempts; nlm = number of levels mastered; rt = response time; gender = female (0), male (1); SEN = special education needs (1); ELL = English language learner (1); ILA = initial literacy ability (at benchmark: reference level); WB = well below benchmark; BB = below benchmark; AB = above benchmark; group = engagement group (group 5: reference); Others = American Indian, Alaskan Native, Black or African American, Hispanic or Latino, Multiracial/Other, and not Specified (reference group: White); DG = decoding game; VG = vocabulary game; HVG = high vocabulary group; HDG = high decoding group.
}}
\label{Stab:betaZ_OR_part1}
{
\begin{tabularx}{\linewidth}{ll*{7}{>{\centering\arraybackslash}X}}
\toprule
\textbf{K} & \textbf{Measure} & \textbf{(Intercept)} & \textbf{nra DG} & \textbf{nra VG} & \textbf{nlm DG} & \textbf{nlm VG} & \textbf{rt DG} & \textbf{rt VG} \\
\midrule
1 & OR & 0.469 & \textbf{0.067} & 1.641 & 1.112 & \textbf{0.450} & \textbf{12.974} &  1.087 \\
  & CI & (0.092, 2.550) & (0.019, 0.233) & (0.864, 3.394) & (0.518, 2.415) & (0.215, 0.914) & (5.830, 33.554) & (0.546, 2.219) \\
\addlinespace
2 & OR & 1.000 & 0.719 & 0.742 & 0.638 & 1.058 & 0.860 & 1.076 \\
  & CI & (0.228, 4.150) & (0.164, 3.222) & (0.194, 2.899) & (0.168, 2.357) & (0.242, 4.127) & (0.197, 3.792) & (0.195, 5.882) \\
\bottomrule
\end{tabularx}}
\end{sidewaystable}

\begin{sidewaystable}[htbp]
\caption*{Table \ref{Stab:betaZ_OR_part1}\ (continued): Part 2 of 2.}
\label{Stab:betaZ_OR_part2}
\setlength{\tabcolsep}{2pt}
{
\begin{tabularx}{\linewidth}{ll*{10}{>{\centering\arraybackslash}X}}
\toprule
\textbf{K} & \textbf{Measure} & \textbf{gender} & \textbf{SEN} & \textbf{ELL} &
\textbf{ILA‐WB} & \textbf{ILA‐BB} & \textbf{ILA‐AB} & \textbf{HVG} &
\textbf{HDG} & \textbf{Asian} & \textbf{Others}\\
\midrule
1 & OR & 1.595 & 15.858 & 0.784 & 0.752 & 1.579 & \textbf{3.816} & 0.782 & 3.555 & 0.583 & 0.404 \\
  & CI & \CI{0.735}{3.648} & \CI{6.519}{45.702} & \CI{0.324}{1.831} & \CI{0.337}{1.627} & \CI{0.670}{4.099} & \CI{1.433}{10.534} & \CI{0.230}{2.659} & \CI{1.072}{11.871} & \CI{0.166}{2.022} & \CI{0.123}{1.274} \\
\addlinespace
2 & OR & 1.046 & 1.090 & 0.936 & 0.557 & 1.021 & 1.233 & 0.519 & 1.004 & 0.655 & 0.746 \\
  & CI & \CI{0.243}{4.551} & \CI{0.273}{4.194} & \CI{0.238}{3.525} & \CI{0.144}{2.137} & \CI{0.241}{4.461} & \CI{0.339}{4.604} & \CI{0.095}{2.967} & \CI{0.190}{5.261} & \CI{0.188}{2.292} & \CI{0.214}{2.661} \\
\bottomrule
\end{tabularx}}
\end{sidewaystable}

\begin{sidewaystable}[htbp]
\centering
\caption{
Posterior means of odds ratios (OR) for $\gamma_{01}$ by attribute ($K$), with 95\% confidence intervals (CI). Statistically significant results (CI excluding 1) are shown in \textbf{bold}. Part 1 of 2.\\
\textit{Note}: $K$ = attribute; (Intercept) = model intercept; nra = number of reattempts; nlm = number of levels mastered; rt = response time; gender = female (0), male (1); SEN = special education needs (1); ELL = English language learner (1); ILA = initial literacy ability (at benchmark: reference level); WB = well below benchmark; BB = below benchmark; AB = above benchmark; group = engagement group (group 5: reference); Others = American Indian, Alaskan Native, Black or African American, Hispanic or Latino, Multiracial/Other, and not Specified (reference group: White); DG = decoding game; VG = vocabulary game; HVG = high vocabulary group; HDG = high decoding group.
}
\label{Stab:gamma01_OR_part1}
{
\begin{tabularx}{\linewidth}{ll*{7}{>{\centering\arraybackslash}X}}
\toprule
\textbf{K} & \textbf{Measure} & \textbf{(Intercept)} & \textbf{nra DG} & \textbf{nra VG} & \textbf{nlm DG} & \textbf{nlm VG} & \textbf{rt DG} & \textbf{rt VG} \\
\midrule
1 & OR & 1.110 & 1.102 & 1.004 & \textbf{4.147} & 1.084 & 2.151 & 2.593  \\
  & CI & (0.193, 5.778) & (0.214,6.053) & (0.167,5.893) & (1.239,10.248) & (0.236,3.858) & (0.515,9.469) & (0.513,11.707) \\
\addlinespace
2 & OR & 1.351 & 1.023 & \textbf{0.371} & 0.471 & 0.421 & 1.011 & \textbf{3.116} \\
  & CI & (0.258,7.096) & (0.168,6.183) & (0.067,0.818) & (0.103,2.299) & (0.072, 2.830) & (0.189,5.483) & (1.198,9.287) \\
\bottomrule
\end{tabularx}}
\end{sidewaystable}

\begin{sidewaystable}[htbp]
\centering
\caption*{Table \ref{Stab:gamma01_OR_part1}\ (continued): Part 2 of 2.}
\label{Stab:gamma01_OR_part2}
\setlength{\tabcolsep}{2pt}
{
\begin{tabularx}{\linewidth}{ll*{10}{>{\centering\arraybackslash}X}}
\toprule
\textbf{K} & \textbf{Measure} & \textbf{gender} & \textbf{SEN} & \textbf{ELL} & \textbf{ILA‐WB} & \textbf{ILA‐BB} & \textbf{ILA‐AB} & \textbf{HVG} & \textbf{HDG} & \textbf{Asian} & \textbf{Others}\\
\midrule
1 & OR & \textbf{0.894} & 1.316 & 0.858 & 1.282 & 1.102 & 1.038 & 0.546 & 1.103 & 0.926 & 1.052 \\
  & CI & \CI{0.785}{0.994} & \CI{0.335}{4.610} & \CI{0.191}{3.351} & \CI{0.376}{5.018} & \CI{0.266}{4.709} & \CI{0.184}{4.953} & \CI{0.091}{3.015} & \CI{0.252}{4.821} & \CI{0.154}{5.573} & \CI{0.226}{4.776} \\
\addlinespace
2 & OR & 0.860 & 0.684& 0.571 & 0.749 & 2.897 & 1.768 & 1.086 & 0.805 & 1.000 & 2.224 \\
  & CI & \CI{0.149}{5.506} & \CI{0.119}{4.359} & \CI{0.109}{2.994} & \CI{0.139}{4.203} & \CI{0.531}{14.151} & \CI{0.343}{9.576} & \CI{0.154}{7.615} & \CI{0.121}{5.265} & \CI{0.195}{5.243} & \CI{0.399}{11.901} \\
\bottomrule
\end{tabularx}}
\end{sidewaystable}

\clearpage
\section{Supplementary Material D}

The true $Q$-matrices are presented under different levels of sparsity and varying numbers of items in Tables~\ref{Stab:qmatrix_j6_sparse_dense}--\ref{Stab:qmatrix_j30_sparse_dense}. Each $Q$-matrix assumes $K = 3$ latent attributes and was held fixed across all simulation replicates. 

\begin{table}[H]
\centering
{
\caption{{True $Q$-matrices for $J_t = 6$ under sparse (left) and dense (right) attribute patterns (Time 1-3).}}
\label{Stab:qmatrix_j6_sparse_dense}
\begin{tabular}{c|ccc|ccc|ccc|ccc|ccc|ccc}
\toprule
\multirow{2}{*}{Item} &
\multicolumn{9}{c|}{\textbf{Sparse Matrix}} &
\multicolumn{9}{c}{\textbf{Dense Matrix}} \\
\cmidrule(lr){2-10} \cmidrule(lr){11-19}
& \multicolumn{3}{c|}{Time 1} & \multicolumn{3}{c|}{Time 2} & \multicolumn{3}{c|}{Time 3}
& \multicolumn{3}{c|}{Time 1} & \multicolumn{3}{c|}{Time 2} & \multicolumn{3}{c}{Time 3} \\
\cmidrule(lr){2-4} \cmidrule(lr){5-7} \cmidrule(lr){8-10}
\cmidrule(lr){11-13} \cmidrule(lr){14-16} \cmidrule(lr){17-19}
& $A_1$ & $A_2$ & $A_3$ & $A_1$ & $A_2$ & $A_3$ & $A_1$ & $A_2$ & $A_3$
& $A_1$ & $A_2$ & $A_3$ & $A_1$ & $A_2$ & $A_3$ & $A_1$ & $A_2$ & $A_3$ \\
\midrule
1 & 1 & 0 & 0 & 1 & 0 & 0 & 1 & 0 & 0 & 1 & 0 & 0 & 1 & 0 & 0 & 1 & 0 & 0 \\
2 & 0 & 1 & 0 & 0 & 1 & 0 & 0 & 1 & 0 & 0 & 1 & 0 & 0 & 1 & 0 & 0 & 1 & 0 \\
3 & 0 & 0 & 1 & 0 & 0 & 1 & 0 & 0 & 1 & 0 & 0 & 1 & 0 & 0 & 1 & 0 & 0 & 1 \\
4 & 1 & 1 & 0 & 1 & 1 & 0 & 1 & 0 & 1 & 1 & 1 & 0 & 1 & 1 & 1 & 0 & 1 & 1 \\
5 & 1 & 0 & 1 & 1 & 1 & 0 & 1 & 0 & 1 & 1 & 0 & 1 & 1 & 1 & 0 & 1 & 1 & 1 \\
6 & 0 & 1 & 1 & 0 & 1 & 1 & 0 & 1 & 1 & 1 & 1 & 1 & 0 & 1 & 1 & 1 & 0 & 1 \\
\bottomrule
\end{tabular} }
\end{table}

\begin{table}[H]
\centering
{
\caption{{True $Q$-matrices for $J_t = 18$ under sparse (left) and dense (right) attribute patterns (Time 1-3).}}
\label{Stab:qmatrix_j18_sparse_dense}
\begin{tabular}{c|ccc|ccc|ccc|ccc|ccc|ccc}
\toprule
\multirow{2}{*}{Item} &
\multicolumn{9}{c|}{\textbf{Sparse Matrix}} &
\multicolumn{9}{c}{\textbf{Dense Matrix}} \\
\cmidrule(lr){2-10} \cmidrule(lr){11-19}
& \multicolumn{3}{c|}{Time 1} & \multicolumn{3}{c|}{Time 2} & \multicolumn{3}{c|}{Time 3}
& \multicolumn{3}{c|}{Time 1} & \multicolumn{3}{c|}{Time 2} & \multicolumn{3}{c}{Time 3} \\
\cmidrule(lr){2-4} \cmidrule(lr){5-7} \cmidrule(lr){8-10}
\cmidrule(lr){11-13} \cmidrule(lr){14-16} \cmidrule(lr){17-19}
& $A_1$ & $A_2$ & $A_3$ & $A_1$ & $A_2$ & $A_3$ & $A_1$ & $A_2$ & $A_3$
& $A_1$ & $A_2$ & $A_3$ & $A_1$ & $A_2$ & $A_3$ & $A_1$ & $A_2$ & $A_3$ \\
\midrule
1  & 1 & 0 & 0 & 1 & 0 & 0 & 1 & 0 & 0 & 1 & 0 & 0 & 1 & 0 & 0 & 1 & 0 & 0 \\
2  & 0 & 1 & 0 & 0 & 1 & 0 & 0 & 1 & 0 & 0 & 1 & 0 & 0 & 1 & 0 & 0 & 1 & 0 \\
3  & 0 & 0 & 1 & 0 & 0 & 1 & 0 & 0 & 1 & 0 & 0 & 1 & 0 & 0 & 1 & 0 & 0 & 1 \\
4  & 1 & 1 & 0 & 1 & 1 & 0 & 1 & 1 & 0 & 1 & 1 & 0 & 1 & 1 & 1 & 1 & 1 & 0 \\
5  & 1 & 0 & 1 & 1 & 0 & 1 & 1 & 0 & 1 & 1 & 0 & 1 & 1 & 0 & 1 & 1 & 1 & 1 \\
6  & 0 & 1 & 1 & 0 & 1 & 1 & 0 & 1 & 1 & 0 & 1 & 1 & 1 & 1 & 0 & 1 & 1 & 1 \\
7  & 1 & 0 & 0 & 0 & 1 & 0 & 1 & 0 & 0 & 1 & 1 & 1 & 1 & 0 & 1 & 1 & 0 & 0 \\
8  & 0 & 1 & 0 & 1 & 0 & 0 & 0 & 1 & 0 & 1 & 1 & 0 & 1 & 1 & 1 & 0 & 1 & 0 \\
9  & 0 & 0 & 1 & 0 & 0 & 1 & 0 & 0 & 1 & 1 & 0 & 1 & 1 & 1 & 1 & 0 & 0 & 1 \\
10 & 1 & 0 & 0 & 0 & 1 & 0 & 1 & 0 & 0 & 0 & 1 & 1 & 0 & 1 & 1 & 0 & 1 & 1 \\
11 & 0 & 1 & 0 & 1 & 0 & 0 & 0 & 1 & 0 & 1 & 1 & 0 & 1 & 1 & 0 & 1 & 1 & 0 \\
12 & 0 & 0 & 1 & 0 & 0 & 1 & 0 & 0 & 1 & 1 & 1 & 0 & 1 & 0 & 0 & 1 & 0 & 1 \\
13 & 1 & 0 & 0 & 0 & 1 & 0 & 0 & 1 & 1 & 1 & 0 & 0 & 0 & 1 & 0 & 0 & 1 & 0 \\
14 & 0 & 1 & 0 & 1 & 0 & 0 & 1 & 0 & 0 & 0 & 1 & 0 & 1 & 1 & 1 & 0 & 1 & 1 \\
15 & 0 & 0 & 1 & 1 & 0 & 1 & 1 & 0 & 0 & 0 & 0 & 1 & 1 & 1 & 0 & 1 & 1 & 0 \\
16 & 1 & 0 & 0 & 1 & 0 & 0 & 1 & 0 & 0 & 1 & 0 & 1 & 1 & 0 & 0 & 1 & 0 & 1 \\
17 & 0 & 1 & 0 & 0 & 1 & 0 & 0 & 1 & 0 & 0 & 1 & 1 & 0 & 1 & 0 & 0 & 1 & 1 \\
18 & 1 & 1 & 0 & 0 & 0 & 1 & 0 & 0 & 1 & 1 & 1 & 1 & 0 & 0 & 1 & 1 & 1 & 1 \\
\bottomrule
\end{tabular} }
\end{table}

\begin{table}[H]
\centering
{
\caption{{True $Q$-matrices for $J_t = 30$ under sparse (left) and dense (right) attribute patterns (Time 1-3).}}
\label{Stab:qmatrix_j30_sparse_dense}
\begin{tabular}{c|ccc|ccc|ccc|ccc|ccc|ccc}
\toprule
\multirow{2}{*}{Item} &
\multicolumn{9}{c|}{\textbf{Sparse Matrix}} &
\multicolumn{9}{c}{\textbf{Dense Matrix}} \\
\cmidrule(lr){2-10} \cmidrule(lr){11-19}
& \multicolumn{3}{c|}{Time 1} & \multicolumn{3}{c|}{Time 2} & \multicolumn{3}{c|}{Time 3}
& \multicolumn{3}{c|}{Time 1} & \multicolumn{3}{c|}{Time 2} & \multicolumn{3}{c}{Time 3} \\
\cmidrule(lr){2-4} \cmidrule(lr){5-7} \cmidrule(lr){8-10}
\cmidrule(lr){11-13} \cmidrule(lr){14-16} \cmidrule(lr){17-19}
& $A_1$ & $A_2$ & $A_3$ & $A_1$ & $A_2$ & $A_3$ & $A_1$ & $A_2$ & $A_3$
& $A_1$ & $A_2$ & $A_3$ & $A_1$ & $A_2$ & $A_3$ & $A_1$ & $A_2$ & $A_3$ \\
\midrule
1  & 1 & 0 & 0 & 1 & 0 & 0 & 1 & 0 & 0 & 1 & 0 & 0 & 1 & 0 & 0 & 1 & 0 & 0 \\
2  & 0 & 1 & 0 & 0 & 1 & 0 & 0 & 1 & 0 & 0 & 1 & 0 & 0 & 1 & 0 & 0 & 1 & 0 \\
3  & 0 & 0 & 1 & 0 & 0 & 1 & 0 & 0 & 1 & 0 & 0 & 1 & 0 & 0 & 1 & 0 & 0 & 1 \\
4  & 1 & 1 & 0 & 1 & 1 & 0 & 1 & 0 & 1 & 1 & 1 & 1 & 1 & 1 & 1 & 1 & 0 & 1 \\
5  & 1 & 0 & 1 & 1 & 1 & 1 & 1 & 1 & 1 & 1 & 0 & 1 & 1 & 1 & 1 & 1 & 1 & 1 \\
6  & 0 & 1 & 1 & 0 & 1 & 0 & 0 & 1 & 1 & 0 & 1 & 1 & 0 & 1 & 0 & 0 & 1 & 1 \\
7  & 1 & 0 & 0 & 1 & 1 & 0 & 1 & 0 & 0 & 1 & 0 & 0 & 1 & 1 & 0 & 1 & 0 & 0 \\
8  & 0 & 0 & 1 & 0 & 0 & 1 & 0 & 1 & 0 & 0 & 1 & 1 & 0 & 1 & 1 & 0 & 1 & 0 \\
9  & 1 & 0 & 1 & 1 & 0 & 1 & 0 & 0 & 1 & 1 & 1 & 1 & 1 & 0 & 1 & 0 & 0 & 1 \\
10 & 1 & 1 & 1 & 1 & 1 & 1 & 1 & 1 & 1 & 1 & 0 & 1 & 1 & 1 & 1 & 1 & 1 & 1 \\
11 & 0 & 1 & 1 & 1 & 0 & 0 & 1 & 1 & 1 & 0 & 1 & 1 & 1 & 1 & 0 & 1 & 1 & 1 \\
12 & 1 & 0 & 0 & 1 & 0 & 0 & 1 & 0 & 0 & 0 & 1 & 1 & 1 & 0 & 0 & 1 & 1 & 1 \\
13 & 1 & 0 & 0 & 0 & 1 & 0 & 1 & 0 & 0 & 1 & 0 & 0 & 1 & 1 & 0 & 1 & 0 & 1 \\
14 & 0 & 1 & 0 & 0 & 0 & 1 & 0 & 1 & 0 & 0 & 1 & 0 & 1 & 0 & 1 & 0 & 1 & 1 \\
15 & 0 & 0 & 1 & 1 & 1 & 1 & 0 & 0 & 1 & 0 & 0 & 1 & 1 & 0 & 1 & 0 & 0 & 1 \\
16 & 1 & 0 & 1 & 1 & 0 & 0 & 1 & 0 & 1 & 1 & 1 & 1 & 1 & 0 & 0 & 1 & 1 & 1 \\
17 & 0 & 0 & 1 & 0 & 1 & 0 & 0 & 1 & 0 & 1 & 0 & 1 & 0 & 1 & 0 & 0 & 1 & 0 \\
18 & 1 & 0 & 1 & 0 & 0 & 1 & 1 & 0 & 1 & 1 & 0 & 1 & 0 & 0 & 1 & 1 & 0 & 1 \\
19 & 0 & 1 & 1 & 1 & 1 & 0 & 1 & 0 & 1 & 0 & 1 & 1 & 1 & 1 & 0 & 1 & 0 & 1 \\
20 & 0 & 1 & 1 & 1 & 1 & 0 & 1 & 0 & 1 & 1 & 1 & 1 & 1 & 1 & 0 & 1 & 0 & 1 \\
21 & 0 & 1 & 1 & 1 & 1 & 1 & 1 & 1 & 1 & 0 & 1 & 1 & 1 & 1 & 1 & 1 & 1 & 1 \\
22 & 1 & 0 & 0 & 1 & 0 & 0 & 0 & 1 & 0 & 1 & 1 & 0 & 0 & 1 & 0 & 0 & 1 & 0 \\
23 & 1 & 1 & 0 & 1 & 1 & 0 & 1 & 1 & 0 & 1 & 0 & 0 & 1 & 1 & 0 & 1 & 0 & 0 \\
24 & 1 & 0 & 1 & 0 & 0 & 1 & 1 & 0 & 1 & 1 & 0 & 1 & 0 & 1 & 1 & 1 & 0 & 1 \\
25 & 0 & 1 & 0 & 0 & 1 & 0 & 0 & 1 & 0 & 0 & 1 & 1 & 0 & 1 & 1 & 0 & 1 & 0 \\
26 & 0 & 0 & 1 & 0 & 0 & 1 & 1 & 0 & 1 & 1 & 0 & 1 & 0 & 1 & 1 & 1 & 0 & 1 \\
27 & 1 & 0 & 1 & 1 & 0 & 0 & 1 & 0 & 1 & 1 & 0 & 1 & 1 & 1 & 0 & 1 & 0 & 1 \\
28 & 0 & 1 & 0 & 0 & 1 & 0 & 0 & 1 & 0 & 0 & 1 & 0 & 0 & 1 & 0 & 0 & 1 & 0 \\
29 & 1 & 0 & 0 & 1 & 0 & 1 & 1 & 0 & 0 & 1 & 0 & 0 & 1 & 0 & 1 & 1 & 0 & 0 \\
30 & 1 & 0 & 1 & 1 & 0 & 0 & 1 & 0 & 1 & 1 & 0 & 1 & 1 & 1 & 0 & 1 & 1 & 1 \\
\bottomrule
\end{tabular}
}
\end{table}

\clearpage
\section{Supplementary Material E}

This section gives the simulation results when $Q$-matrix is known to show our recovery of parameters did not lost accuracy brings with $Q$-matrix is unknown. The attribute profile classification accuracy is provided in Table \ref{Stab:aar}. Estimation accuracy of item parameters ($g_{jt}$ and $s_{jt}$), initial covariate effects on attribute mastery ($\beta$) and attribute acquisition ($\gamma_{01}$) and attribute lossing ($\gamma_{10}$) are given in Tables \ref{Stab:gs_mae_rmse}-\ref{Stab:gamma}.

\begin{table}[!h]
\centering
\caption{{Average Attribute Recovery (AAR) across different numbers of items.}}
\label{Stab:aar}
{
\begin{tabular}{cc|ccc|ccc|ccc}
\toprule
 &  & \multicolumn{3}{c|}{Time 1} 
   & \multicolumn{3}{c|}{Time 2}
   & \multicolumn{3}{c}{Time 3} \\
\cmidrule(lr){3-5}\cmidrule(lr){6-8}\cmidrule(lr){9-11}
$N$ & $J$ 
& K1 & K2 & K3 
& K1 & K2 & K3 
& K1 & K2 & K3 \\
\midrule
200 & 6  
& 0.910 & 0.940 & 0.920 
& 0.910 & 0.950 & 0.925 
& 0.915 & 0.925 & 0.955 \\
200 & 18 
& 0.995 & 0.980 & 0.995 
& 1.000 & 0.995 & 1.000 
& 1.000 & 0.990 & 1.000 \\
200 & 30 
& 0.985 & 0.995 & 0.990 
& 1.000 & 0.995 & 1.000 
& 1.000 & 1.000 & 1.000 \\
\bottomrule
\end{tabular}}
\end{table}

\begin{table}[!h]
\centering
\caption{{Mean absolute error (MAE) for guessing ($g$) and slipping ($s$) parameters.}}
\label{Stab:gs_mae_rmse}
{\begin{tabular}{cc|ccc|ccc}
\toprule
 &  & \multicolumn{3}{c|}{$g$ MAE} 
   & \multicolumn{3}{c}{$s$ MAE} \\
\cmidrule(lr){3-5}\cmidrule(lr){6-8}
$N$ & $J$ 
& T1 & T2 & T3 
& T1 & T2 & T3 \\
\midrule
200 & 6  
& 0.030 & 0.040 & 0.060 
& 0.216 & 0.054 & 0.017 \\
200 & 18 
& 0.020 & 0.021 & 0.036 
& 0.048 & 0.025 & 0.019 \\
200 & 30 
& 0.018 & 0.031 & 0.033 
& 0.073 & 0.032 & 0.028 \\
\bottomrule
\end{tabular}}
\end{table}

\begin{table}[!h]
\centering
\caption{{Mean absolute error (MAE) of $\beta_0$ and $\beta_Z$ parameters.}}
\label{Stab:beta}
{\begin{tabular}{cc|ccc|ccc}
\toprule
 &  & \multicolumn{3}{c|}{$\beta_0$ MAE}
   & \multicolumn{3}{c}{$\beta_Z$ MAE} \\
\cmidrule(lr){3-5}\cmidrule(lr){6-8}
$N$ & $J$ & K1 & K2 & K3 & K1 & K2 & K3 \\
\midrule
200 & 6  
& 0.403 & 0.226 & 0.185 
& 0.403 & 0.226 & 0.185 \\
200 & 18 
& 0.158 & 0.136 & 0.109 
& 0.158 & 0.136 & 0.109 \\
200 & 30 
& 0.163 & 0.187 & 0.172 
& 0.163 & 0.187 & 0.172 \\
\bottomrule
\end{tabular}}
\end{table}

\begin{table}[!h]
\centering
\caption{{Mean absolute error (MAE) of transition parameters.}}
\label{Stab:gamma}
{
\begin{tabular}{cc|ccc|ccc}
\toprule
 &  & \multicolumn{3}{c|}{$\gamma_{01}^{12}$ MAE}
   & \multicolumn{3}{c}{$\gamma_{01}^{23}$ MAE} \\
\cmidrule(lr){3-5}\cmidrule(lr){6-8}
$N$ & $J$ & K1 & K2 & K3 & K1 & K2 & K3 \\
\midrule
200 & 6  
& 0.401 & 0.156 & 0.241 
& 0.234 & 0.248 & 0.192 \\
200 & 18 
& 0.076 & 0.195 & 0.165 
& 0.309 & 0.235 & 0.138 \\
200 & 30 
& 0.288 & 0.203 & 0.124 
& 0.200 & 0.352 & 0.169 \\
\bottomrule
\end{tabular}}
\end{table}

\clearpage
\section{Supplementary Material F}
For sparsity in the $Q$-matrix, 
\begin{align*}
Q_{jk} &\sim \mathrm{Bernoulli}(\theta), \\
\theta &\sim \mathrm{Beta}(\alpha,\beta),
\end{align*}
where the prior mean $\frac{\alpha}{\alpha+\beta}$ equals the non–zero proportion of the true $Q$-matrix and the concentration defined as $\alpha+\beta$. The sparse $Q$-matrix in simulations {included approximately 40-50\%} nonzero entries, and the dense $Q$-matrix includes about {56-60\%} nonzero entries. {For $J=6$, the dense condition was set to round 40\%, and for larger $J$, it was set to round 50\%.} Here, we examined the robustness of the model performance to alternative prior specifications for $\theta$, the sparsity parameter of the $Q$-matrix. {To examine sensitivity to the prior specification of the sparsity parameter, we conducted additional simulations under the smallest design condition ($N=200$, $J=6$). For sparse $Q$-matrices, the baseline prior mean was approximately 0.3, whereas for dense $Q$-matrices it was approximately 0.5. We varied the prior mean around these baseline values to represent mild prior misspecification. Specifically, prior means of 0.2, 0.3, and 0.4 were considered for sparse conditions, and 0.4, 0.5, and 0.6 for dense conditions. For each prior mean, the concentration parameter $c=\alpha+\beta$ was set to 5, 10, and 20 to assess sensitivity to prior strength. Under each setting, we evaluated $Q$-matrix recovery and item parameter estimation (Table~\ref{Stab:sensitivity_simulation_qgs}), as well as attribute classification performance (Table~\ref{Stab:sensitivity_simulation_aar}).
}

Across 25 replications under each prior condition, we assessed the recovery of the $Q$-matrix, the bias in the posterior means of item parameters ($g$ and $s$), and the attribute classification accuracy (Tables \ref{Stab:sensitivity_simulation_qgs} and \ref{Stab:sensitivity_simulation_aar}). Results showed that the changes in posterior estimates remained small across different prior settings, with absolute biases for $g$ and $s$ parameters generally below 0.04, with most differences below 0.02. The $Q$-matrix recovery rates ranged from 75\% to 100\%, and the attribute-level accuracy rates exceeded 84\% in most cases. These findings demonstrate that the model's estimation of $Q$-matrix, item parameters, and attribute profiles was robust to reasonable prior misspecification.
\begin{table}[htbp]
\centering
\caption{{Q-matrix recovery and item-parameter bias under alternative priors for the sparsity parameter $\theta$.\\
\textit{Note: PriMean = Prior mean ($\frac{\alpha}{\alpha+\beta}$); Conc. = Concentration ($\alpha+\beta$); $Q_{\text{Acc}}$ = recovery rate of Q-matrix; $g_{\text{MAE}}$ = mean absolute error of guessing; $s_{\text{MAE}}$ = mean absolute error of slipping.} }}
\label{Stab:sensitivity_simulation_qgs}
{
\begin{tabular}{ccc|ccc|ccc|ccc}
\toprule
\multicolumn{3}{c|}{\textbf{Prior}}
& \multicolumn{3}{c|}{\textbf{Time 1}}
& \multicolumn{3}{c|}{\textbf{Time 2}}
& \multicolumn{3}{c}{\textbf{Time 3}} \\
\cmidrule(lr){1-3}
\cmidrule(lr){4-6}
\cmidrule(lr){7-9}
\cmidrule(lr){10-12}
PriMean & Conc. & $\mathrm{Beta}(\alpha,\beta)$
& $Q_{\text{Acc}}$ & $g_{\text{MAE}}$ & $s_{\text{MAE}}$
& $Q_{\text{Acc}}$ & $g_{\text{MAE}}$ & $s_{\text{MAE}}$
& $Q_{\text{Acc}}$ & $g_{\text{MAE}}$ & $s_{\text{MAE}}$ \\
\midrule
\multicolumn{12}{l}{\textit{Sparse $Q$ matrices}}\\
\midrule
0.2 & 10 & $\mathrm{Beta}(2,8)$ 
& 0.957 & 0.033 & 0.150
& 0.920 & 0.060 & 0.080
& 0.907 & 0.104 & 0.059 \\
0.2 & 15 & $\mathrm{Beta}(3,12)$
& 0.958 & 0.032 & 0.158
& 0.917 & 0.053 & 0.076
& 0.944 & 0.079 & 0.047 \\
0.2 & 20 & $\mathrm{Beta}(4,16)$
& 0.963 & 0.030 & 0.154
& 0.926 & 0.052 & 0.067
& 0.938 & 0.082 & 0.057 \\
0.3 & 10 & $\mathrm{Beta}(3,7)$
& 0.963 & 0.034 & 0.151
& 0.932 & 0.053 & 0.070
& 0.932 & 0.078 & 0.054 \\
0.3 & 15 & $\mathrm{Beta}(9,21)$
& 0.933 & 0.037 & 0.183
& 0.933 & 0.052 & 0.068
& 0.944 & 0.083 & 0.052 \\
0.3 & 20 & $\mathrm{Beta}(6,14)$
& 0.961 & 0.032 & 0.155
& 0.911 & 0.054 & 0.077
& 0.933 & 0.083 & 0.056 \\
0.4 & 10 & $\mathrm{Beta}(4,6)$
& 0.972 & 0.032 & 0.151
& 0.933 & 0.058 & 0.074
& 0.933 & 0.084 & 0.055 \\
0.4 & 15 & $\mathrm{Beta}(6,9)$
& 0.972 & 0.037 & 0.196
& 0.917 & 0.039 & 0.079
& 0.944 & 0.096 & 0.048 \\
0.4 & 20 & $\mathrm{Beta}(8,12)$
& 0.944 & 0.033 & 0.157
& 0.928 & 0.056 & 0.078
& 0.939 & 0.079 & 0.053 \\
\bottomrule
\end{tabular}}
\end{table}

\begin{table}[htbp]
\centering
\caption{{Attribute-level accuracy rate (AAR) under varying $\theta$ priors.\\
\textit{Note: PriMean $=\frac{\alpha}{\alpha+\beta}$; Conc. $=\alpha+\beta$; $Tt$ = Time $t$; $Ak$ = Attribute $k$. Values are percentages; bold indicates the column-wise maximum.}}}
\label{Stab:sensitivity_simulation_aar}
{
\begin{tabular}{ccc|ccc|ccc|ccc}
\toprule
\multicolumn{3}{c|}{\textbf{Prior}} & \multicolumn{9}{c}{\textbf{AAR (\%)}} \\
\cmidrule(lr){1-3}\cmidrule(lr){4-12}
PriMean & Conc. & $\mathrm{Beta}(\alpha,\beta)$
& \multicolumn{3}{c|}{T1} & \multicolumn{3}{c|}{T2} & \multicolumn{3}{c}{T3} \\
\cmidrule(lr){4-6}\cmidrule(lr){7-9}\cmidrule(lr){10-12}
& & & A1 & A2 & A3 & A1 & A2 & A3 & A1 & A2 & A3 \\
\midrule
0.2 & 10 & $\mathrm{Beta}(2,8)$
& 91.2 & \textbf{91.6} & 89.6
& 92.8 & 93.0 & 88.8
& 94.3 & 92.2 & 95.1 \\
0.2 & 15 & $\mathrm{Beta}(3,12)$
& 91.6 & 90.8 & 89.4
& 93.0 & 93.2 & 89.7
& 94.1 & \textbf{92.6} & \textbf{95.6} \\
0.2 & 20 & $\mathrm{Beta}(4,16)$
& 91.5 & \textbf{91.6} & 89.4
& 93.5 & 93.2 & 89.6
& 94.2 & 92.2 & \textbf{95.6} \\
0.3 & 10 & $\mathrm{Beta}(3,7)$
& 91.6 & \textbf{91.6} & 89.4
& \textbf{93.6} & 93.0 & 89.6
& \textbf{94.4} & 92.2 & \textbf{95.7} \\
0.3 & 20 & $\mathrm{Beta}(6,14)$
& \textbf{92.4} & 90.5 & \textbf{90.8}
& 93.2 & 93.2 & 89.2
& 93.8 & 92.5 & 95.8 \\
0.3 & 30 & $\mathrm{Beta}(9,21)$
& 91.6 & \textbf{91.7} & 89.7
& 92.9 & 92.9 & \textbf{90.7}
& \textbf{94.5} & 92.1 & \textbf{95.6} \\
0.4 & 10 & $\mathrm{Beta}(4,6)$
& 91.8 & 91.4 & 89.7
& 93.1 & 93.3 & 90.2
& 94.2 & 92.0 & 95.8 \\
0.4 & 15 & $\mathrm{Beta}(6,9)$
& 91.4 & 90.9 & 88.0
& 93.2 & 92.9 & 89.4
& 94.4 & 91.9 & \textbf{95.6} \\
0.4 & 20 & $\mathrm{Beta}(8,12)$
& 92.2 & 89.8 & 89.5
& 92.9 & \textbf{93.5} & 87.8
& 94.3 & 91.9 & \textbf{95.9} \\
\bottomrule
\end{tabular}}
\end{table}

\clearpage
\section{Supplementary Material G}
Table \ref{Stab:gs_bias_dense} includes the estimation bias of item parameters for dense $Q$-matrix.

\begin{table}[!h]
\centering
{
\caption{{Estimation accuracy of item parameters ($g_{jt}$ and $s_{jt}$), evaluated by MAE and RMSE, under Dense $Q$-matrix, sample sizes ($N$), and number of items ($J_t$).}}
\label{Stab:gs_bias_dense}
\centering
\begin{tabular}[t]{cccc|cc|cc}
\toprule
$N$ & $J_t$ & $T$ & Sparsity & $g_{\text{MAE}}$ (SE) & $g_{\text{RMSE}}$ (SE) & $s_{\text{MAE}}$ (SE) & $s_{\text{RMSE}}$ (SE) \\
\midrule
200 & 6  & 1 & Dense & .029 (.002) & .036 (.002) & .208 (.012) & .260 (.012) \\
    &    & 2 &       & .045 (.003) & .059 (.003) & .077 (.006) & .110 (.006) \\
    &    & 3 &       & .059 (.004) & .079 (.004) & .051 (.004) & .071 (.004) \\
\addlinespace[0.3em]
    & 18 & 1 &       & .023 (.002) & .030 (.002) & .065 (.008) & .095 (.008) \\
    &    & 2 &       & .028 (.003) & .038 (.003) & .035 (.004) & .047 (.004) \\
    &    & 3 &       & .029 (.003) & .037 (.003) & .023 (.002) & .030 (.002) \\
\addlinespace[0.3em]
    & 30 & 1 &       & .026 (.002) & .035 (.002) & .063 (.005) & .090 (.005) \\
    &    & 2 &       & .040 (.003) & .054 (.003) & .035 (.002) & .045 (.002) \\
    &    & 3 &       & .039 (.003) & .054 (.003) & .023 (.001) & .029 (.001) \\
\midrule
400 & 6  & 1 & Dense & .023 (.001) & .031 (.001) & .153 (.007) & .198 (.007) \\
    &    & 2 &       & .040 (.002) & .055 (.002) & .062 (.004) & .092 (.004) \\
    &    & 3 &       & .065 (.003) & .084 (.003) & .049 (.003) & .073 (.003) \\
\addlinespace[0.3em]
    & 18 & 1 &       & .018 (.001) & .022 (.001) & .046 (.003) & .066 (.003) \\
    &    & 2 &       & .021 (.001) & .028 (.001) & .024 (.001) & .032 (.001) \\
    &    & 3 &       & .021 (.001) & .028 (.001) & .017 (.001) & .021 (.001) \\
\addlinespace[0.3em]
    & 30 & 1 &       & .024 (.002) & .035 (.002) & .054 (.004) & .085 (.004) \\
    &    & 2 &       & .038 (.002) & .054 (.002) & .023 (.001) & .029 (.001) \\
    &    & 3 &       & .026 (.002) & .035 (.002) & .017 (.001) & .022 (.001) \\
\midrule
600 & 6  & 1 & Dense & .019 (.001) & .027 (.001) & .132 (.005) & .178 (.005) \\
    &    & 2 &       & .035 (.001) & .048 (.002) & .070 (.003) & .103 (.004) \\
    &    & 3 &       & .056 (.002) & .077 (.003) & .039 (.002) & .060 (.002) \\
\addlinespace[0.3em]
    & 18 & 1 &       & .015 (.001) & .020 (.001) & .039 (.002) & .057 (.002) \\
    &    & 2 &       & .020 (.001) & .027 (.001) & .018 (.001) & .023 (.001) \\
    &    & 3 &       & .019 (.001) & .025 (.001) & .014 (.001) & .018 (.001) \\
\addlinespace[0.3em]
    & 30 & 1 &       & .021 (.001) & .032 (.001) & .041 (.001) & .058 (.001) \\
    &    & 2 &       & .036 (.001) & .054 (.001) & .019 (.000) & .025 (.000) \\
    &    & 3 &       & .025 (.001) & .037 (.001) & .015 (.000) & .019 (.000) \\
\bottomrule
\end{tabular}
}
\end{table}

\clearpage
\section{Supplementary Material H}

\begin{figure}[ht]
  \caption{Root Mean Squared Errors (RMSE) of item-level guessing and slipping parameter estimates. Panels A and B correspond to the guessing parameters under conditions with $(N=200, J=6)$ and $(N=600, J=30)$, respectively. Panels C and D show the slipping parameters under the same conditions.}
  \label{Sfig:rmse_panel}
  \centering
  \includegraphics[width=\textwidth,
                   height=0.8\textheight,
                   keepaspectratio]{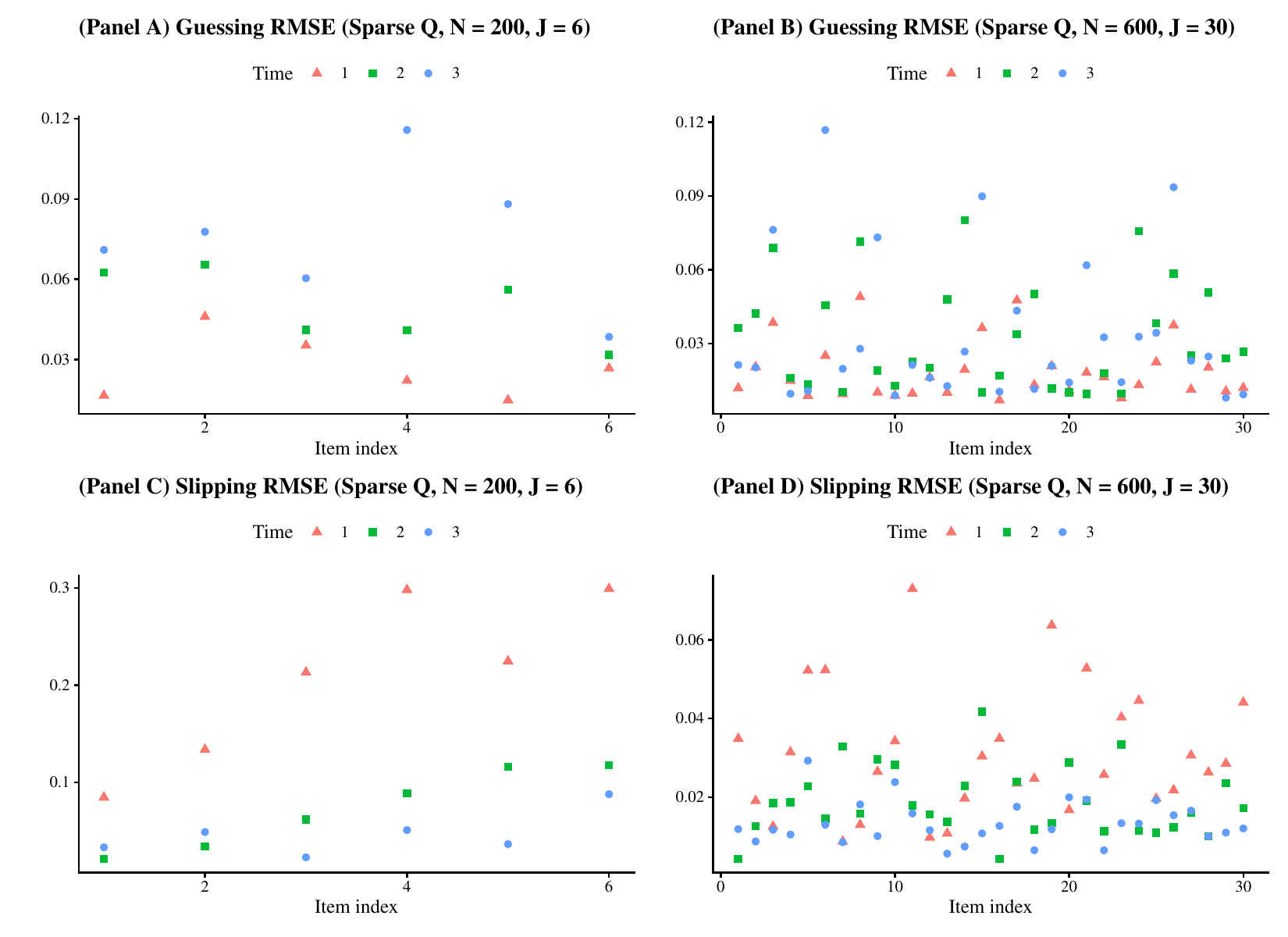}
\end{figure}

The detailed item parameters for the guessing ($g$) and slipping ($s$) parameters are presented in Figure \ref{Sfig:rmse_panel}. {The figure reports the root mean squared error (RMSE) for each item.}

Panels A and B display the item-level RMSE of guessing parameters from two test conditions: $(N=200, J=6)$ and $(N=600, J=30)$, respectively. Panels C and D show the corresponding RMSEs for slipping parameters under the same conditions. Triangular, square{, and circular} markers represent time points $T=1$, $T=2$, {and $T = 3$,} respectively.

\begin{figure}[ht]
  \centering
  \caption{RMSE of item-level guessing and slipping parameter estimates under different $\theta$ values. Panels A and B show guessing RMSEs for $\theta=0.5$ and $\theta=0.7$, respectively. Panels C and D show slipping RMSEs under the same conditions.}
  \label{Sfig:rmse_theta_panel}
  \includegraphics[width=\textwidth,
                   height=0.8\textheight,
                   keepaspectratio]{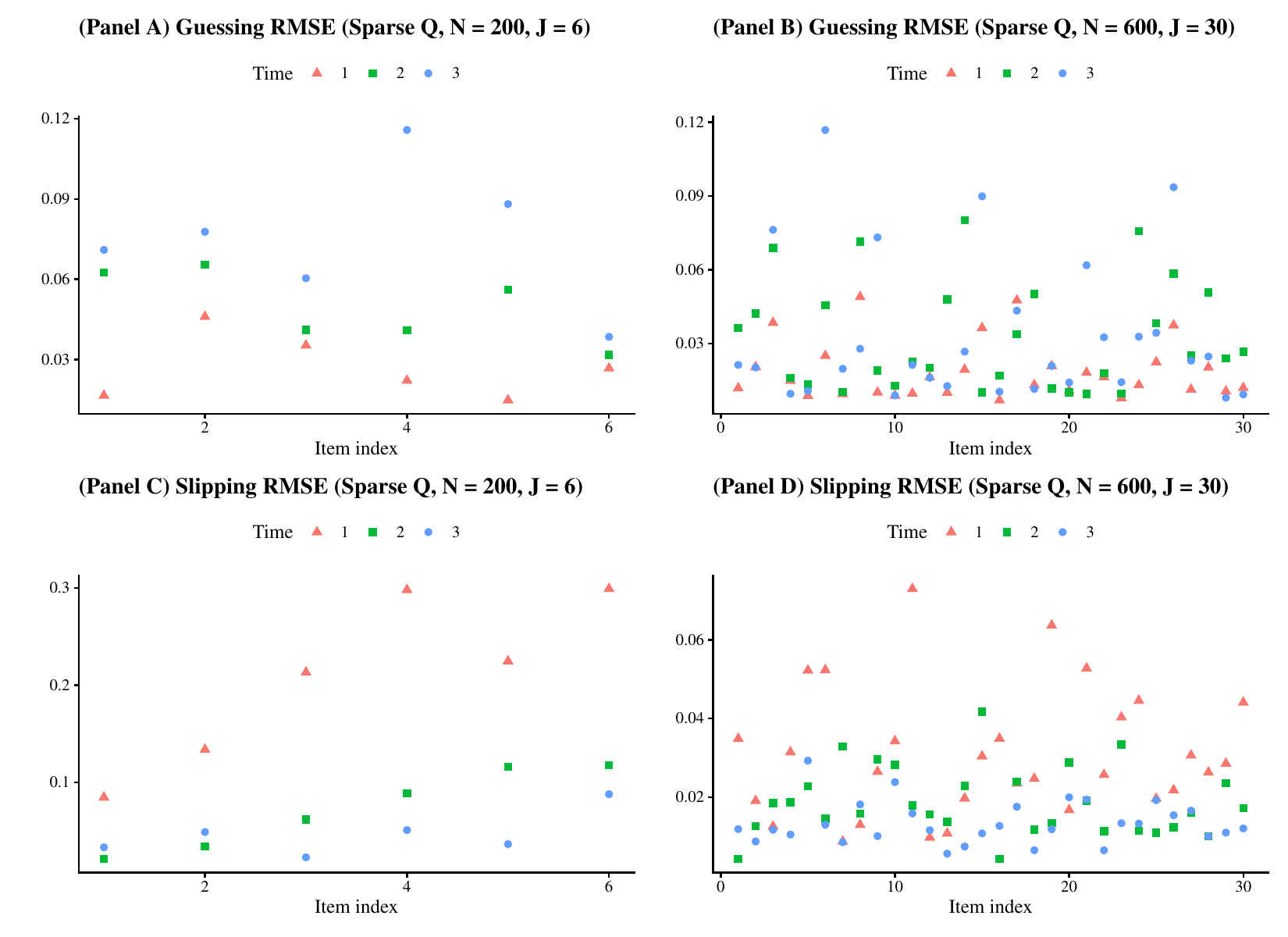}
\end{figure}

Figure~\ref{Sfig:rmse_theta_panel} compares the item-level RMSE of the estimated guessing and slipping parameters under two different prior settings for $\theta$ {(sparse and dense)}, with fixed test conditions $(N = 600, J = 30)$. Panels A and B present the RMSEs for guessing parameters, while Panels C and D present the results for slipping parameters. The triangular{, square, and circular} markers indicate estimates at time points $T=1${, $T=2$, and $T=3$} respectively. The comparison shows how prior informativeness impacts parameter recovery, particularly in terms of consistency across items and time points.

\clearpage

\clearpage
\section{Supplementary Material I}
{Under the sparse $Q$, we report 95\% credible intervals under the smallest condition ($N=200$, $J_t=6$) and the largest condition ($N=600$, $J_t=30$). For the first attribute, we present intervals for the intercept ($\beta_{0,1}$) and for the first covariate effect ($\beta_{Z,11}$), together with the corresponding transition effects $\gamma_{01,1}^{12}$ and $\gamma_{01,1}^{23}$ for the same covariate. In addition, to illustrate uncertainty in item parameters, we report credible intervals for the guessing and slipping parameters of two representative items: item 1 (a single-attribute item) and item 4 (a multiple-attribute item), at each time point ($t=1,2,3$).}

\begin{table}[H]
\centering
\caption{{95\% Credible intervals for selected parameters under sparse $Q$}}
\label{Stab:ci_sparse_small_large}
{
\begin{tabular}{lccc|ccc}
\toprule
\multirow{2}{*}{Parameter} 
& \multicolumn{3}{c|}{\textbf{$(N, J_t) = (200, 6)$}} 
& \multicolumn{3}{c}{\textbf{$(N, J_t) = (600, 30)$}} \\
\cmidrule(lr){2-4} \cmidrule(lr){5-7}
& Lower & Mean & Upper & Lower & Mean & Upper \\
\midrule
\multicolumn{7}{c}{\textit{Initial and Transition Regression Parameters (Attribute 1)}} \\
\midrule
$\beta_{0,1}$      
& -1.60 & -0.943 & -0.316 
& -1.53 & -1.15  & -0.814 \\
$\beta_{Z,11}$      
& -1.38 & -0.802 & -0.286  
& -0.927 & -0.578 & -0.230  \\
$\gamma_{01,1}^{12}$ 
& -1.65 & -0.880 & -0.183 
& -1.14 & -0.743 & -0.370 \\
$\gamma_{01,1}^{23}$ 
& -1.02 & -0.118 &  0.740  
& -0.582 & -0.142 &  0.285 \\
\midrule
\multicolumn{7}{c}{\textit{Item Parameters (Item 1)}} \\
\midrule
$g_{1,1}$ 
& 0.075 & 0.146 & 0.231 
& 0.103  & 0.151 & 0.209 \\
$g_{2,1}$ 
& 0.060 & 0.157 & 0.285 
& 0.083 & 0.137 & 0.204 \\
$g_{3,1}$ 
& 0.075 & 0.210 & 0.376 
& 0.075 & 0.139 & 0.222 \\
$s_{1,1}$ 
& 0.039 & 0.173 & 0.329 
& 0.084 & 0.149 & 0.230 \\
$s_{2,1}$ 
& 0.082 & 0.164 & 0.253 
& 0.087 & 0.133 & 0.189 \\
$s_{3,1}$ 
& 0.092 & 0.147 & 0.212 
& 0.086 & 0.127 & 0.176 \\
\midrule
\multicolumn{7}{c}{\textit{Item Parameters (Item 4)}} \\
\midrule
$g_{1,4}$ 
& 0.096 & 0.150 & 0.230 
& 0.086 & 0.123 & 0.166 \\
$g_{2,4}$ 
& 0.091 & 0.196 & 0.313 
& 0.094 & 0.138 & 0.189 \\
$g_{3,4}$ 
& 0.088 & 0.226 & 0.404 
& 0.088 & 0.143 & 0.208 \\
$s_{1,4}$ 
& 0.068 & 0.396 & 0.787 
& 0.037 & 0.148 & 0.337 \\
$s_{2,4}$ 
& 0.043 & 0.208 & 0.451 
& 0.080 & 0.142 & 0.221 \\
$s_{3,4}$ 
& 0.069 & 0.177 & 0.339 
& 0.084 & 0.130 & 0.187 \\
\bottomrule
\end{tabular}}
\end{table}

\section{Appendix}

See the file \textit{Game content} for the full table of participant content codes across 12 items.


\end{document}